\documentclass[prd, aps, twocolumn, showpacs,superscriptaddress,nofootinbib]{revtex4-1}

\usepackage{amsmath}
\usepackage{amsfonts}
\usepackage{amssymb}	
\usepackage{graphicx}
\usepackage{bm}
\usepackage{color}
\usepackage[utf8]{inputenc}
\usepackage[normalem]{ulem}
\usepackage[dvipsnames]{xcolor}

\usepackage{hyperref}

\def\lm{{\ell m}}   

\newcommand{\mc}[1]{\mathcal{#1}}

\begin{document}

\title{Factorization and resummation: A new paradigm to improve\\ 
gravitational wave amplitudes. III: the spinning test-body terms.}

\author{Alessandro \surname{Nagar}}
\affiliation{Centro Fermi - Museo Storico della Fisica e Centro Studi e Ricerche ``Enrico Fermi'', 00184 Roma, Italy}
\affiliation{INFN Sezione di Torino, Via P.~Giuria 1, 10125 Torino, Italy}
\affiliation{Institut des Hautes Etudes Scientifiques, 91440 Bures-sur-Yvette, France}
\author{Francesco \surname{Messina}}
\affiliation{Dipartimento di Fisica, Università degli studi di Milano Bicocca, Piazza della Scienza 3, 20126 Milano, Italy}
\affiliation{INFN, Sezione di Milano Bicocca, Piazza della Scienza 3, 20126 Milano, Italy}
\author{Chris \surname{Kavanagh}}
\affiliation{
Max Planck Institute for Gravitational Physics (Albert Einstein Institute),
Am M\"uhlenberg 1,
Potsdam 14476,
Germany}
\affiliation{Institut des Hautes Etudes Scientifiques, 91440 Bures-sur-Yvette, France}
\author{Georgios \surname{Lukes-Gerakopoulos}}
\affiliation{Astronomical Institute of the Academy of Sciences of the Czech Republic, Bo\v{c}n\'{i} II 1401/1a, CZ-141 00 Prague, Czech Republic}
\author{Niels \surname{Warburton}}
\affiliation{School of Mathematics and Statistics, University College Dublin, Belfield, Dublin 4, Ireland}
\author{Sebastiano \surname{Bernuzzi}}
\affiliation{Theoretisch-Physikalisches Institut,  Friedrich-Schiller-Universit{\"a}t Jena, 07743, Jena, Germany}
\author{Enno \surname{Harms}}
\affiliation{Theoretisch-Physikalisches Institut,  Friedrich-Schiller-Universit{\"a}t Jena, 07743, Jena, Germany}

\begin{abstract}

  We present new calculations of the energy flux of a spinning test-body on circular orbits
  around a Schwarzschild black hole at linear order in the particle
  spin. We compute the multipolar fluxes up 
  to $\ell=m=6$ using two independent numerical solvers of the
  Teukolsky equation, one in the time domain and the 
  other in the frequency domain. After linearization in the spin of the particle, we 
  obtain an excellent agreement ($\sim 10^{-5}$) between the two numerical results.
  The calculation of the multipolar fluxes is also performed analytically (up to $\ell=7$)
  using the post-Newtonian (PN) expansion of the Teukolsky equation solution;
  each mode is obtained at 5.5PN order beyond the corresponding 
  leading-order contribution. From the analytical fluxes we obtain the PN-expanded analytical waveform amplitudes. 
  These quantities are then resummed using new procedures either
  based on the factorization of 
  the orbital contribution (and resumming it independently from the
  spin-dependent factor) or on the factorization of the tail
  contribution solely for odd-parity multipoles. 
  We compare these prescriptions and the resummation procedure proposed in Pan et
  al.~\href{https://doi.org/10.1103/PhysRevD.83.064003}{[Phys.~Rev.~D~83~(2011)~064003]}
  to the numerical data.
  We find that the new procedures significantly improve over the
  existing one that, notably, is inconsistent with the numerical data
  for $\ell+m=\text{odd}$ multipoles already at low orbital frequencies.
  Our study suggests that the approach to waveform resummation 
  used in current effective-one-body-based waveform models should be modified to improve its robustness and
  accuracy all over the binary parameter space.   
\end{abstract}
\date{\today}
\maketitle

\section{Introduction}

Test-mass results have been crucial to devise robust resummation techniques
for the truncated post-Newtonian expansion that give access to analytical gravitational waveform 
and fluxes for circularized, nonprecessing, binaries~\cite{Damour:1997ub,Damour:2008gu,Pan:2010hz,Nagar:2016ayt,Messina:2018ghh}.
Such resummed waveform, and related fluxes, are one of the crucial building blocks of 
effective-one-body (EOB) waveform models for coalescing relativistic 
binaries~\cite{Nagar:2018plt,Akcay:2018yyh,Nagar:2018zoe,Cotesta:2018fcv,Hinderer:2017jcs,Bohe:2016gbl}.
Up to now, resummation of PN-expanded analytical result is the only approach that can be
adopted to improve the behavior of the PN-expansions in the strong-field, fast velocity
regime up to  merger~\cite{Damour:2009kr}. From the very beginning of this endeavor~\cite{Damour:1997ub}
the development (and testing) of resummation techniques has been driven by comparisons
between some analytically resummed waveform and numerical waveforms (or fluxes) generated
by a {\it nonspinning} particle inspiralling and plunging into a Schwarzschild or a 
Kerr black hole~\cite{Damour:2007xr,Harms:2014dqa}. By contrast, none of the resummation
approaches routinely used in state-of-the-art EOB models~\cite{Nagar:2018zoe,Bohe:2016gbl}
has been tested in the special case where the particle (which models a test black-hole)
is spinning. This was not done up to now for at least two reasons: (i) on the one hand,
robust and accurate numerical computations of the energy fluxes from a spinning particle
on circular orbits around a Kerr black became available only recently~\cite{Lukes-Gerakopoulos:2017vkj,Harms:2016ctx,Harms:2015ixa}; 
(ii) on the other hand, the analytical PN knowledge of the fluxes of a spinning particle around a Kerr
black hole was only known at global 2.5PN order~\cite{Tanaka:1996ht} and only recently pushed 
to 3.5PN accuracy~\cite{Cotesta:2018fcv}.
This paper builds on previous works and improves them along two directions: (i) the numerical fluxes
of Refs.~\cite{Harms:2015ixa,Harms:2016ctx} are recomputed, in the time-domain, at an improved 
accuracy and increasing the number of multipoles. In addition, they are compared with an analogous 
calculation performed with a completely independent numerical code in the frequency domain, 
finding excellent consistency between the two methods once the results are linearized in the particle
spin; (ii) though we here only consider the case of a spinning particle around a Schwarzschild
black hole, the 2.5PN accurate results of Ref.~\cite{Tanaka:1996ht} are pushed to much higher PN order,
namely {\it relative} 5.5PN accuracy for all multipoles of the flux up to $\ell=7$. The availability of such 
new PN information, at high order, allows us to extensively test the standard waveform resummation 
techniques of Refs.~\cite{Damour:1997ub,Damour:2008gu,Pan:2010hz} in a corner of the binary 
parameter space that had not been covered before. Similarly, we use these new numerical data to
check new resummation approaches proposed recently in Refs.~\cite{Nagar:2016ayt,Messina:2018ghh}
and that are going to be partly incorporated in the next generation of EOB waveform
models~\cite{Riemenschneider:2019}.

The paper is organized as follows. In Sec.~\ref{sec:dEdt} we summarize
the analytical and numerical approaches to compute the gravitational
wave fluxes from a spinning particle around a Schwarzschild black hole.
From the energy fluxes, decomposed in multipoles, we obtain the gravitational
waveform amplitudes, both numerically and analytically. The aim of Sec.~\ref{sec:resum}
is to compare the numerical waveform amplitudes with several analytical representations,
either in PN-expanded form or using some resummation technique. Concluding remarks
are reported in Sec.~\ref{sec:conclusions}. The paper is completed
by technical Appendixes that explicitly report the outcome of the PN calculations.
Throughout this work we use geometrized units such that $G=c=1$.
We also define $M$ as the mass of the primary black-hole and $\mu$ as the mass
of the secondary black hole such that $M\gg\mu$.

\section{Energy fluxes emitted by a spinning particle around a Schwarzschild black hole}
\label{sec:dEdt}

In this section we consider the energy flux radiated in gravitational waves by
a spinning particle on a circular orbit of radius $r_0$ around a Schwarzschild black hole. 
We will restrict our attention to the case where the particle's spin axis is aligned with the orbital angular momentum.
We compute the radiated flux both analytically, via high-order post-Newtonian calculations,
and numerically, using two independent approaches.

\subsection{Post-Newtonian results}

\subsubsection{Dynamics of a spinning particle}

The equations of motion of a spinning test body moving on a curved
spacetime background (Schwarzschild) are given
by the Mathisson-Papapetrou-Dixon equations (MPD)~\cite{Mathisson:1937zz, Papapetrou:1951pa, Dixon:1970zza}
\begin{align}\label{eq:MPD}
	\frac{{\rm D} p^{\mu}}{d \tau}  = &
	- \frac12 \, R^\mu{}_{\nu \alpha \beta} \, u^\nu \, S^{\alpha \beta}
	\,,
	\\
	\frac{{\rm D}S^{\mu\nu}}{d \tau}  = & 
	2 \, p^{[\mu} u^{\nu]}
	\,,
\end{align}
where $p^\mu$ is the four-momentum, $u^\mu$ is the four-velocity ($\tau$ is the proper time), $R^\mu{}_{\nu \alpha \beta}$ is the Riemann tensor of the spacetime and $S^{\alpha\beta}$ is the spin-tensor.
From the spin tensor we can define the spin magnitude, $s = (\frac12 S_{\mu\nu}S^{\mu\nu})^{1/2}$,
from which we define the spin variable
\begin{align}
  \label{eq:def_sigma}
	\sigma\equiv \dfrac{s}{\mu M}\, .
\end{align}
As in previous work~\cite{Lukes-Gerakopoulos:2017vkj,Harms:2016ctx,Harms:2015ixa},
we consider $-1\leq \sigma \leq 1$. We will comment in Appendix~\ref{sec:Spint} on
the meaning of these limits and on the interpretation of the test-mass results
presented here as part of the general case where the masses of the two bodies are comparable.
The MPD equations do not form a closed systems of evolution equations and a closure,
called spin supplementary condition (SSC), is required. The choice of a SSC amounts to choosing
a centre of mass for the spinning body \cite{Costa:2014nta}. For astrophysically relevant values of the spin,
there are strong indications that any physically meaningful quantity will not depend upon the
SSC choice~\cite{Harms:2016ctx,Lukes-Gerakopoulos:2017vkj}.
In this paper we choose the Tulczyjew-Dixon condition $S^{\mu\nu} p{}_\nu=0$~\cite{Tulczyjew:1965iez}.

Throughout this section, we will work to linear order in $\sigma$ for
which we have $p^{\alpha} = \mu u^\alpha +\mc{O}(\sigma^2)$. The orbital
frequency is then given by~\cite{Tanaka:1996ht}
\begin{align}\label{eq:orb_freq}	
	\Omega 		&= u^{3/2}\left(1-\frac32{\sigma}u^{3/2}\right) 			
\end{align}
where $u = M/r_0$ and $r_0$ is the radius of a circular orbit expressed in
Schwarzschild coordinates.

\subsubsection{Calculation of the energy fluxes}
\label{sec:TEUKpn}

In the Teukolsky approach, the problem of computing the energy and angular momentum
fluxes at infinity from perturbations to either a Kerr or Schwarzschild black hole
has been well laid out in the literature. Since the main aim of this work is to use
the output of such calculations, we will only give an overview of the main 
ideas and refer the reader to Misao Sasaki and Hideyuki Tagoshi's
living review~\cite{Sasaki:2003xr} for an in-depth discussion of the topic. 

We begin with the spin-weight $s=-2$ Teukolsky equation
\begin{align}\label{eq:teuk}
  \mc{O}{}{}_{-2}\psi(t,r,\theta,\phi)=4\pi T,
\end{align}
where $\mc{O}$ is a second-order differential operator and $T$ is
formed from the stress energy of the perturbation $T_{\mu\nu}$. When
working in the frequency domain, the problem is simplified greatly,
and by separating
${}_{-2}\psi$
into radial and angular components, namely
\begin{align}
  {}_{-2}\psi =\frac{1}{2\pi}\int e^{-i \omega t}\psi_{\ell m }(r){}_{-2}S_{\ell m}(\theta,\phi)d\omega,
\end{align}
where $_{-2}S_{\ell m}(\theta,\phi)$ are the spin-weighted spheroidal harmonics which reduce to the spin-weighted spherical harmonics in the Schwarzschild limit. The radial functions $\psi_{\ell m }(r)$ then satisfy a second order ordinary differential equation
\begin{align}
	\mc{D}\psi_{\ell m}(r)=4 \pi T_{\ell m}(r).
\end{align}
In the case of a spinning particle on a circular orbit around a Schwarzschild black hole, the source term here takes the schematic form
\begin{align}\label{eq:teuk:src}
	 T_{\ell m}=&A_0 \delta(r-r_p)+A_1 \delta'(r-r_p)+A_2 \delta''(r-r_p) \nonumber \\
	 \qquad &+A_3 \delta'''(r-r_p).
\end{align}
If we have a pair of homogeneous solutions satisfying retarded boundary conditions at the horizon and at infinity, $R_{\ell m}^{-}$ and $R_{\ell m}^+$ respectively, then the general solution to this equation is given by
\begin{align}
	\psi_{\ell m}(r)=C_{\ell m}^{+}R_{\ell m}^+(r) + C_{\ell m}^- R_{\ell m}^-(r),
\end{align}
where
\begin{align}
C_{\ell m}^\pm&=\frac{1}{W}\int R_{\ell m}^{\mp}T_{\ell m} \Delta^{-2}dr \nonumber \\
&=B_0 R_{\ell m}^{\mp}(r_p)+B_1 R_{\ell m}^{\mp}{}'(r_p)+B_2 R_{\ell m}^{\mp}{}''(r_p) \nonumber \\
&\qquad+B_3 R_{\ell m}^{\mp}{}'''(r_p). \label{Eq:NormConst}
\end{align}
The $B_i=B_i(r_p,\sigma)$ are determined by completing the integral, $W$ is the
invariant Wronskian of the two solutions and $\Delta=r(r-2 M)$. 
Normalizing the homogeneous solutions so that
$R_+\sim r^3e^{i \omega r^*}$ as $r\rightarrow\infty$, the energy flux emitted at radial infinity is given simply by
\begin{align}
	\frac{dE^\infty}{dt}=\sum_{\ell m} F_\lm =\sum_{\ell m}\frac{|C_{\ell m}^{+}|^2}{4 \pi m^2\Omega^2} \label{Eq:FluxFormula}
\end{align}
where $\Omega$ is the orbital frequency given in Eq.~\eqref{eq:orb_freq}.

We now briefly describe the calculation of the $F_{\lm}$ functions appearing in Eq.~\eqref{Eq:FluxFormula} as a PN expansion, i.e
an expansion for large orbital radius and small frequency. This calculation
was first presented for a spinning particle on a circular orbit by
Tanaka et al.~\cite{Tanaka:1996ht}, though truncated at 2.5PN order.
We will extend their calculation to a higher order in the expansion.
To evaluate Eq.~\eqref{Eq:NormConst} we first need PN expansions for the homogeneous
radial functions $R^{\pm}_{\ell m}$.  The well established method for systematically
computing these uses the solutions of Mano, Suzuki and Takasugi (MST) ~\cite{Mano:1996vt,Mano:1996mf}. 
The MST solutions to the homogeneous radial Teukolsky equation are given as infinite series of either hypergeometric ${}_2F_1$ functions or irregular confluent 
hypergeometric $U$ functions. In the PN limit, restricting the frequencies to the allowed harmonics of the orbital frequency, these infinite series truncate at 
finite orders and, with modern algebraic software, are methodically Taylor expanded to a desired PN order. For in depth discussion of our calculations see \cite{Kavanagh:2016idg}. With these in hand the evaluation of  Eq.~\eqref{Eq:NormConst} 
is straightforwardly accomplished by also Taylor expanding the $B_i$'s, thus giving PN expansions for the energy flux modes. At each order in the PN expansion we also work to linear order in the spin $\sigma$.
In practice, we found it easy to consider modes up to $\ell=7$ and obtain each multipolar flux
at 5.5PN accuracy {\it beyond} the leading-order contribution.
More precisely, each PN-expanded multipolar contribution is factorized as
\begin{equation}
	F_\lm = F_\lm^{(N,\epsilon)}\hat{F}_\lm \ ,
\end{equation}
where $F_\lm^{(N,\epsilon)}$ is the Newtonian (or leading-order) prefactor, while $\hat{F}_\lm$ is the PN correction. Here, $\epsilon=0,1$ depending
on the parity of $\ell+m$.  For each $(\ell,m)$, $\hat{F}_\lm$ is given by 
a 5.5PN-accurate polynomial, i.e. it has the structure $1+ x + x^{3/2}+x^2+x^{5/2}+\dots+x^{11/2}$, where
$x = (GM\Omega/c^3)^{2/3}={\cal O}(c^{-2})$ is the PN-ordering
frequency parameter. 
The defining formulas for the $F_\lm^{(N,\epsilon)}$ are given explicitly in Sec.~\ref{sec:Newt}; the Newton-normalized PN-expanded multipolar
fluxes, $\hat{F}_\lm$, in Sec.~\ref{app:fluxes}.

\subsection{Numerical results}\label{sec:NR}

For computing numerically the radiated fluxes we employ two codes that solve the
Teukolsky equation. One is a 2+1 time-domain code of Ref.~\cite{Harms:2015ixa,Harms:2016ctx}
and the other a frequency domain code~\cite{akcay:2019}. The details of each of these codes
are presented elsewhere and so here we will only give a brief overview of each code
and show that their results are consistent with one another.

\subsubsection{Frequency domain approach}

The method employed by the frequency domain numerical code follows
closely the description of the PN calculation given in Sec.~\ref{sec:TEUKpn}
with the exception that the homogeneous solutions are computed numerically.
This is done using the semi-analytic MST method (see~\cite{Sasaki:2003xr}
for a review and~\cite{Casals:2015nja, Casals:2016soq} for extensions we use).
A similar version of the code to compute the homogeneous solutions
is publicly available as part of the Black Hole Perturbation Toolkit~\cite{BHPToolkit}.
With the homogeneous solutions in hand the inhomogeneous solutions are
computed by convolving them with the Teukolsky source. From the inhomogeneous
solutions we can compute the radiated fluxes per mode from the complex asymptotic
amplitudes, $C_\lm^\pm$, of the radial Teukolsky solutions via Eq.~\eqref{Eq:FluxFormula}.

Both in the source and in the orbital dynamics we linearize with respect
to $\sigma$. Our resulting fluxes though are not linear in $\sigma$.
This is because the radial Teukolsky equation contains a term proportional
to the square of the mode frequency, $\omega = m\Omega$. In principle it
would be possible to solve this equation to linear order in $\sigma$ but
we have not attempted to do so in this work. Instead, at a range of fixed 
orbital frequencies, we numerically compute
the fluxes for various values of $\sigma$ and fit this data using a polynomial
and extract the linear in $\sigma$ piece. It is important to make this
step as the quadratic and higher contributions to the raw frequency-domain
flux data are not complete as we are not including, e.g., higher order corrections
to the orbital dynamics.

\begin{figure}[t]
	\includegraphics[width=8.5cm]{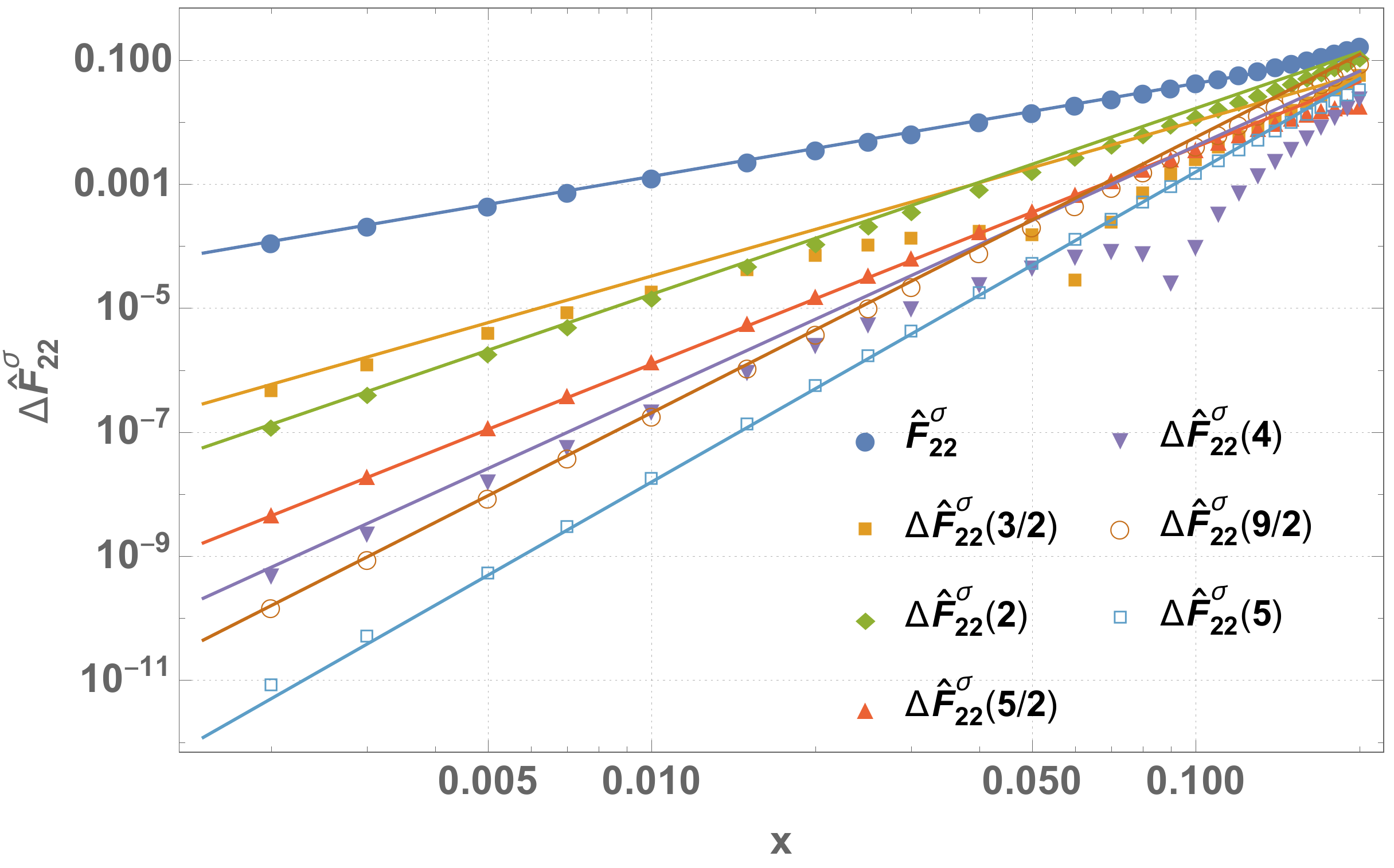}
	\caption{Consistency test between FD calculation and PN-expanded fluxes.
          Comparison of the linear-in-sigma contribution of the $\ell=m=2$
          (Newtonian normalized) flux to infinity between the numerical frequency domain results and the
          PN series. Here the markers show the difference $\Delta \hat{F}^\sigma_{22}(n) = |\hat{F}_{22}^{\text{FD}\sigma} - \hat{F}_{22}^{\text{PN}\sigma}(n)|$
          where $\hat{F}_{22}^{\text{PN}\sigma}(n)$ is the $\mathcal{O}(\sigma)$ piece of the PN series truncated at $x^{n}$.
		  The full PN series is give by Eq.~\eqref{eq:PN_flux22_normalized} and successively higher order truncations of this are shown
		  by the colored curves.
          The $\mathcal{O}(\sigma)$ piece of the numerical results is extracted from 
		  the numerical data using the method described in the main text.
          At large radii we observe that subtracting successively higher order PN series
          improves the agreement with the numerical data, as expected.}
        \label{fig:flux_inf_FD_vs_PN}
\end{figure}

The code is written in C++ and internally (particularly for the MST part of the calculation)
it uses extended precision. We have high confidence in the results for three
reasons: i) when $\sigma=0$ we recover known flux results non-spinning bodies
to $\sim$14 significant digits ii) for $\sigma \neq0$ and after extracting the
linear in $\sigma$ contribution we see good agreement with the PN results -- see Fig.~\ref{fig:flux_inf_FD_vs_PN} -- and iii) we have reconstructed the metric perturbation at the particle
using the standard CCK procedure~\cite{Chrzanowski:1975wv,Cohen:1974cm,Kegeles:1979an}
and from this we have shown that the local self-force experienced by
the particle is balanced (to a relative error of $10^{-9}$) by the radiated
fluxes through the spacetime boundaries. Details on each of these checks
will be presented in~\cite{akcay:2019}.

\subsubsection{Numerical results: time-domain approach with Hamiltonian dynamics}

The time-domain method solves the 2+1 Teukolsky equation obtained by
separating Eq.~\eqref{eq:teuk} in the azimuthal ($\phi$)
direction \cite{Harms:2015ixa,Harms:2016ctx}. The resulting wave
equations for each $m$-mode are written in the scri-fixing
hyperboloidal and 
horizon penetrating co-ordinates
developed in~\cite{Zenginoglu:2010cq,Zenginoglu:2011jz,Harms:2016ctx}.
The source of Teukolsky equation is implemented in a general way and
then specified to Eq.~\eqref{eq:teuk:src} 
for a spinning test body (in the pole-dipole approximation) moving on an 
arbitrary trajectory~\cite{Harms:2015ixa,Zelenka:2019uke}. 
The Teukolsky equation is numerically solved using the method-of-lines with a
4th order accurate Runge-Kutta time integrator and 6th order accurate finite-differencing
operators. 
The delta functions in the source can be discretized either using a
narrow Gaussian or discrete delta functions; the
former option is used in case of circular orbits as discussed in~\cite{Harms:2015ixa}.
The code was extensively tested and delivers multipolar waveforms and
GW fluxes at null infinity with an accuracy well below the $1\%$ level up
to $m=4$ modes ~\cite{Nagar:2014kha,Harms:2015ixa,Harms:2016ctx,Lukes-Gerakopoulos:2017vkj}.
The data used for this work are produced exactly as described in 
\cite{Harms:2015ixa,Harms:2016ctx}. Circular equatorial trajectories
for the test body are computed using the Hamiltonian formalism detailed
in Sec.~III of~\cite{Harms:2016ctx}. The Hamiltonian of a spinning particle at linear
order in the spin was originally obtained in Ref.~\cite{Barausse:2009aa}.
The same Hamiltonian can be recasted in certain specific effective-one-body 
(EOB) coordinates as illustrated in Ref.~\cite{Bini:2015xua}. The EOB coordinates 
are transformed to the hyperboloidal ones via a transformation linear in $\sigma$.
The EOB dynamics is compatible with the MPD dynamics with Tulczyjew and
the Pirani~\cite{Pirani:1956} SSCs across almost the whole spin and frequency 
range for the Schwarzschild background~\cite{Harms:2016ctx}.
One must, however, be aware that that the Hamiltonian circular dynamics
and the MPD dynamics described above are fully equivalent and compatible
{\it only} when the spin of the particle is small. We address the reader
to Ref.~\cite{Harms:2016ctx} for additional details.
\begin{figure*}[t]
  \includegraphics[width=0.4\textwidth]{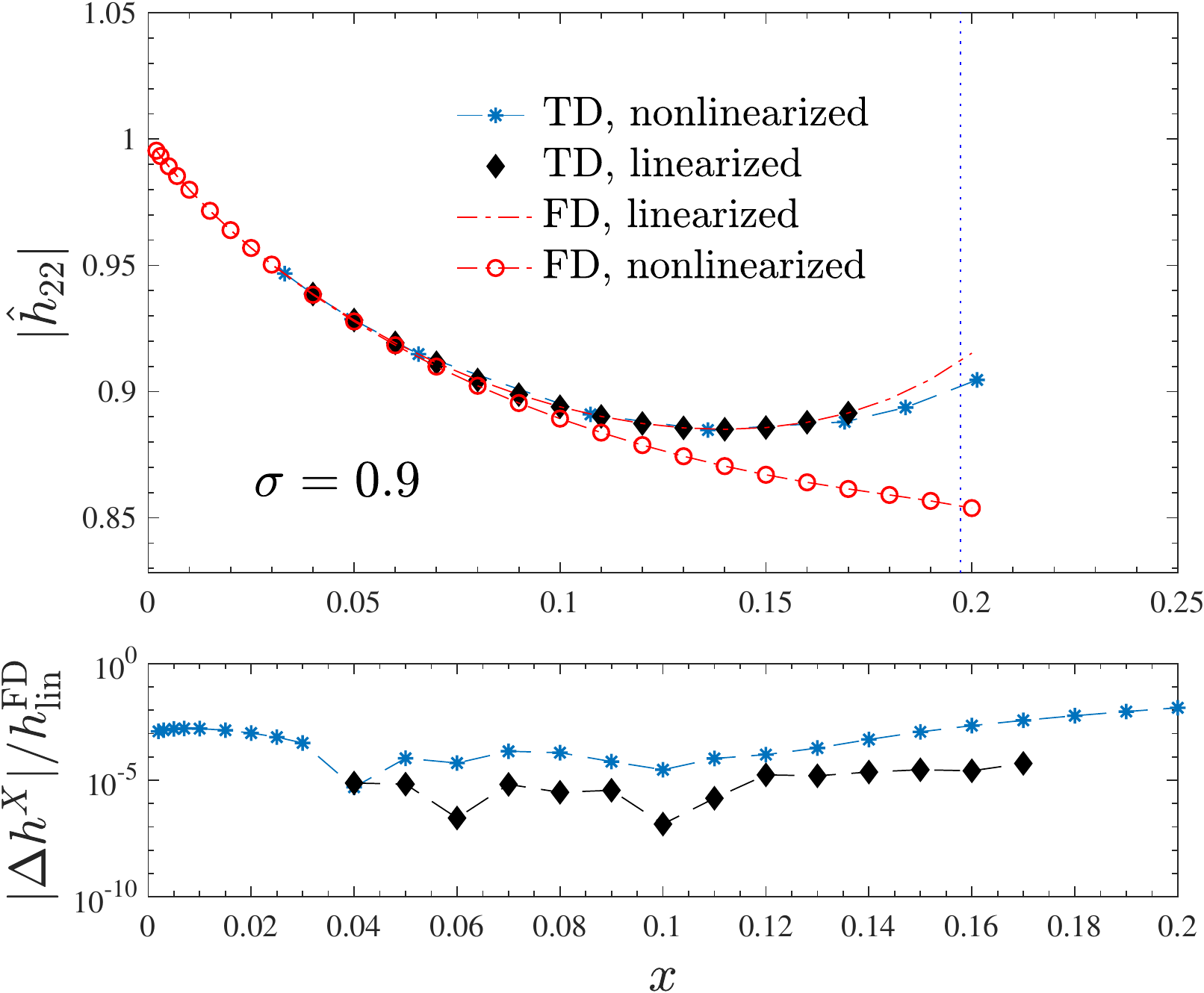}
  \hspace{5mm}
  \includegraphics[width=0.4\textwidth]{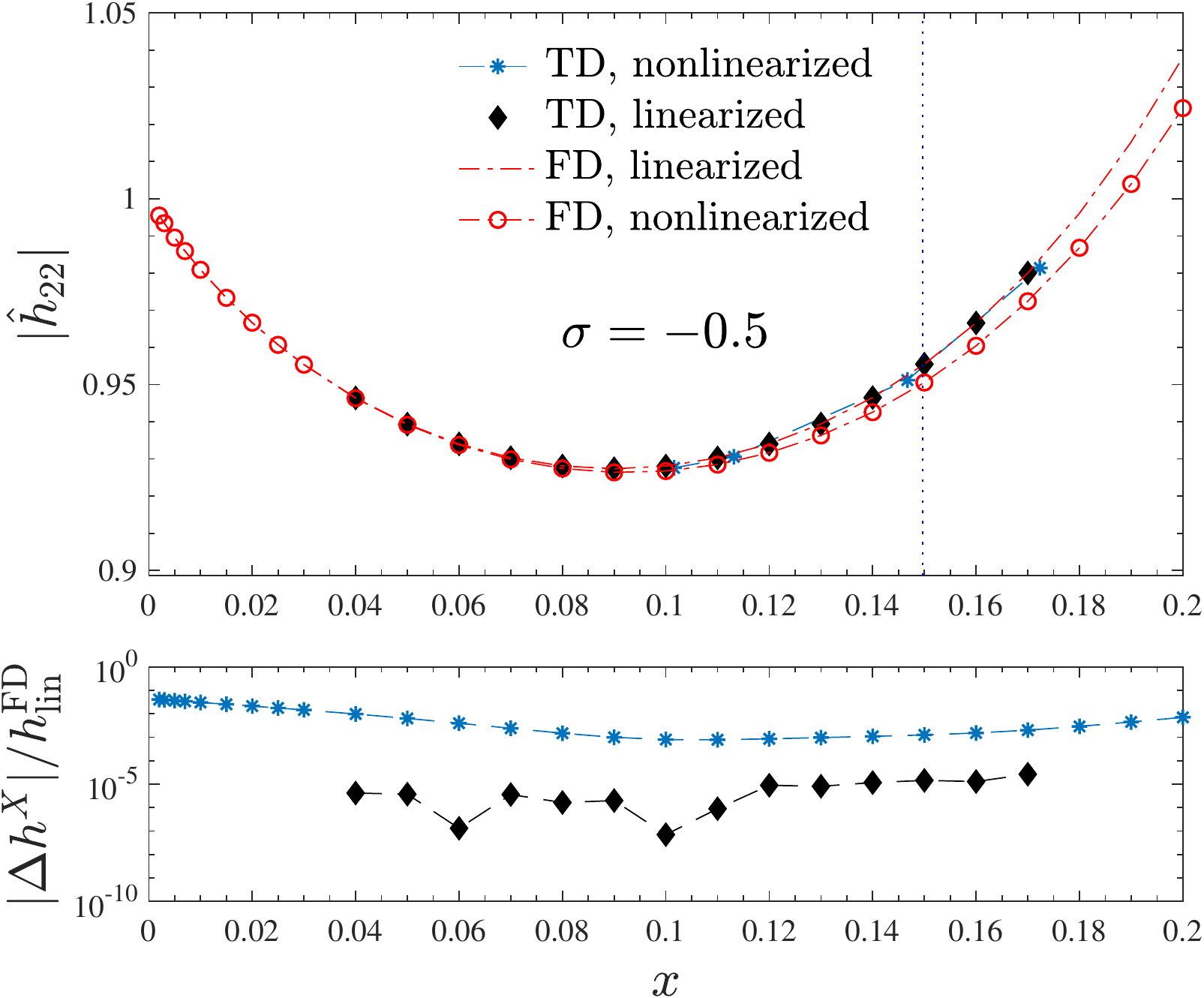}\\
    \vspace{2.5mm}
  \includegraphics[width=0.4\textwidth]{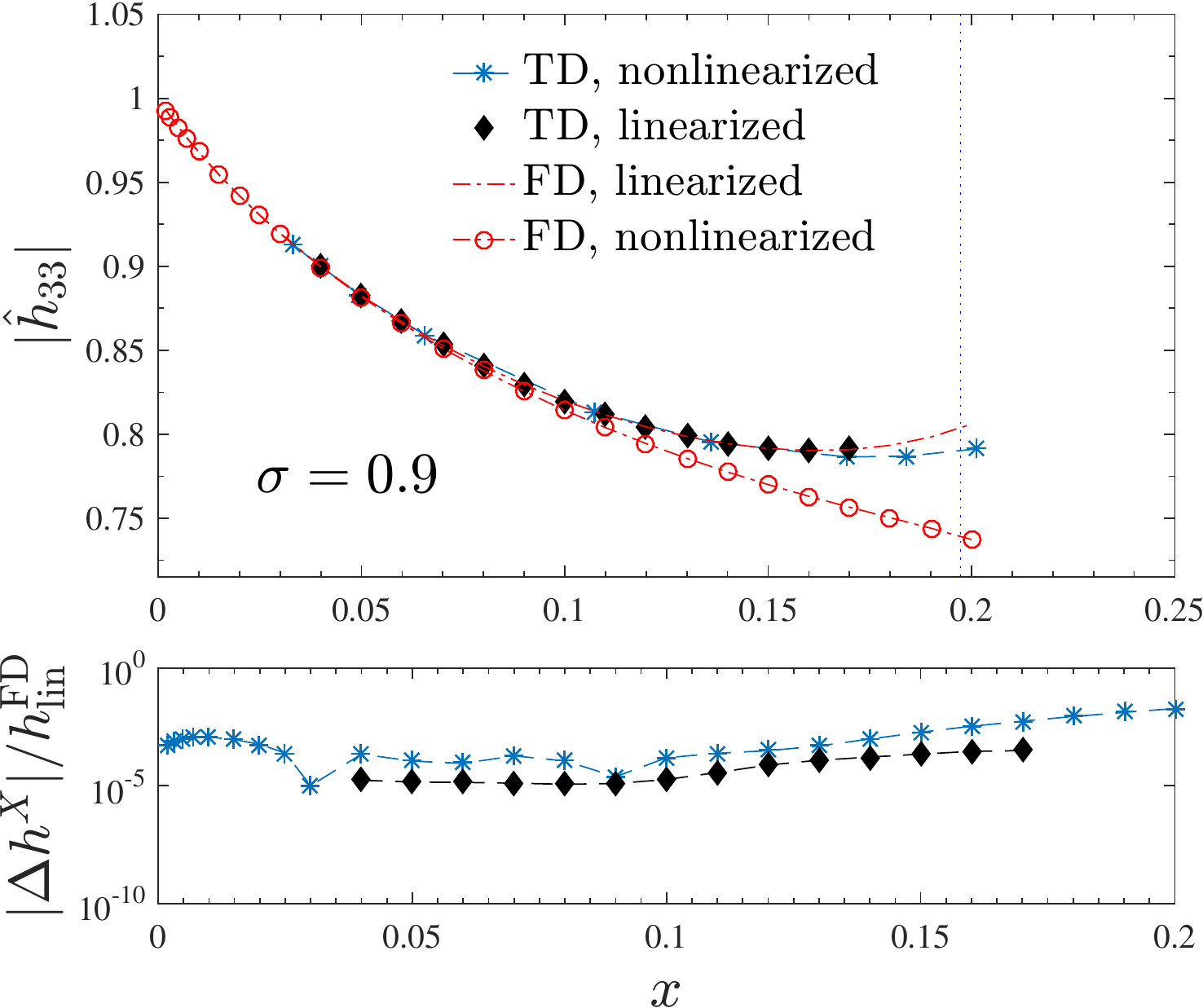}
    \hspace{5mm}
  \includegraphics[width=0.4\textwidth]{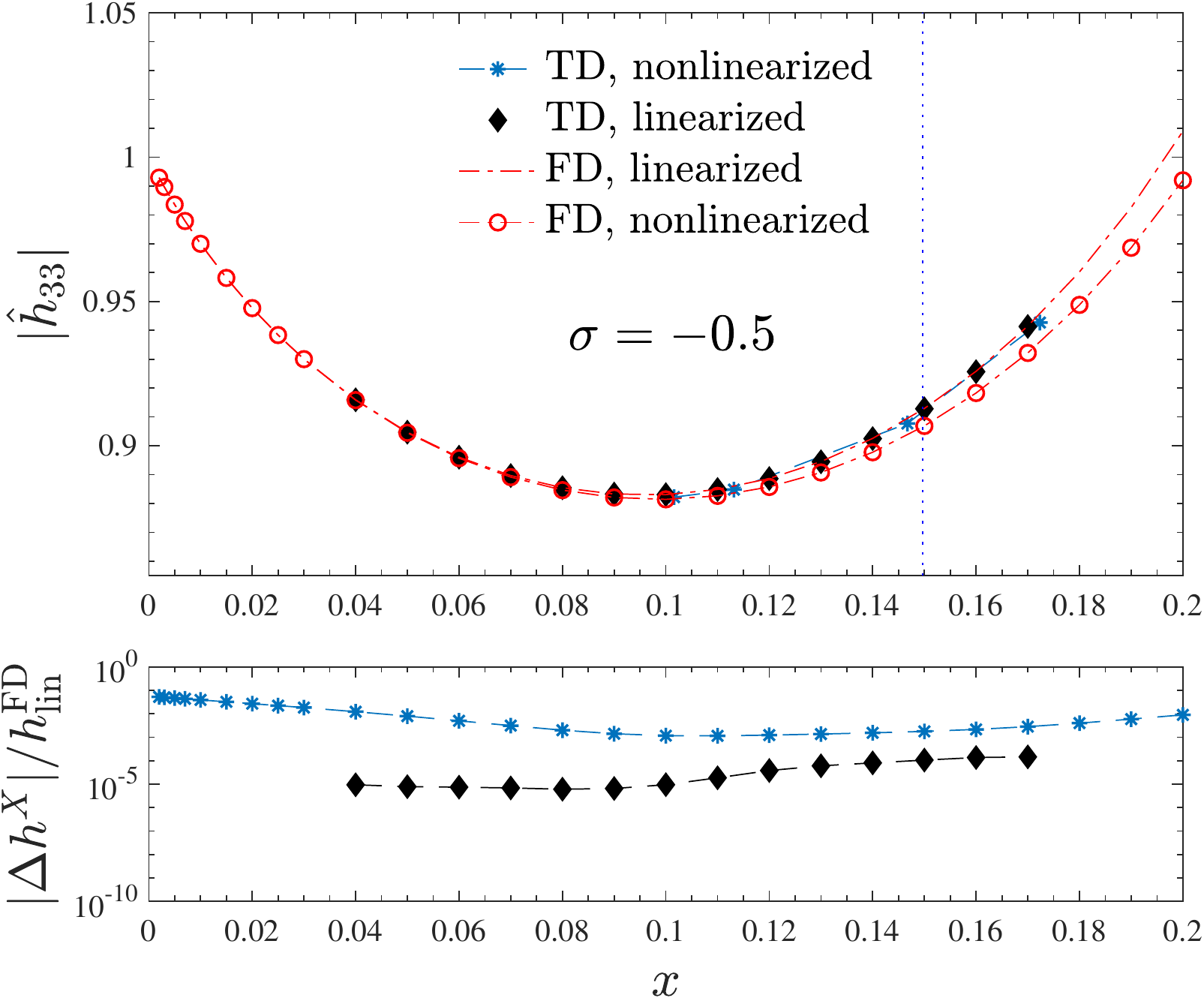}\\
      \vspace{2.5mm}
  \includegraphics[width=0.4\textwidth]{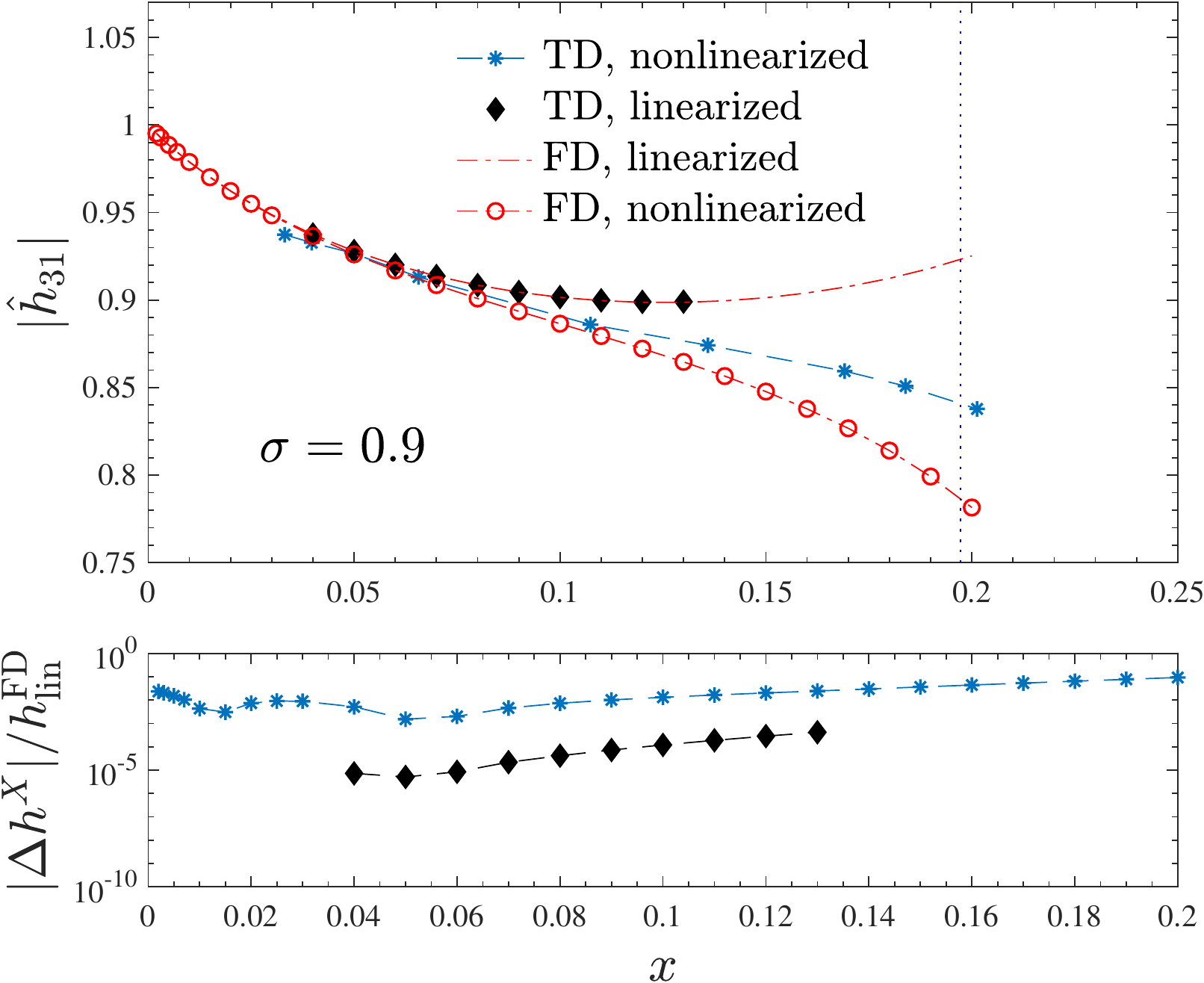}
    \hspace{5mm}
  \includegraphics[width=0.4\textwidth]{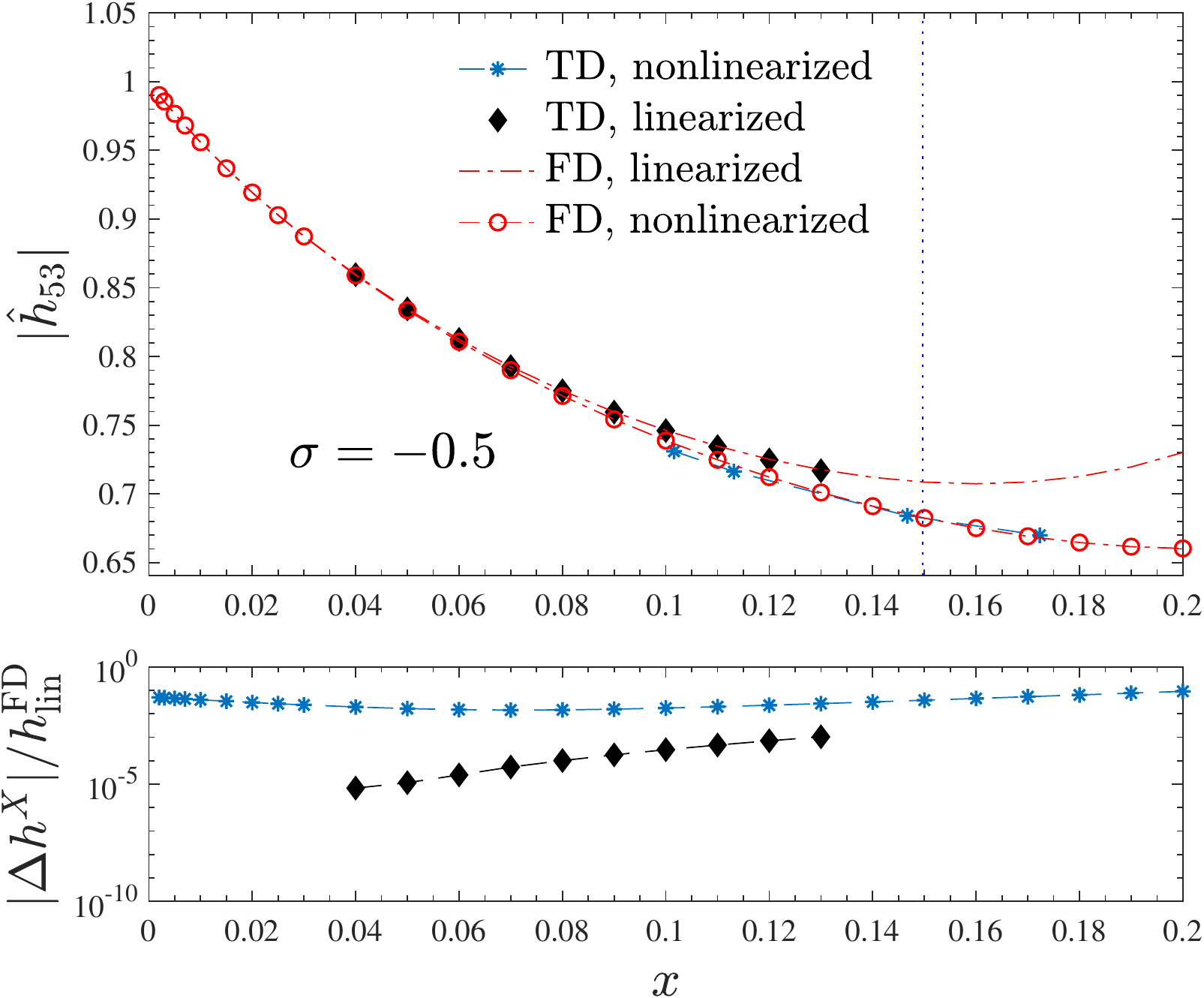}
  \caption{\label{fig:compare_TD_FD}Comparing FD (with MPD dynamics) and TD calculation (with Hamiltonian dynamics)
    for a few multipoles and values of the particle spin. Each plot shows four curves: the TD and FD result with all the contributions
    nonlinear in the particle spin and the same data linearized in $\sigma$. In the latter case, the agreement between TD and
    FD results is excellent. The bottom part of each panel illustrate the fractional difference between: $X=$TD-nonlinearized and FD-linearized
    and the fractional difference between $X=$TD-linearized and FD-linearized. The vertical line marks the location of the LSO, Eq.~\eqref{eq:x_LSO}.}
    \end{figure*}
Following previous work of some of us~\cite{Harms:2016ctx,Lukes-Gerakopoulos:2017vkj}, 
we want to accurately compute numerical fluxes for values of the spins that are {\it large},
i.e., $\sigma\sim 1$ and for orbits that are near the spin-dependent last-stable orbit.
Past work~\cite{Harms:2016ctx} suggests that 
this can be done using the Hamiltonian formalism. Like the FD results mentioned
above, our results will not be linear in $\sigma$, essentially for the same reasons mentioned
above. Thus, to provide a consistent comparison with FD results, we also extracted the linear-in-sigma
piece out of the numerical results.

\subsection{Comparing numerical waveform amplitudes}
Since the main aim of the paper is to check resummation procedure for gravitational 
waveform amplitudes, it is convenient to directly use these quantities for comparing 
the results obtained with the two numerical approaches. In this respect, the extraction
of the linear piece mentioned in the previous section is now performed at the level of the
waveform amplitudes (defined below) and not on the fluxes.
Likewise for the fluxes, each waveform multipole is factorized in the product of 
a Newtonian (leading-order) contribution and a relativistic correction
\begin{equation}
\label{eq:wfmultipole}
h_{\ell m}(x)={\cal R}h^{(N,\epsilon)}_\lm(x)\hat{h}_{\ell m}^{(\epsilon)}(x),
\end{equation}
where ${\cal R}h_\lm^{(N,\epsilon)}$ is the Newtonian prefactor that is analytically 
known (see Appendix~\ref{sec:Newt}). In practice, it is convenient to focus only
on the relativistic correction, since it is a function of order unity whose PN expansion 
has the structure $\hat{h}_\lm^{(\epsilon)}\simeq 1+x+x^{3/2}+\dots$.
Figure~\ref{fig:compare_TD_FD} offers a comprehensive comparison of the various
numerical data at our disposal for an illustrative sample of multipoles and spins.
The top part of each panel of the figure shows 4 curves: (i) the outcome of the time-domain
code with Hamiltonian dynamics; (ii) the same quantity where one has subtracted the
nonlinar-in-spin part; (iii) the outcome of the frequency-domain code with MPD dynamics:
(iv) the same quantity where one has subtracted the nonlinear-in-spin part. 
The vertical line in the plot marks the location of the $\sigma$-dependent frequency
of the last-stable-orbit (LSO), as obtained, for example, in  Ref.~\cite{Harms:2016ctx}
\begin{equation}
  \label{eq:x_LSO}
  x^{\rm LSO}=\frac{1}{6}+\frac{\sigma}{12\sqrt{6}}.
\end{equation}
As can be seen from Fig.~\ref{fig:compare_TD_FD}, the linear-in-spin results from the two codes are in the excellent agreement 
reaching a fractional accuracy of $\sim 10^{-5}$ (this is limited by the precision of the TD code and the need to extract the linear-in-spin contribution). Note that we do not expect agreement between the nonlinear in spin results
as each code effectively includes different pieces of the nonlinear contribution (either through their dynamics 
or other aspects of the calculation).
We emphasize that, to the best of our knowledge, this is the first successful comparison between two 
completely independent Teukolsky codes with a spinning secondary object.
The excellent agreement between the two codes (after linearizing in spin) and the agreement between the FD and PN results presented in Fig.~\ref{fig:flux_inf_FD_vs_PN} gives us a high degree of confidence in our results.

Since the PN results are linear in the spin, we will use the linearized FD data as the
target points to verify the accuracy of the various analytical approximations to the waveform
amplitudes.

\section{Comparing analytical and numerical results}
\label{sec:resum}
\begin{figure*}[t]
  \center
  \includegraphics[width=0.18\textwidth]{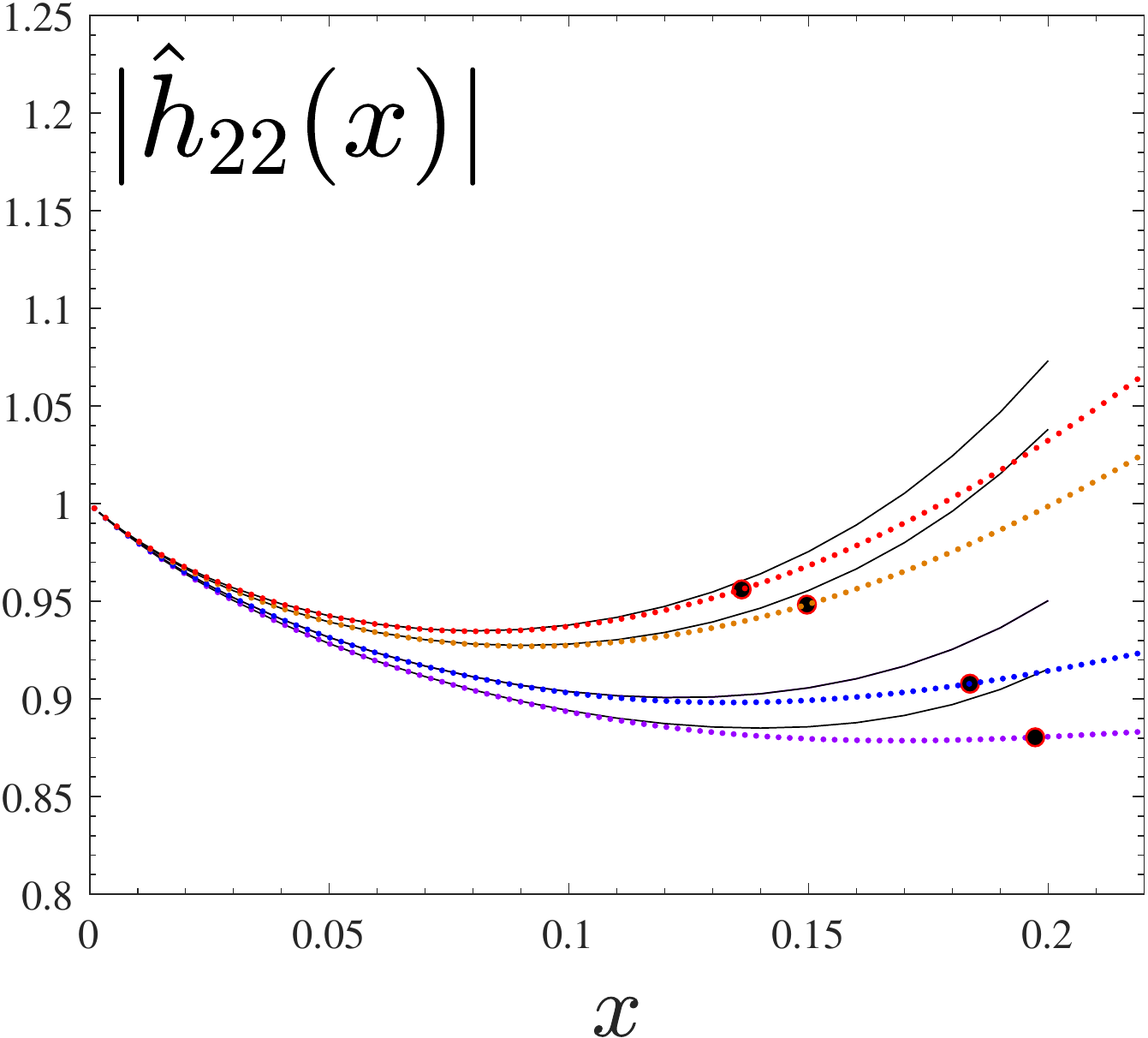}
  \includegraphics[width=0.17\textwidth]{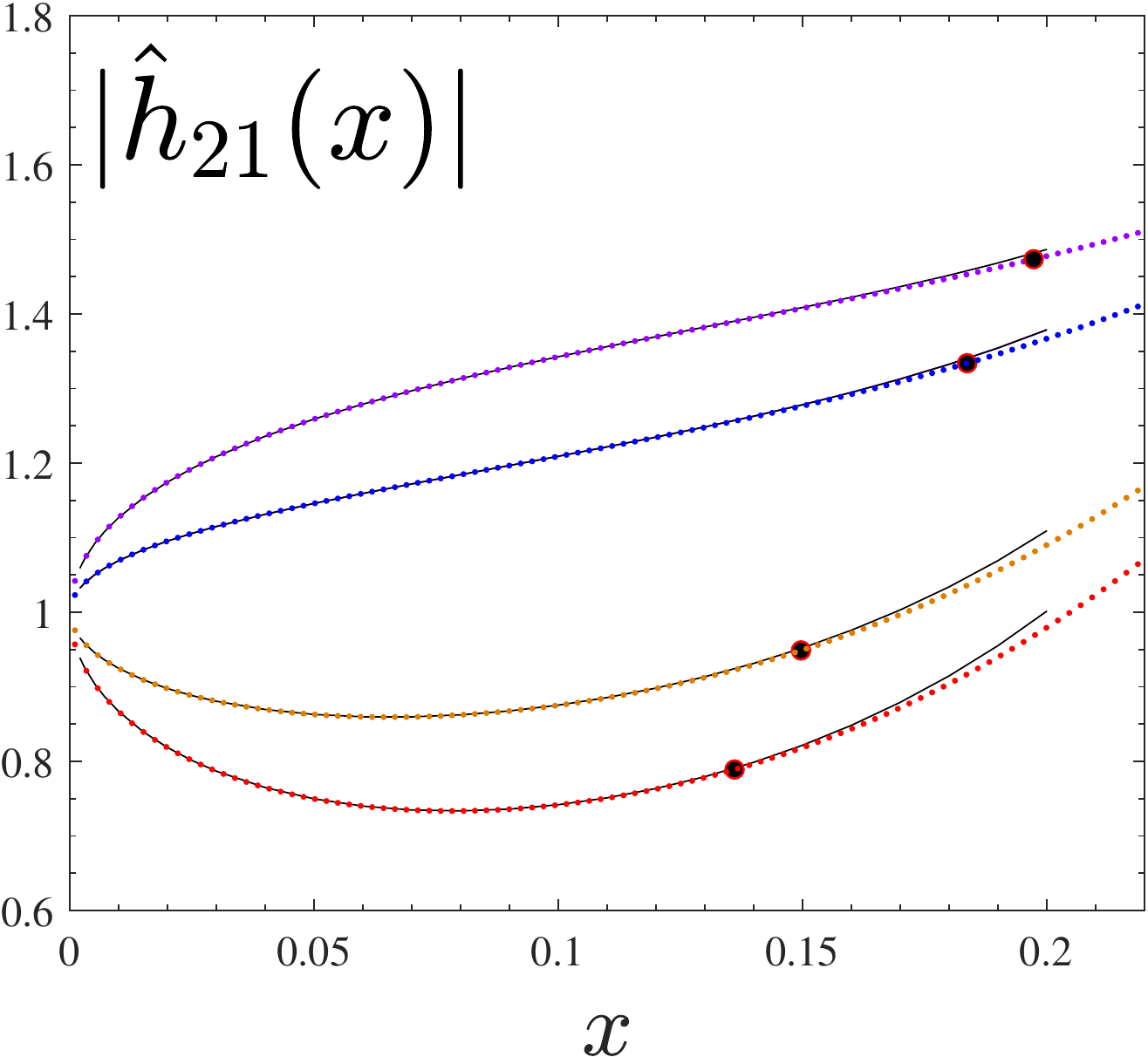}
  \includegraphics[width=0.178\textwidth]{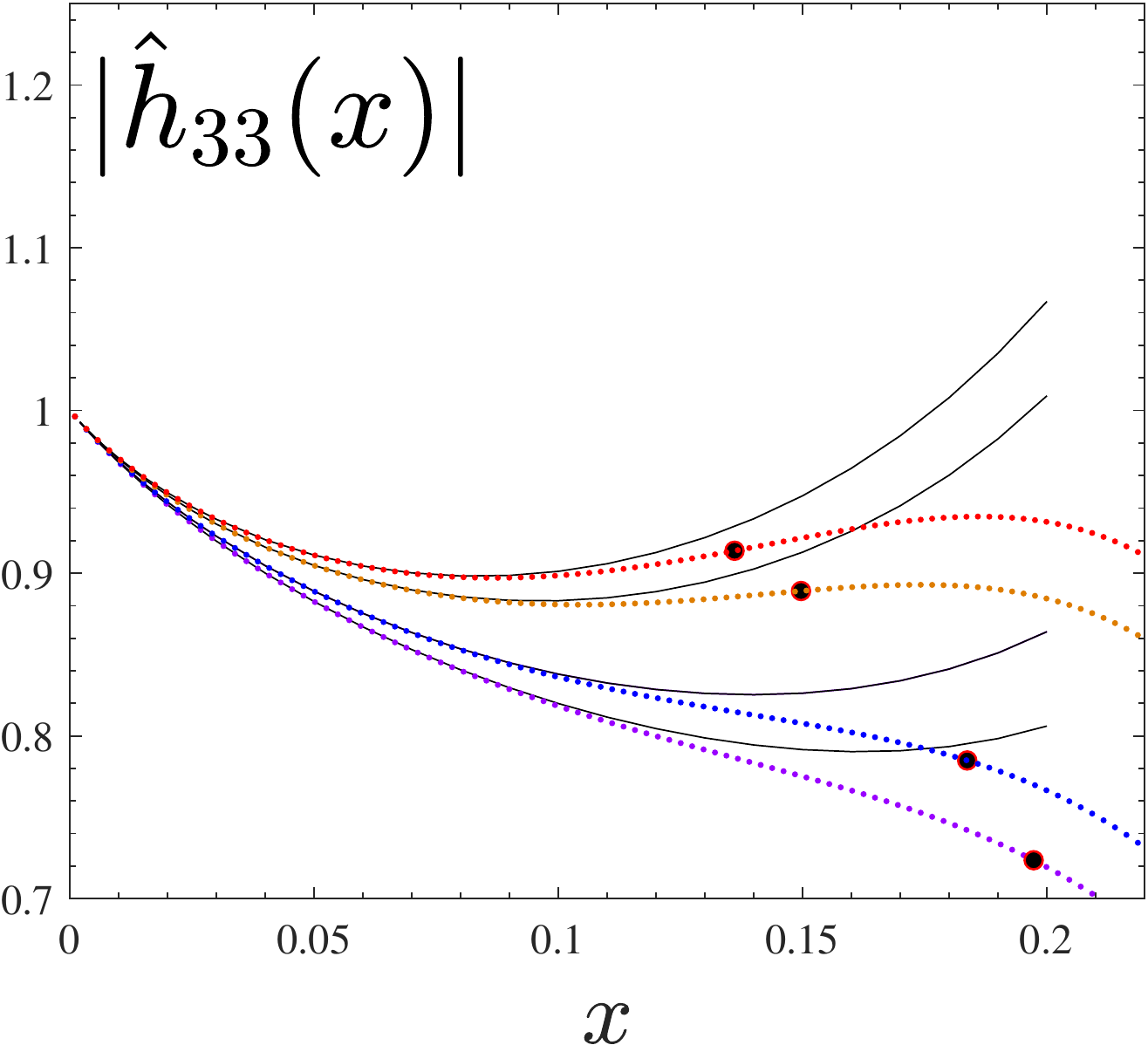}
  \includegraphics[width=0.17\textwidth]{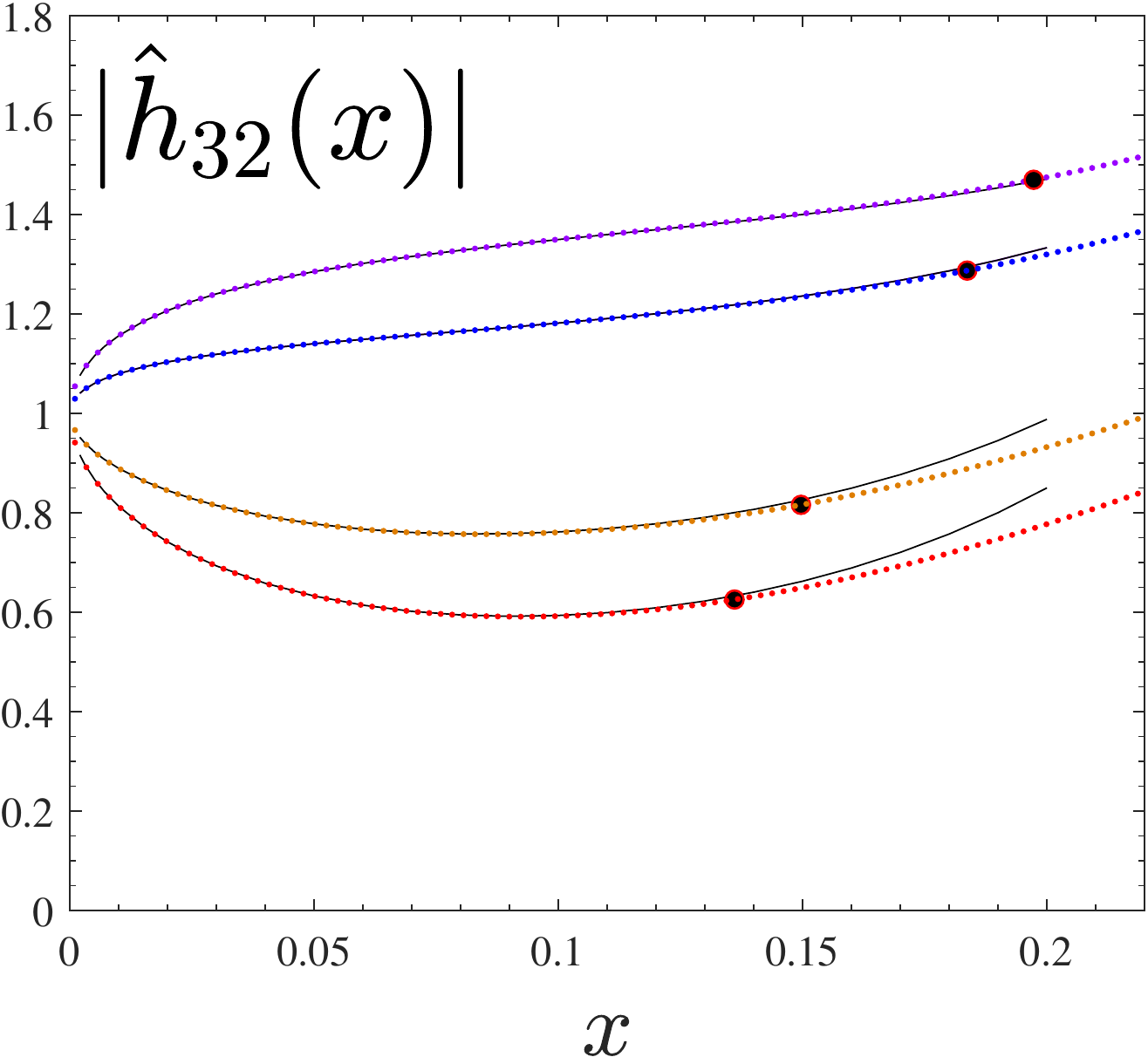}
  \includegraphics[width=0.178\textwidth]{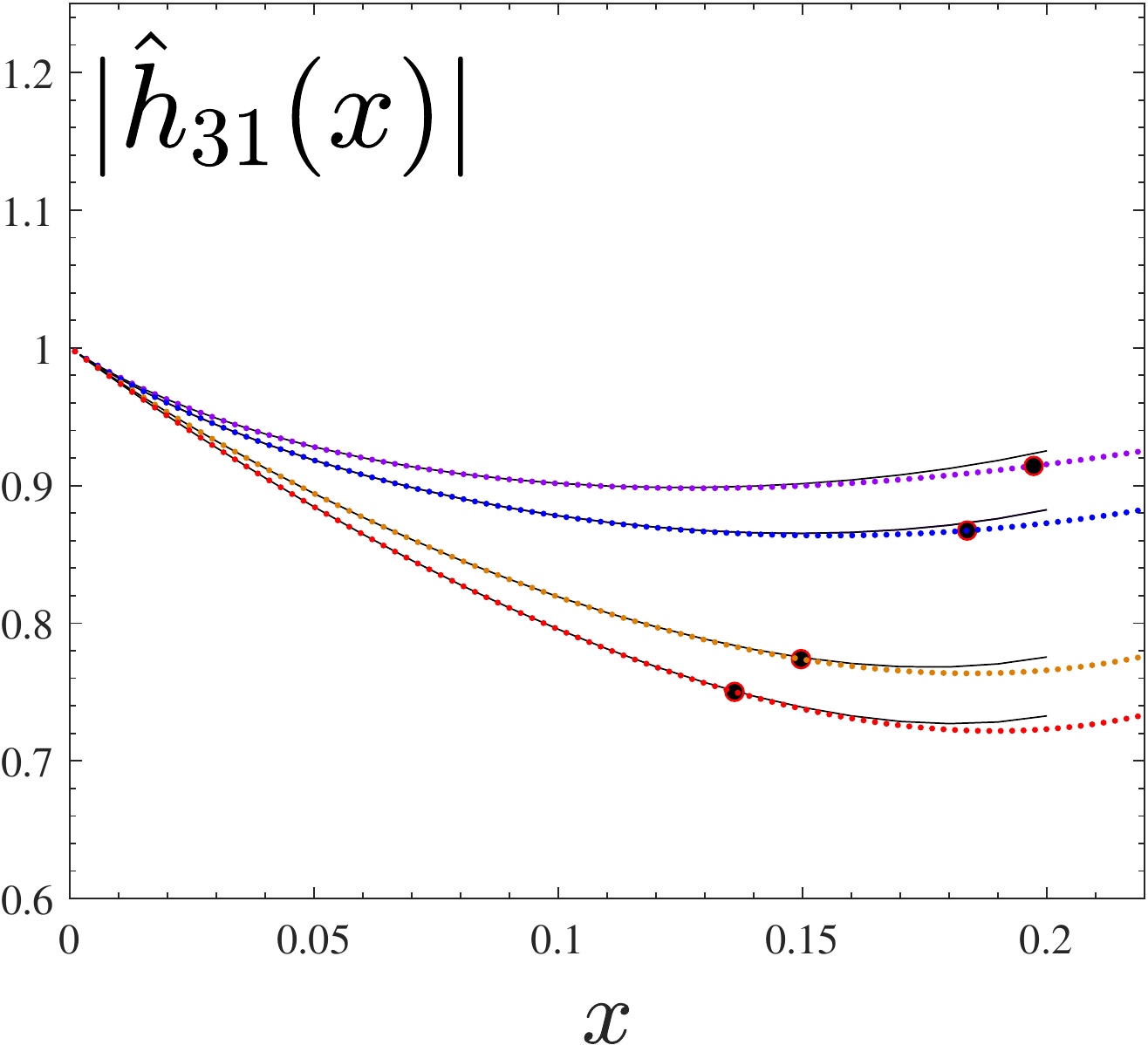}\\
  \includegraphics[width=0.17\textwidth]{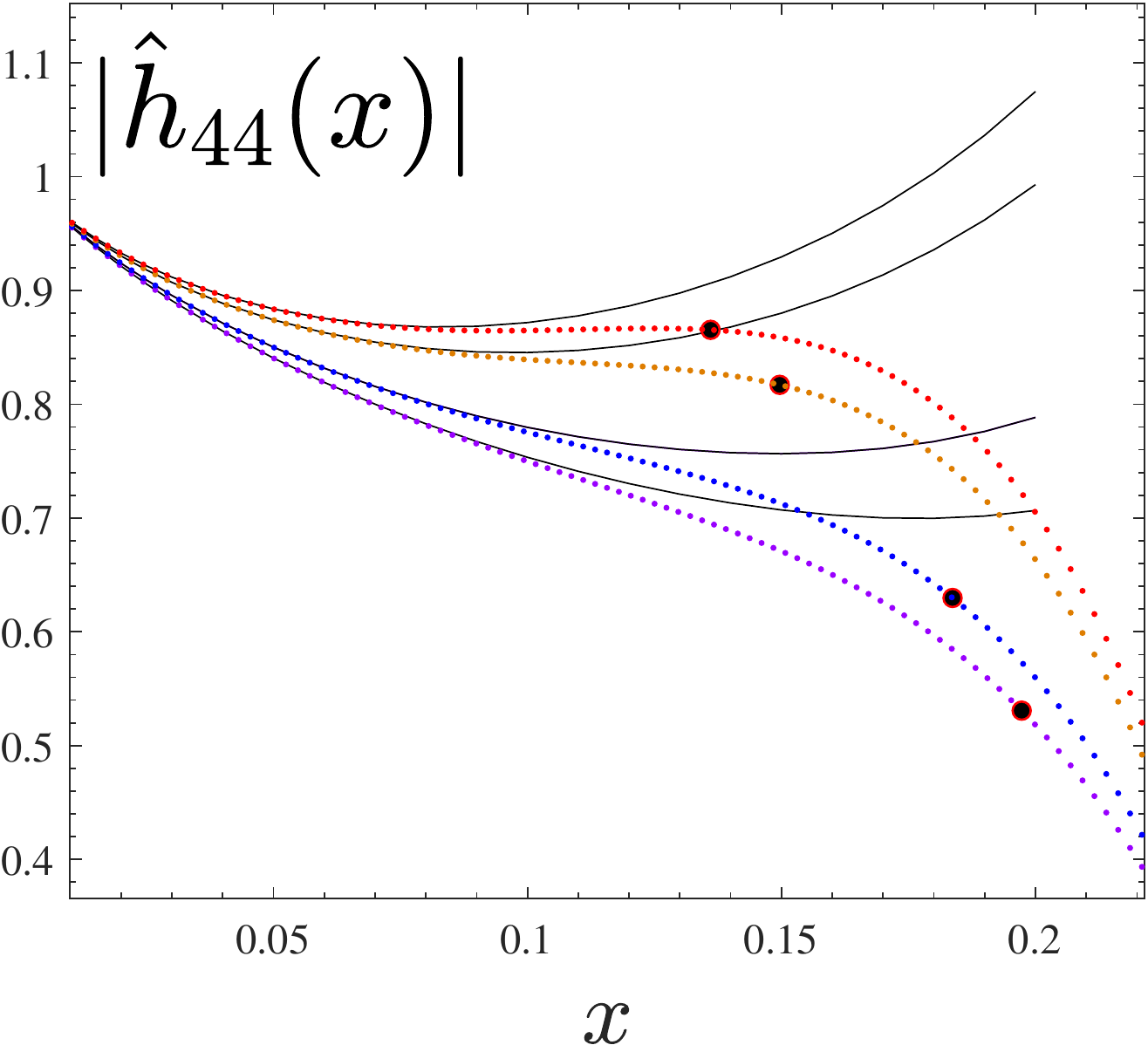}
  \includegraphics[width=0.17\textwidth]{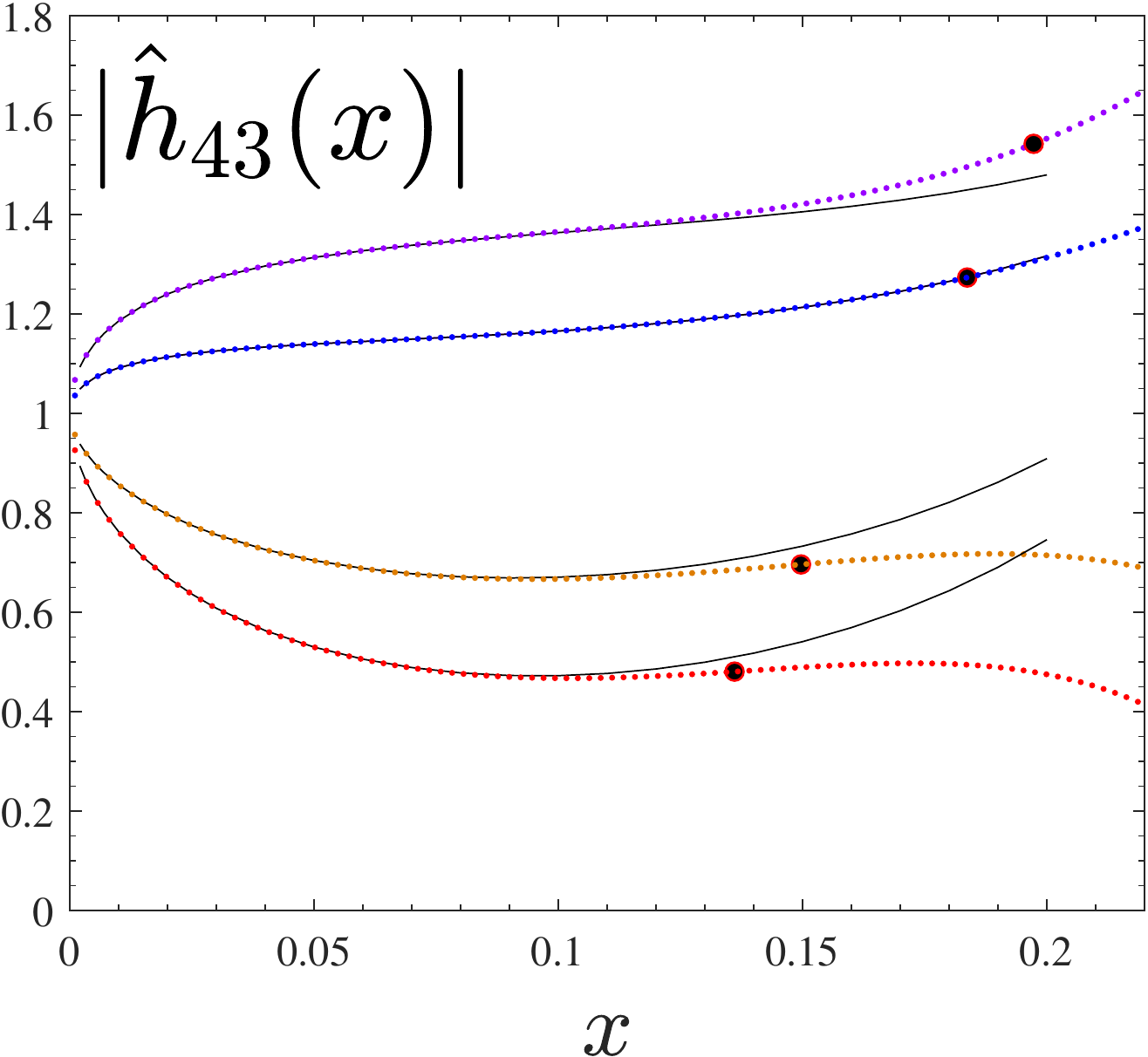}
  \includegraphics[width=0.17\textwidth]{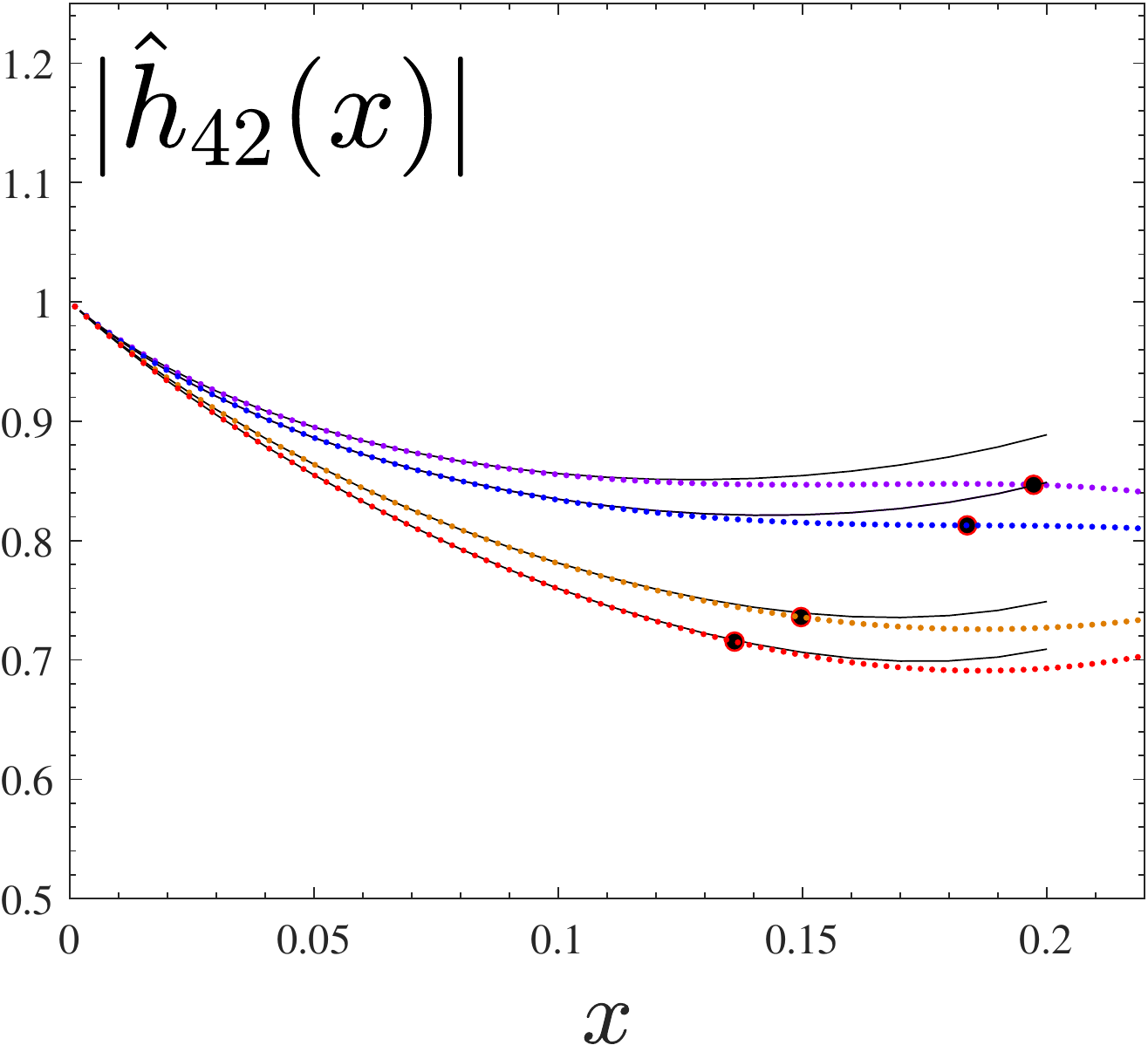}
  \includegraphics[width=0.18\textwidth]{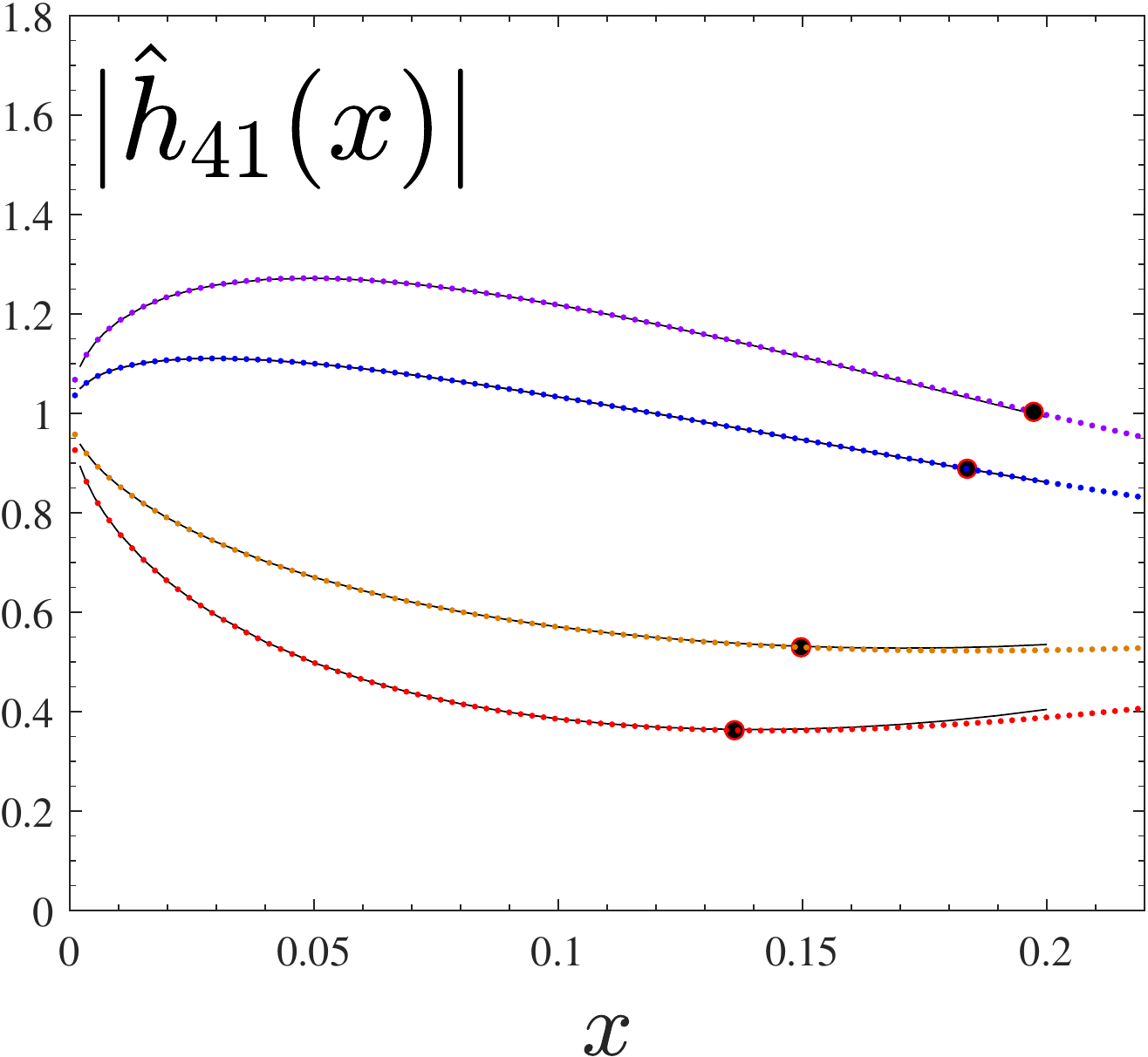}
  \includegraphics[width=0.18\textwidth]{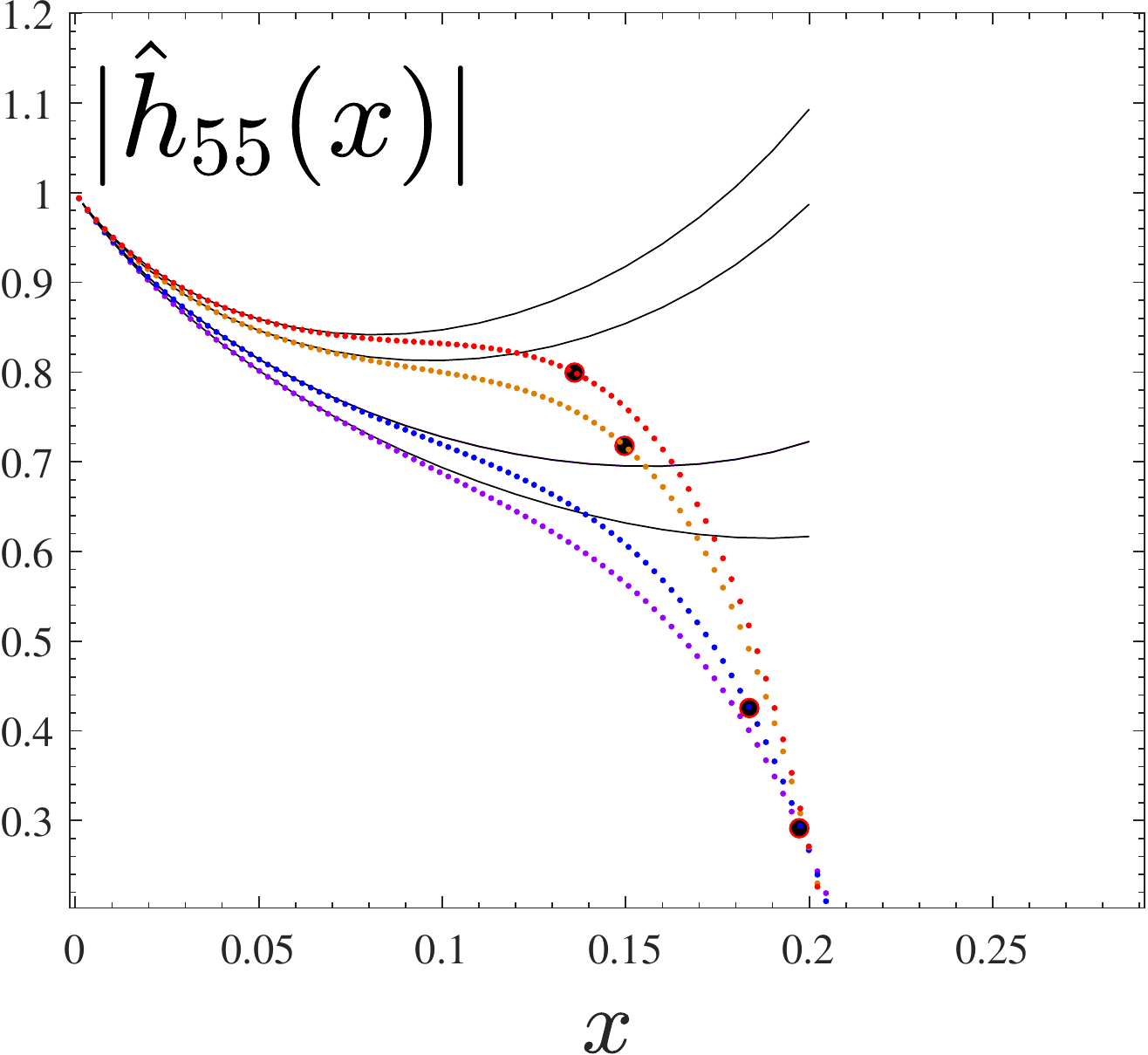}\\
  \includegraphics[width=0.17\textwidth]{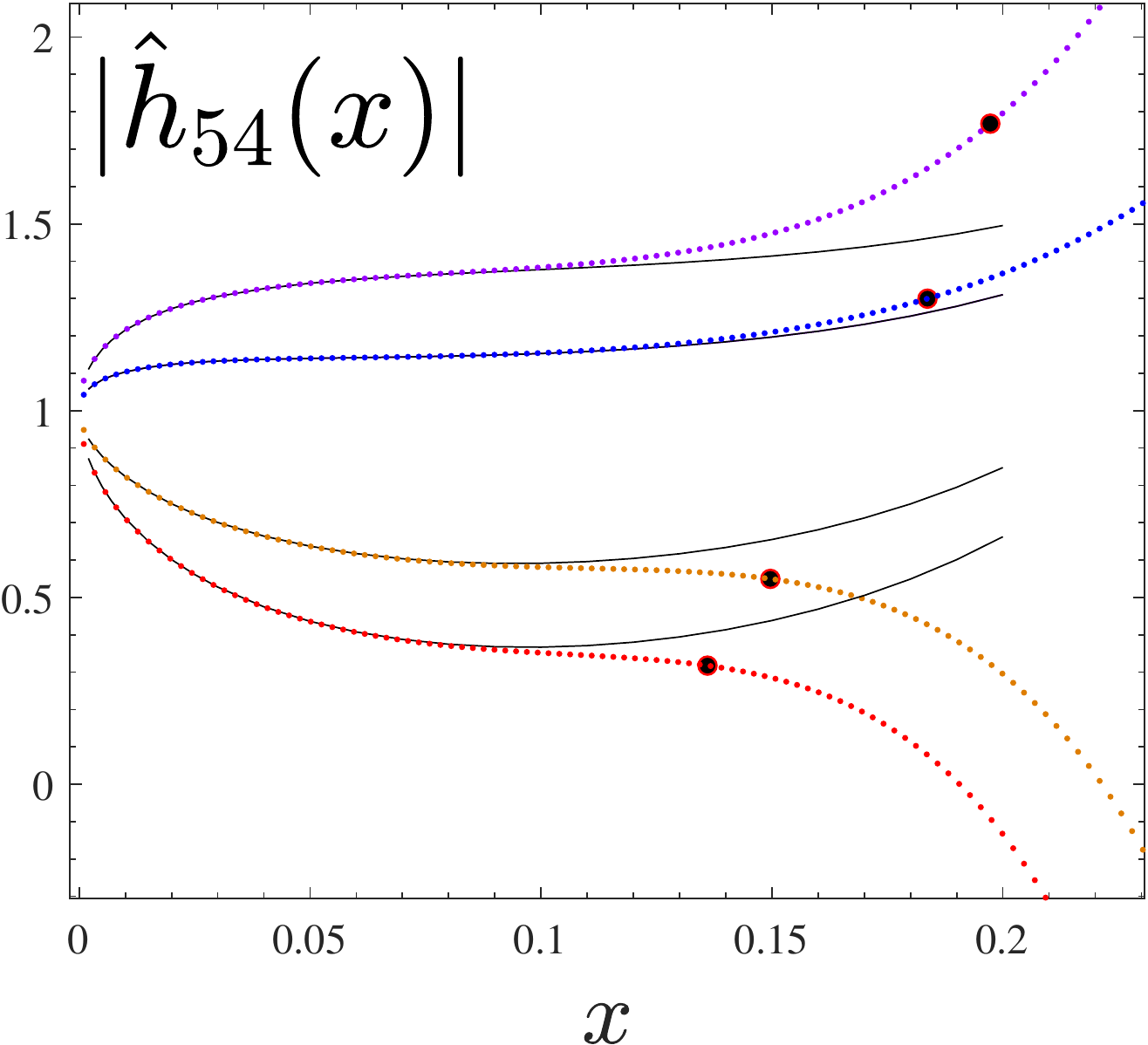}
  \includegraphics[width=0.17\textwidth]{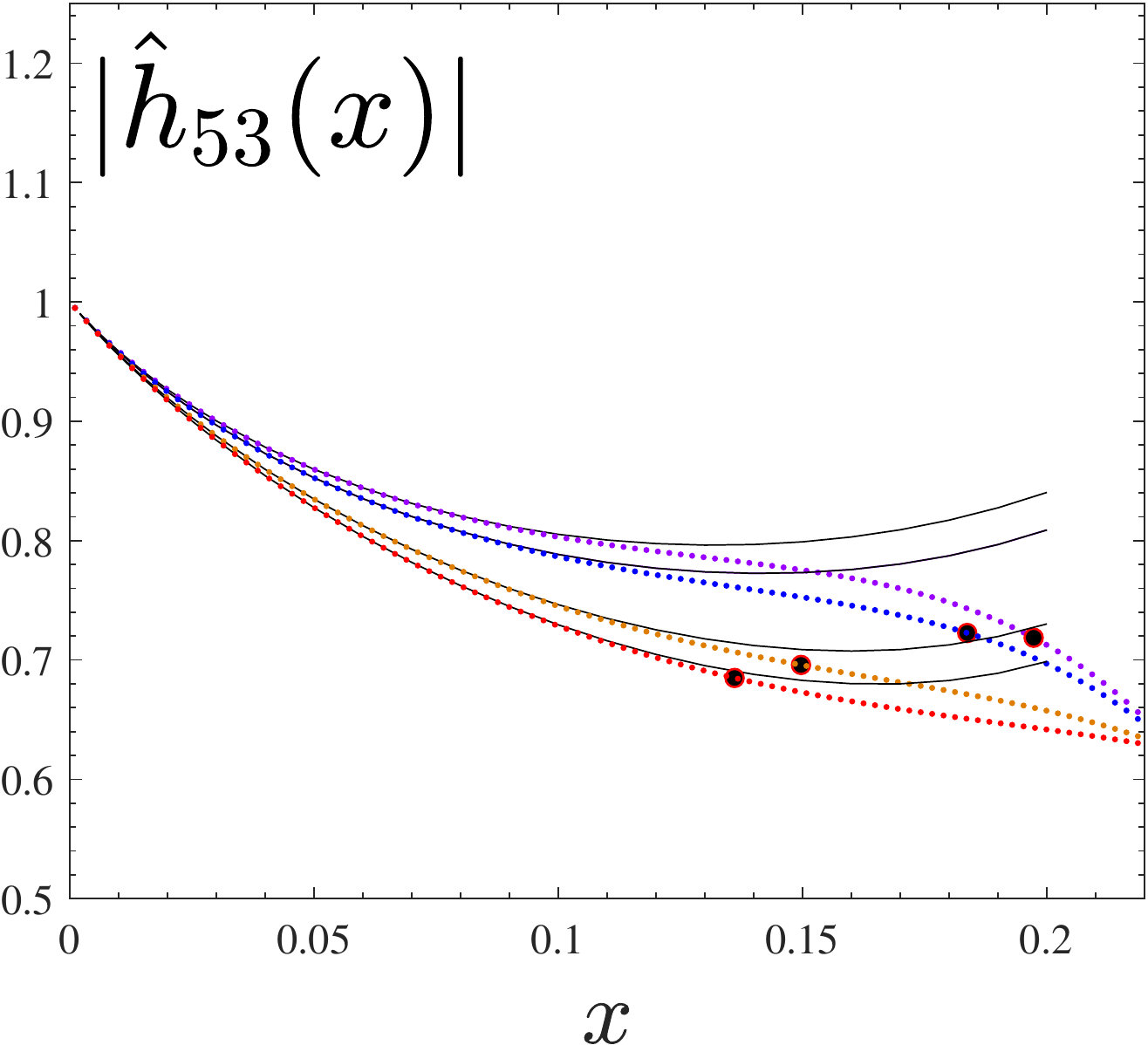}
  \includegraphics[width=0.17\textwidth]{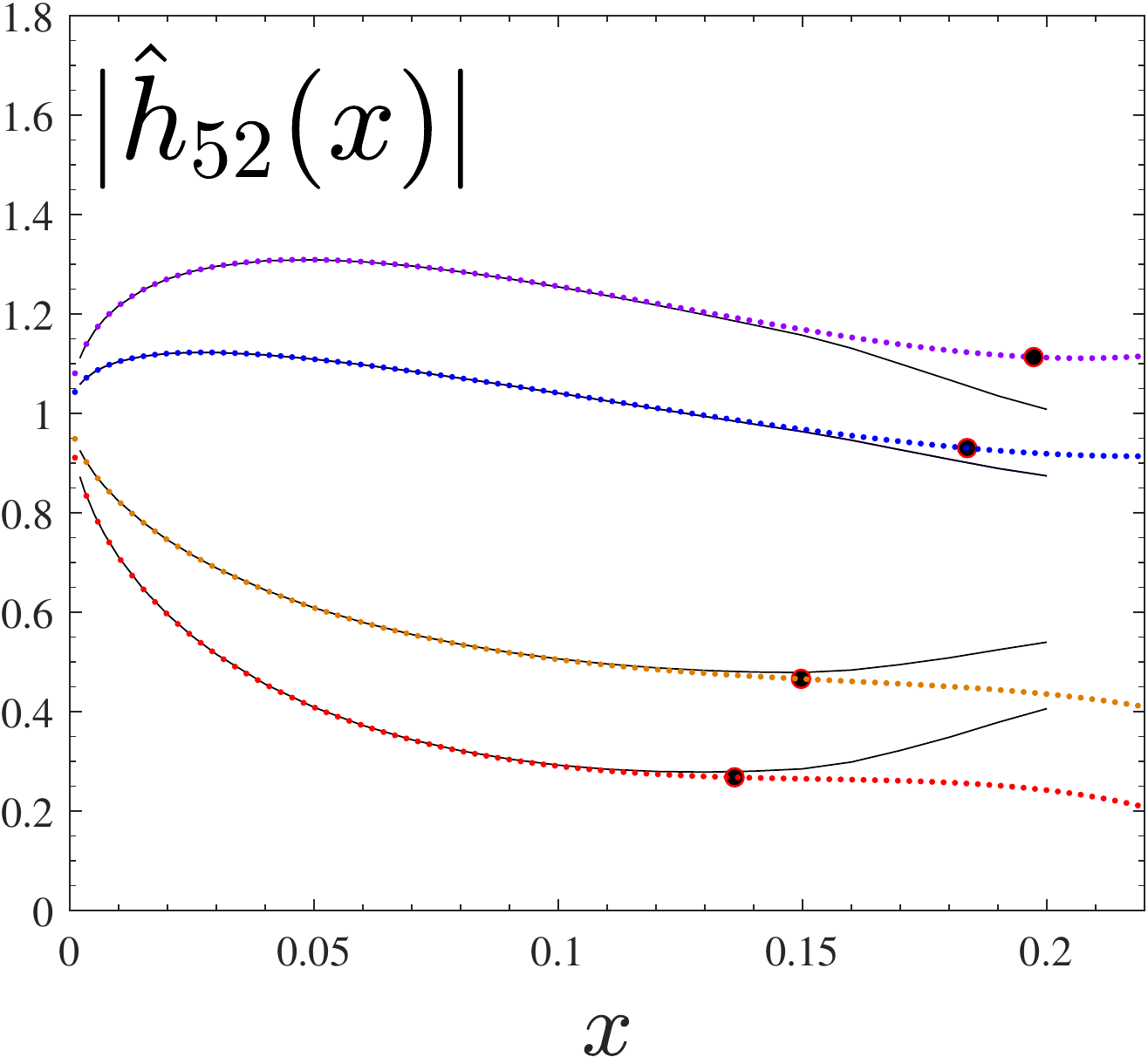}
  \includegraphics[width=0.17\textwidth]{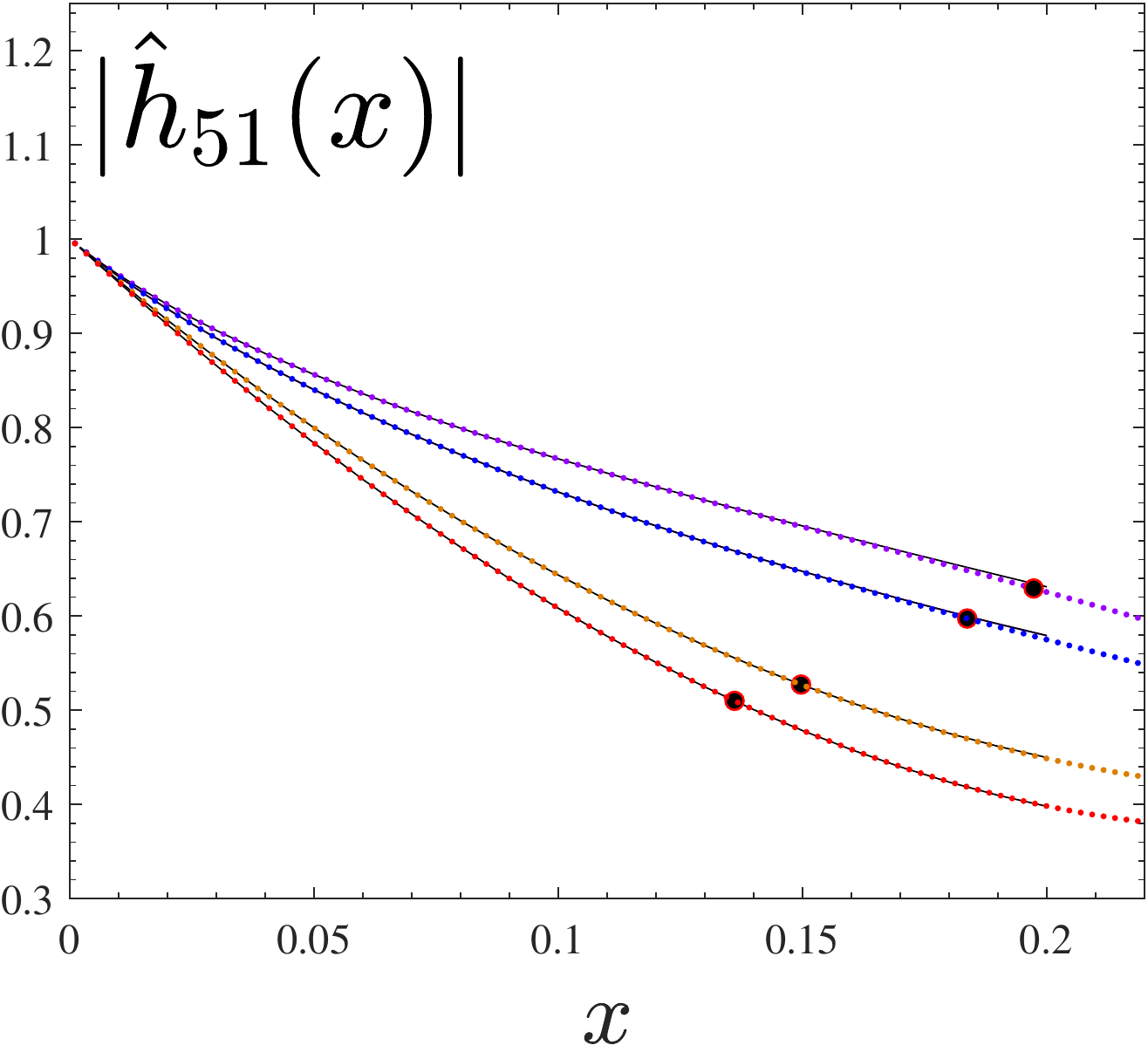}
  \includegraphics[width=0.17\textwidth]{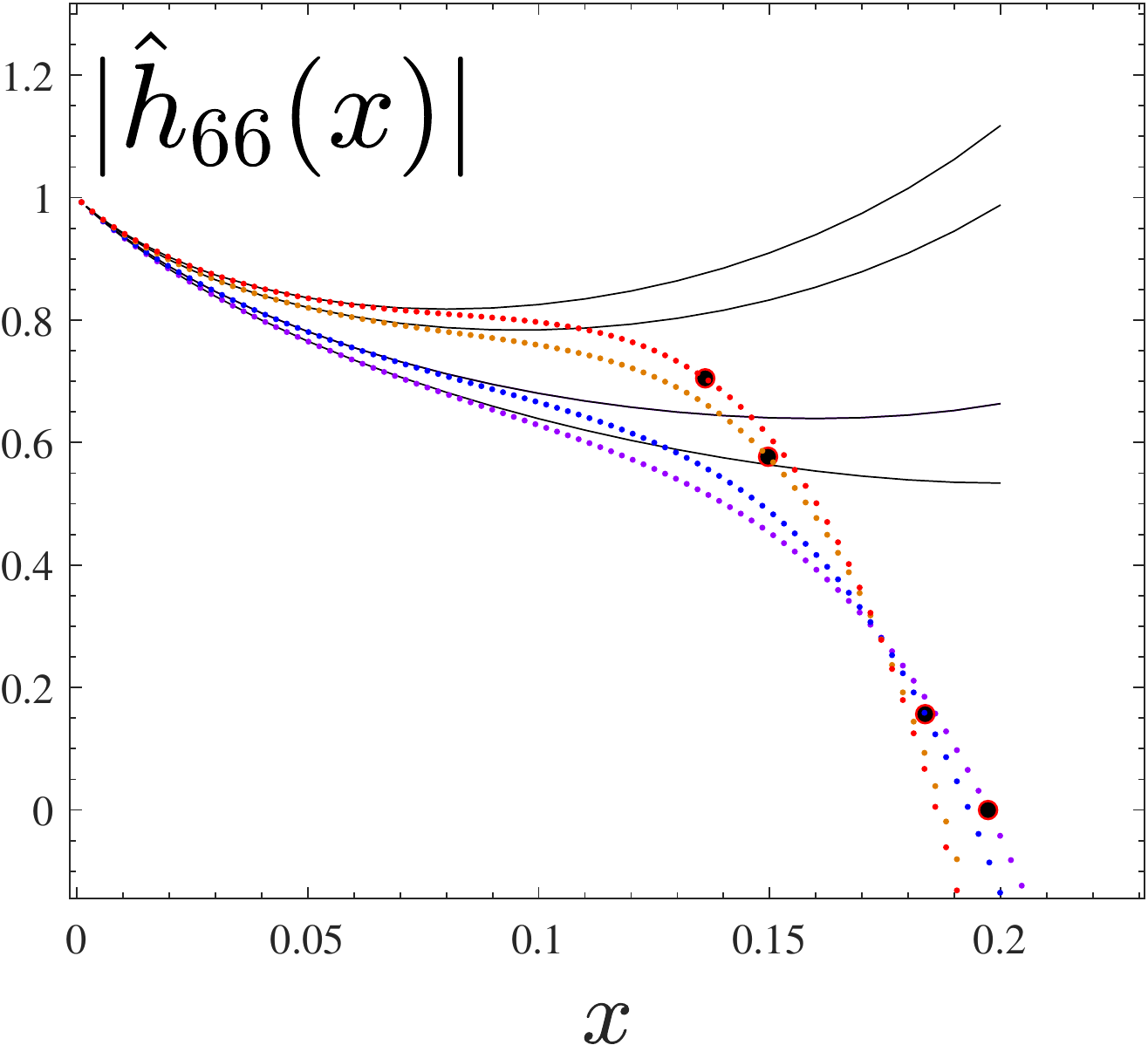}
  \caption{\label{fig:hlmsvsPN}Comparison between the PN-expanded, Newton-normalized,
    waveform amplitudes (dotted lines) and the corresponding numerical ones (black lines).
    The values of the particle spin $\sigma=(-0.90, -0.5, +0.5,+0.90)$
    are respectively indicated by the colors red, orange, blue and purple.
    The colored markers indicate the location of the LSO, Eq.~\eqref{eq:x_LSO}.} 
\end{figure*}
We turn now to comparing the ``exact'' numerical data computed in
the previous section to different analytical approximations of 
$|\hat{h}_\lm^{(\epsilon)}|$. More precisely:
\begin{itemize}
\item[(i)] In Sec.~\ref{subsec:PN}, we consider the straightforward PN-expanded 
expressions of $|\hat{h}_\lm^{(\epsilon)}|$ up to 5.5PN order.
\item[(ii)] In Sec.~\ref{subsec:pan_et_al} we consider the, now standard,
  factorization and resummation of Refs.~\cite{Damour:2008gu,Pan:2010hz}.
  Note that this analytical approach is adopted in state-of-the-art EOB waveform 
  models for coalescing binaries~\cite{Bohe:2016gbl,Nagar:2018zoe}.
\item[(iii)] In Sec.~\ref{sec:iResum} we explore the performance of a hybrid approach
that consists of: implementing the orbital factorization, and subsequent resummation,
of Refs.~\cite{Nagar:2016ayt,Messina:2018ghh} for the $\epsilon=0$ modes, while
the $\epsilon=1$ only benefit the factorization of the tail contribution.
\item[(iv)] In Sec.~\ref{sec:iResum_odd} we test the special orbital-factorization and resummation 
of odd-$m$ modes that is adopted in the $\nu\neq 0$ case and that was recently shown to 
yield excellent consistency between EOB and state-of-the-art NR waveform amplitudes.
\end{itemize}

\subsection{PN-expanded waveform amplitudes}
\label{subsec:PN}
The comparison with the plain PN-expanded waveform amplitudes 
is exhibited in Fig.~\ref{fig:hlmsvsPN}. One observes
good consistency between the numerical and analytical results at low frequencies, as expected.
This progressively worsens towards the LSO, that now is indicated by
colored markers on the figure. This worsening is prominent
for higher modes. It is, however, interesting to note
that the PN-expanded $m=1$ modes already offer an excellent
representation of the numerical data up to the LSO and for
any value of $\ell$. 
%
\begin{figure*}[t]
  \center
  \includegraphics[width=0.17\textwidth]{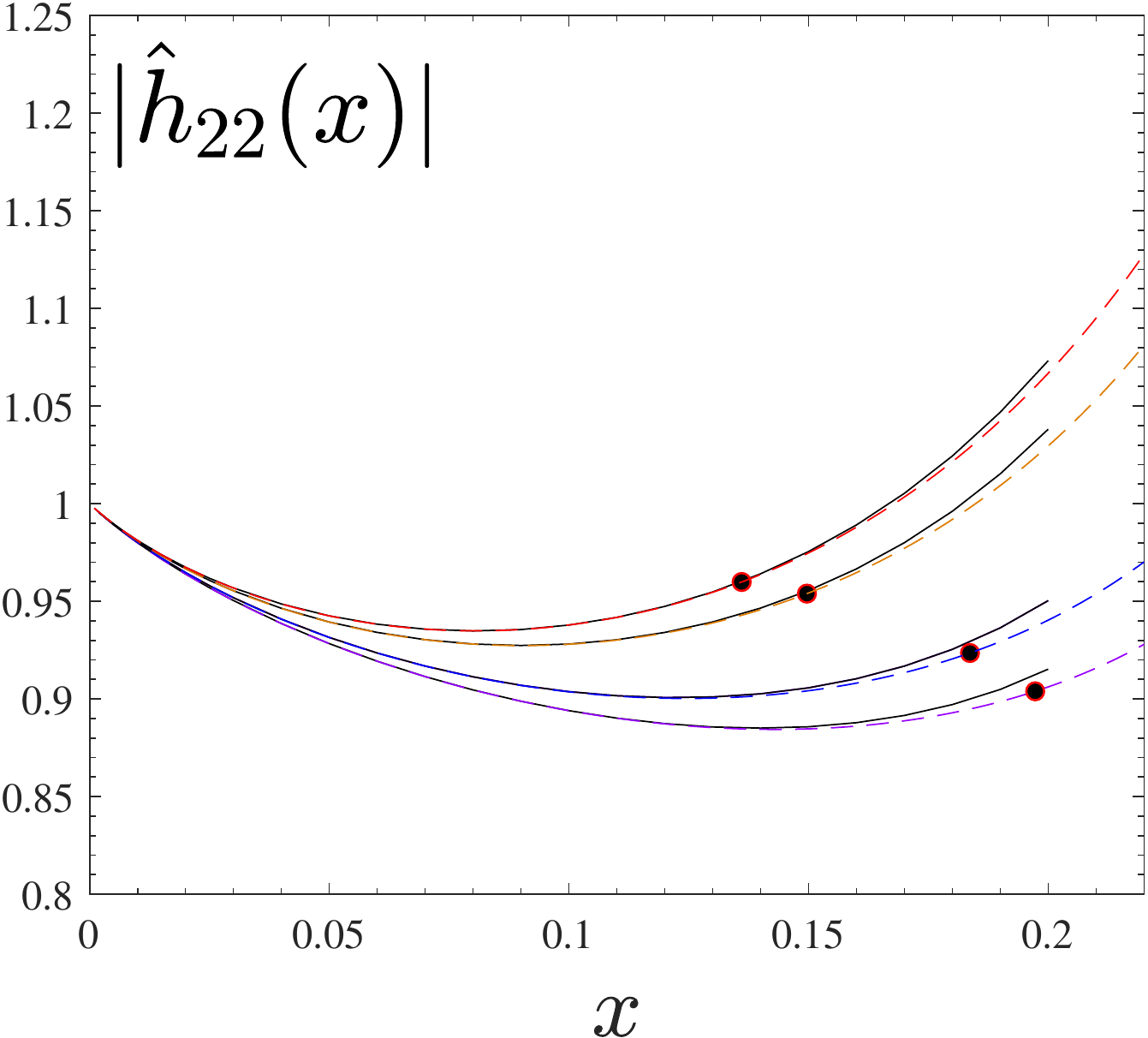}
  \includegraphics[width=0.17\textwidth]{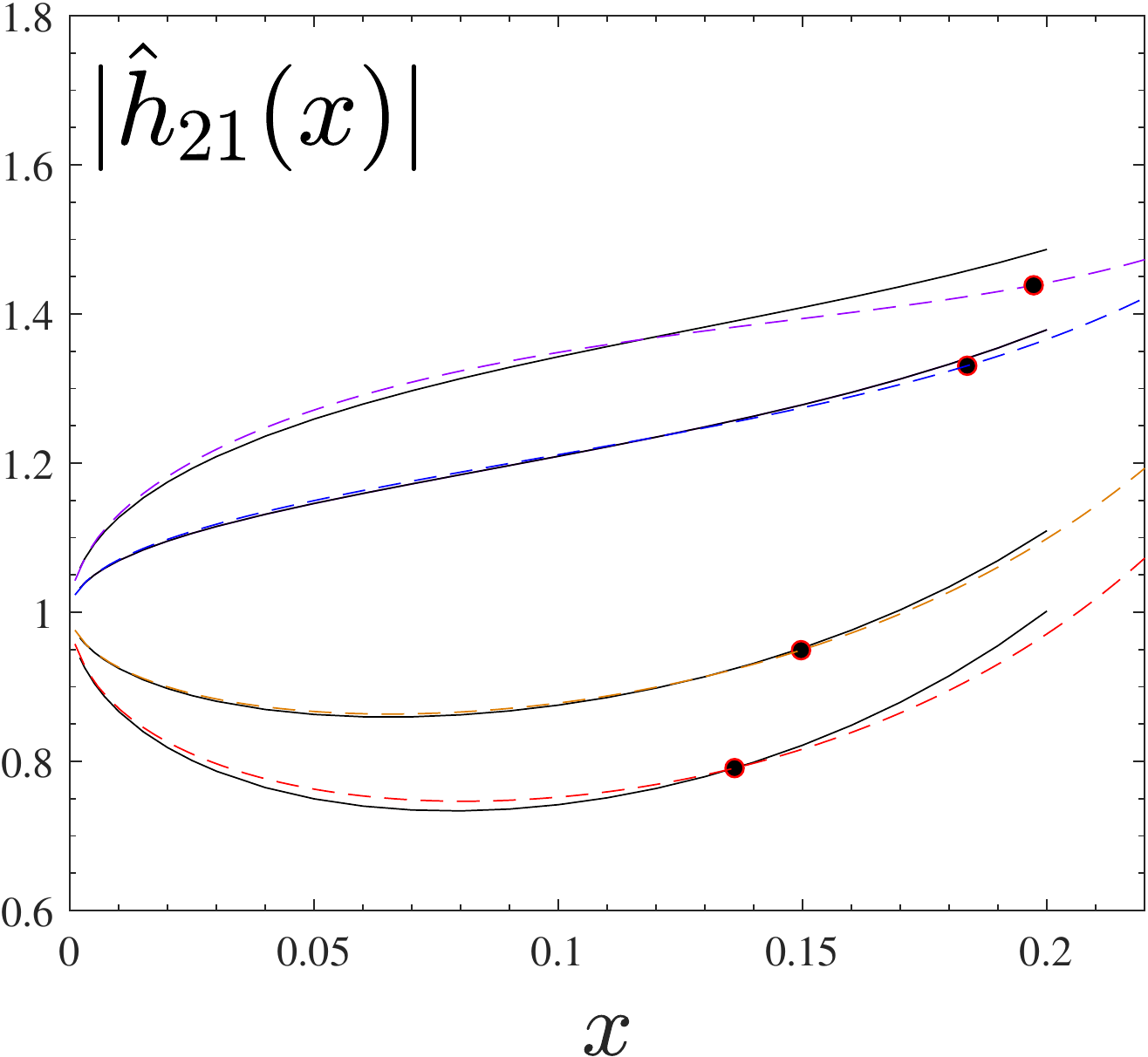}
  \includegraphics[width=0.17\textwidth]{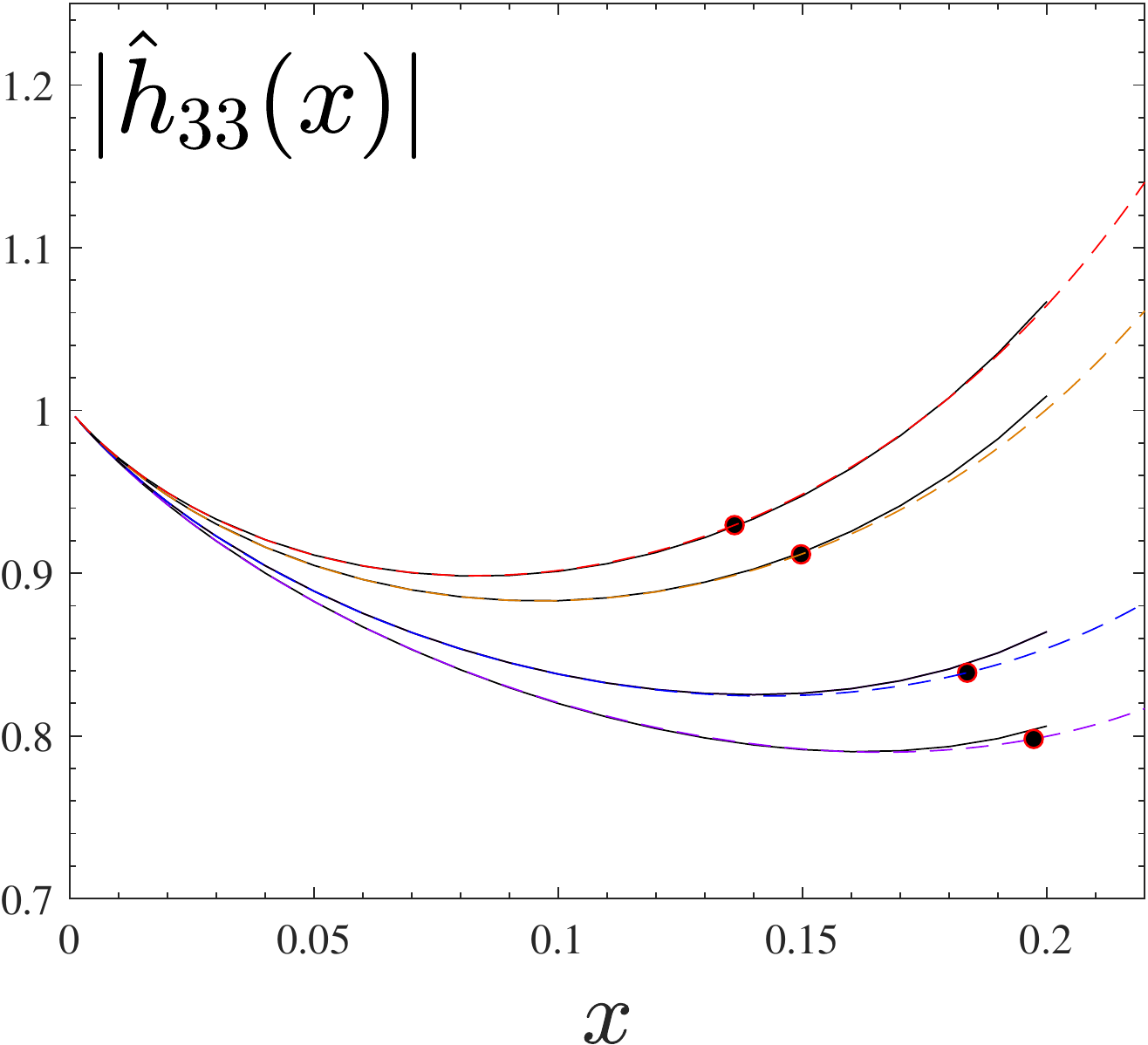}
  \includegraphics[width=0.17\textwidth]{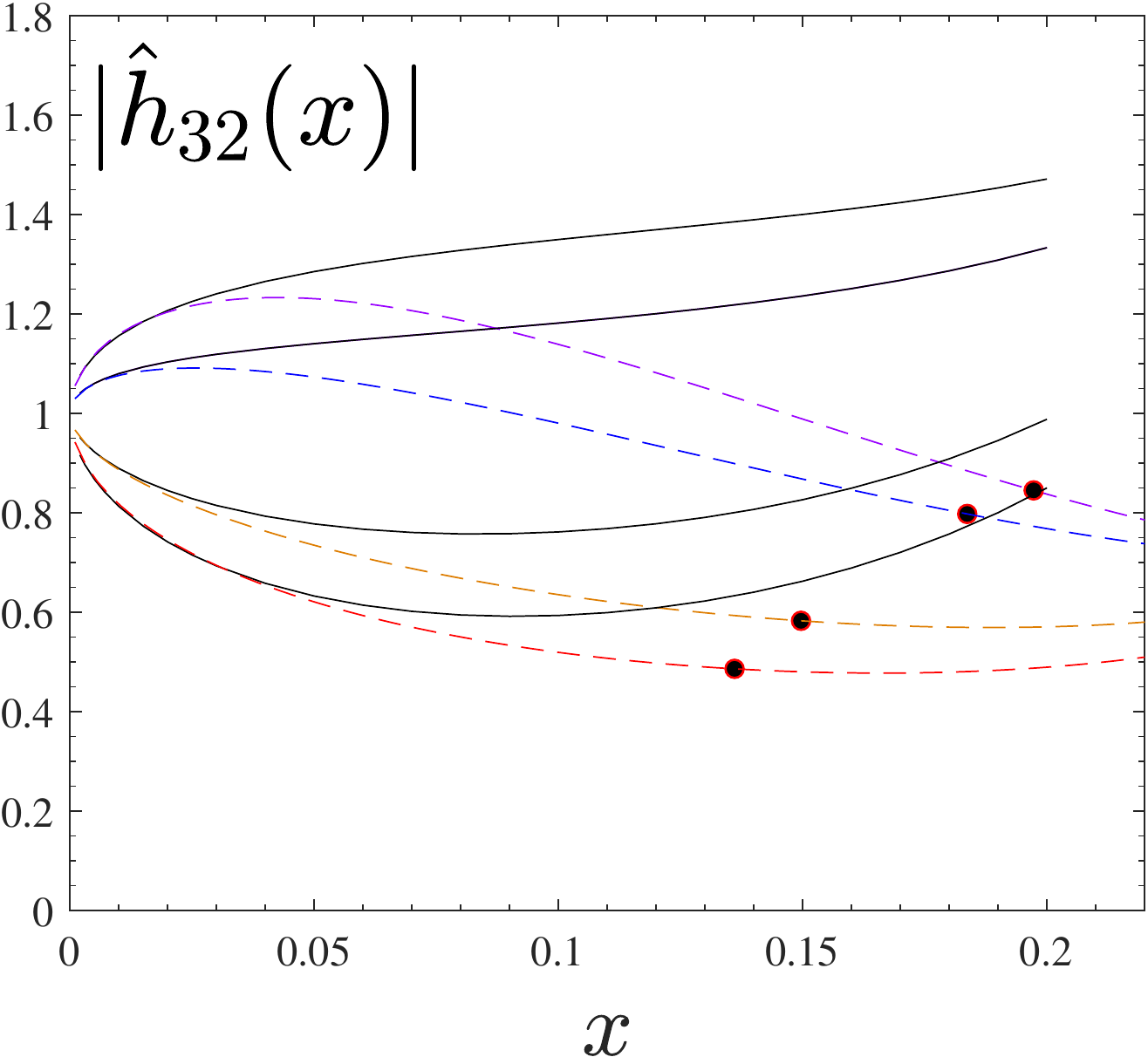}
  \includegraphics[width=0.17\textwidth]{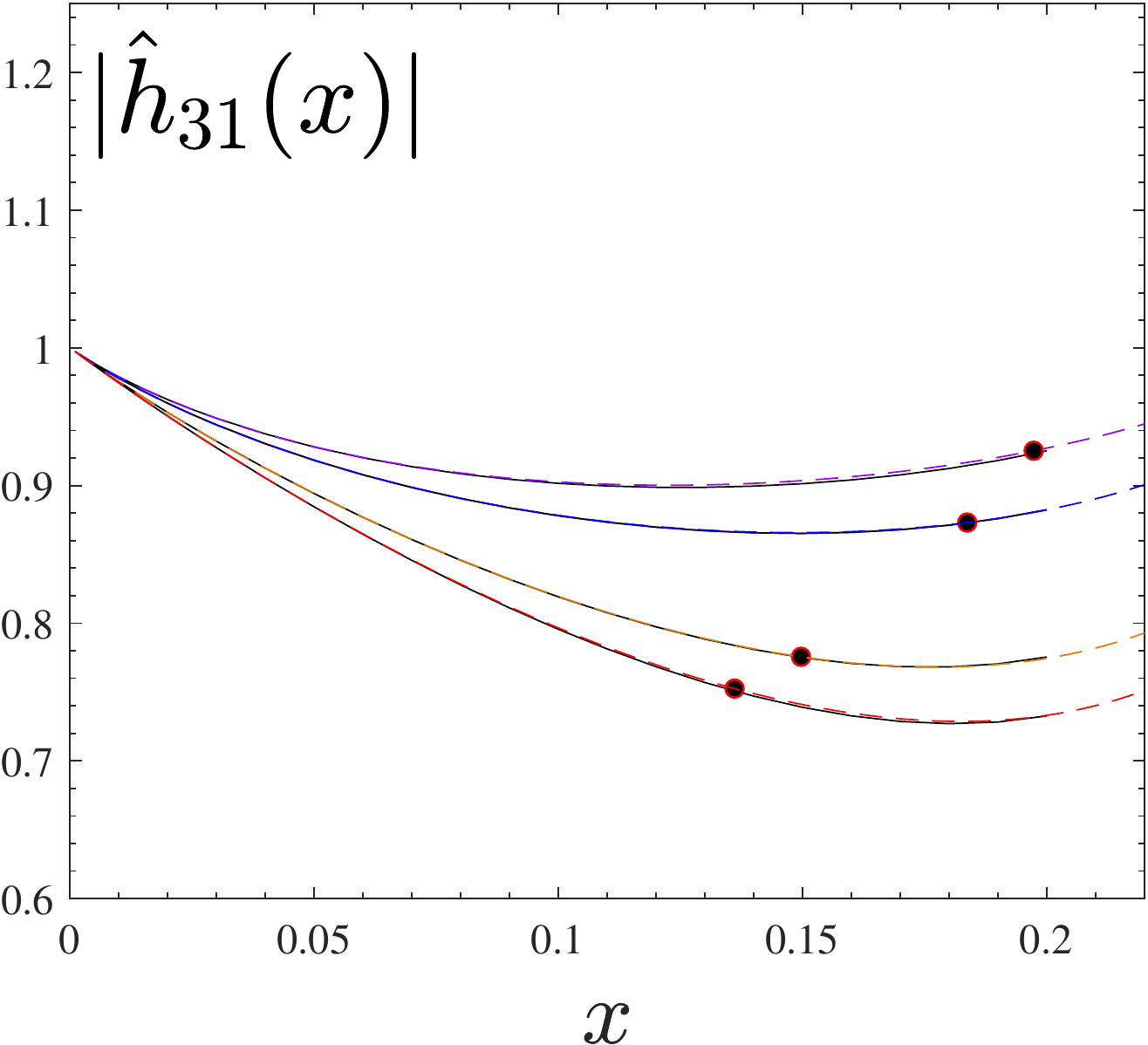}\\
  \includegraphics[width=0.17\textwidth]{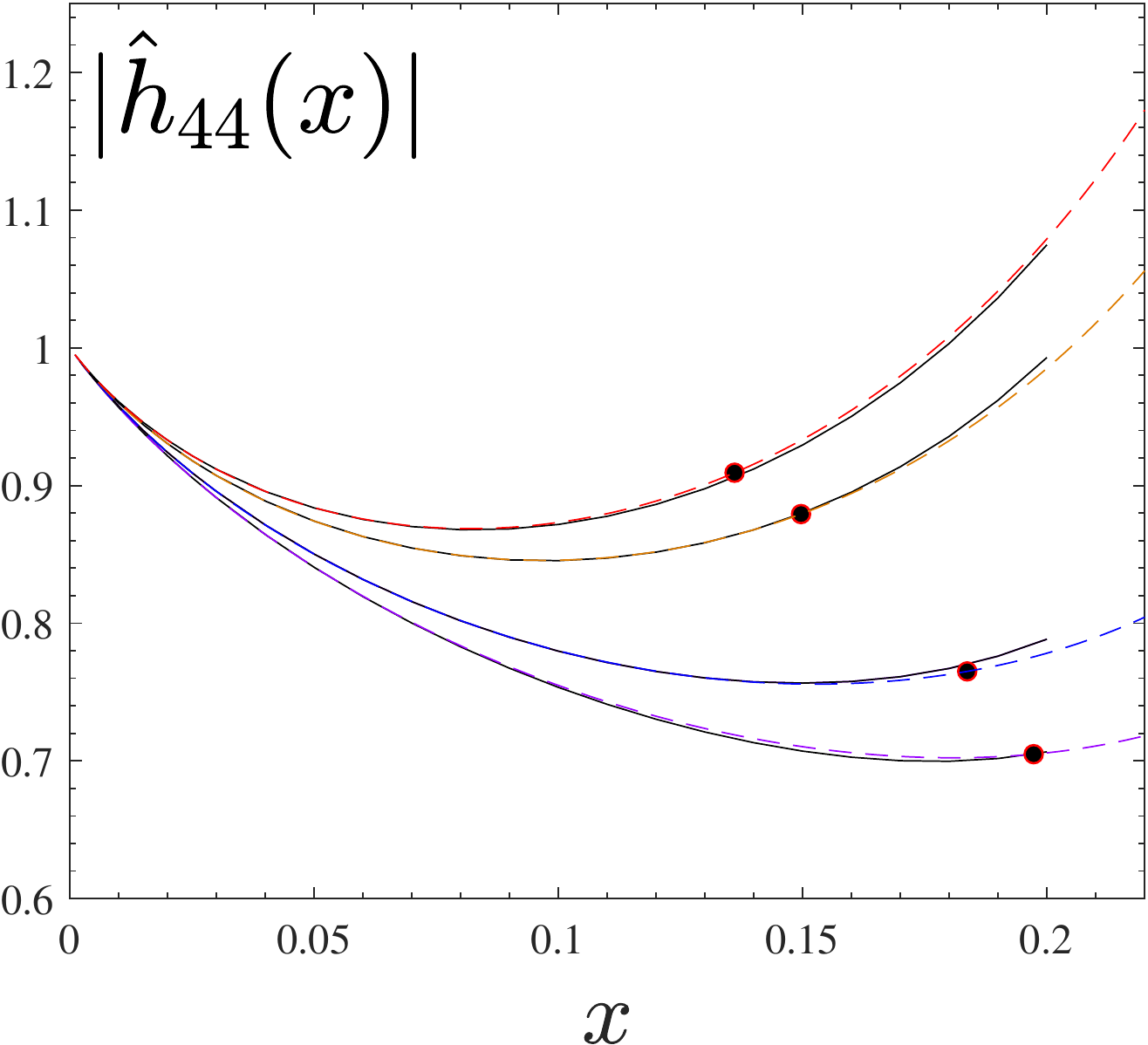}
  \includegraphics[width=0.17\textwidth]{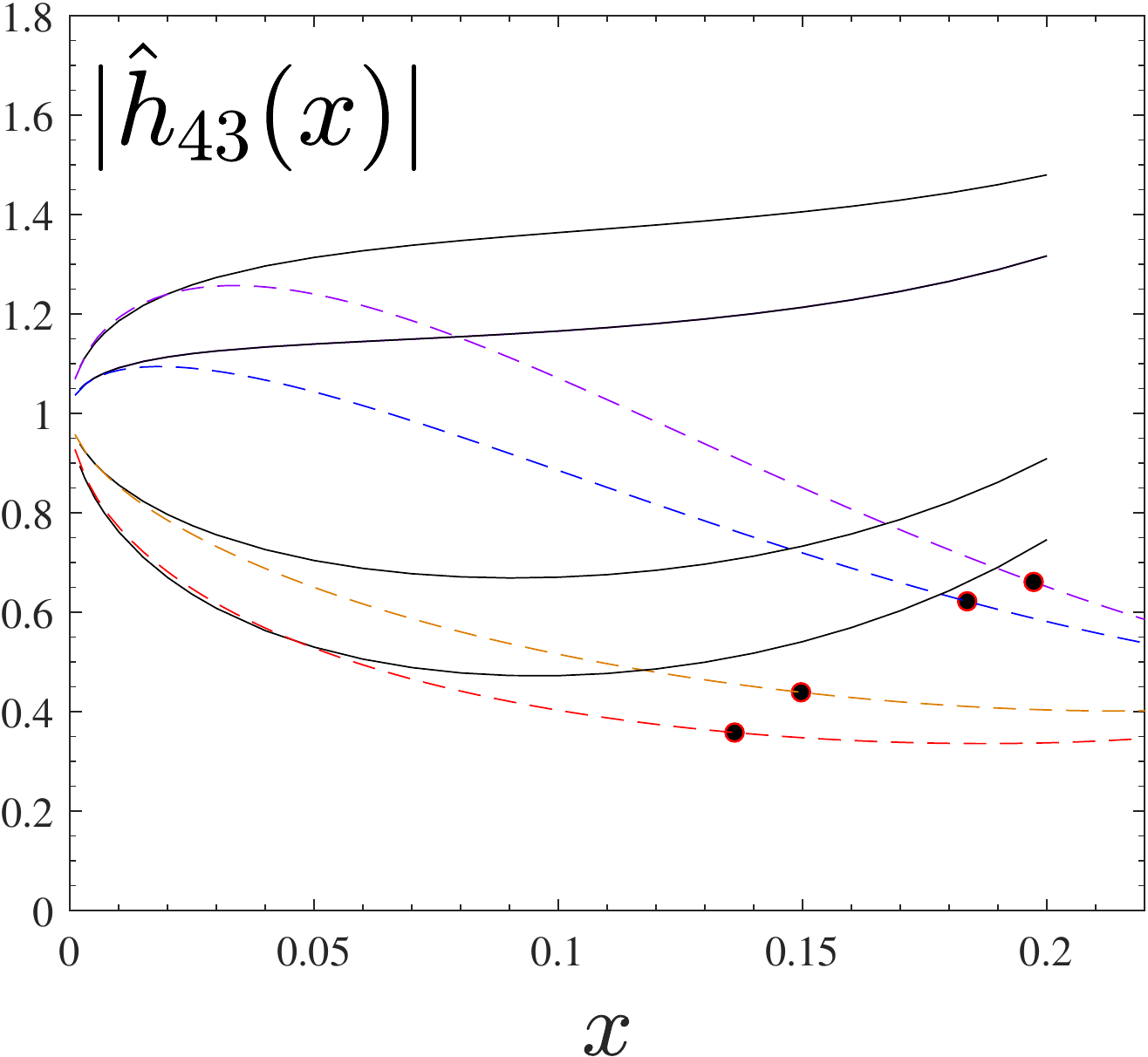}
  \includegraphics[width=0.17\textwidth]{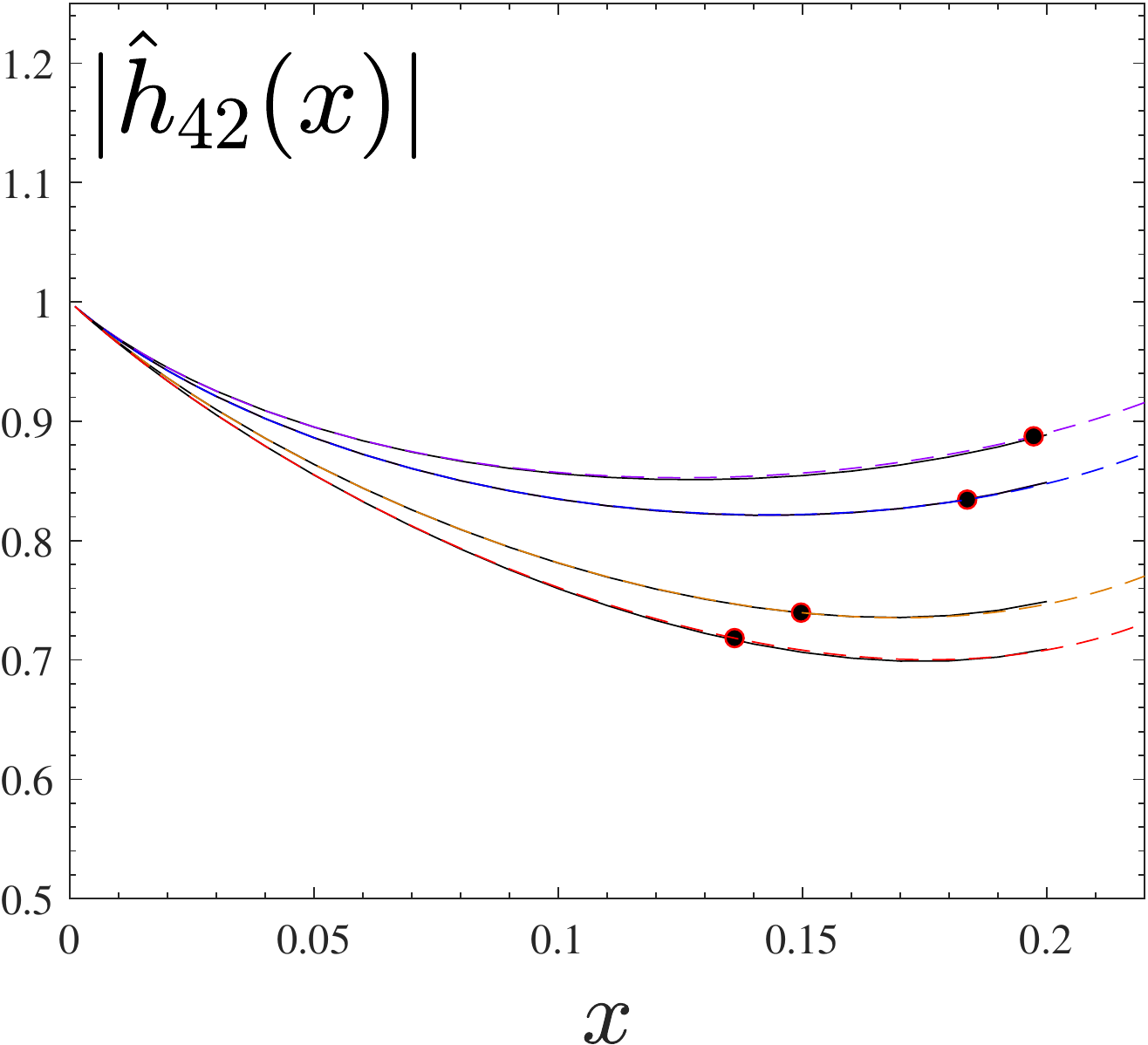}
  \includegraphics[width=0.17\textwidth]{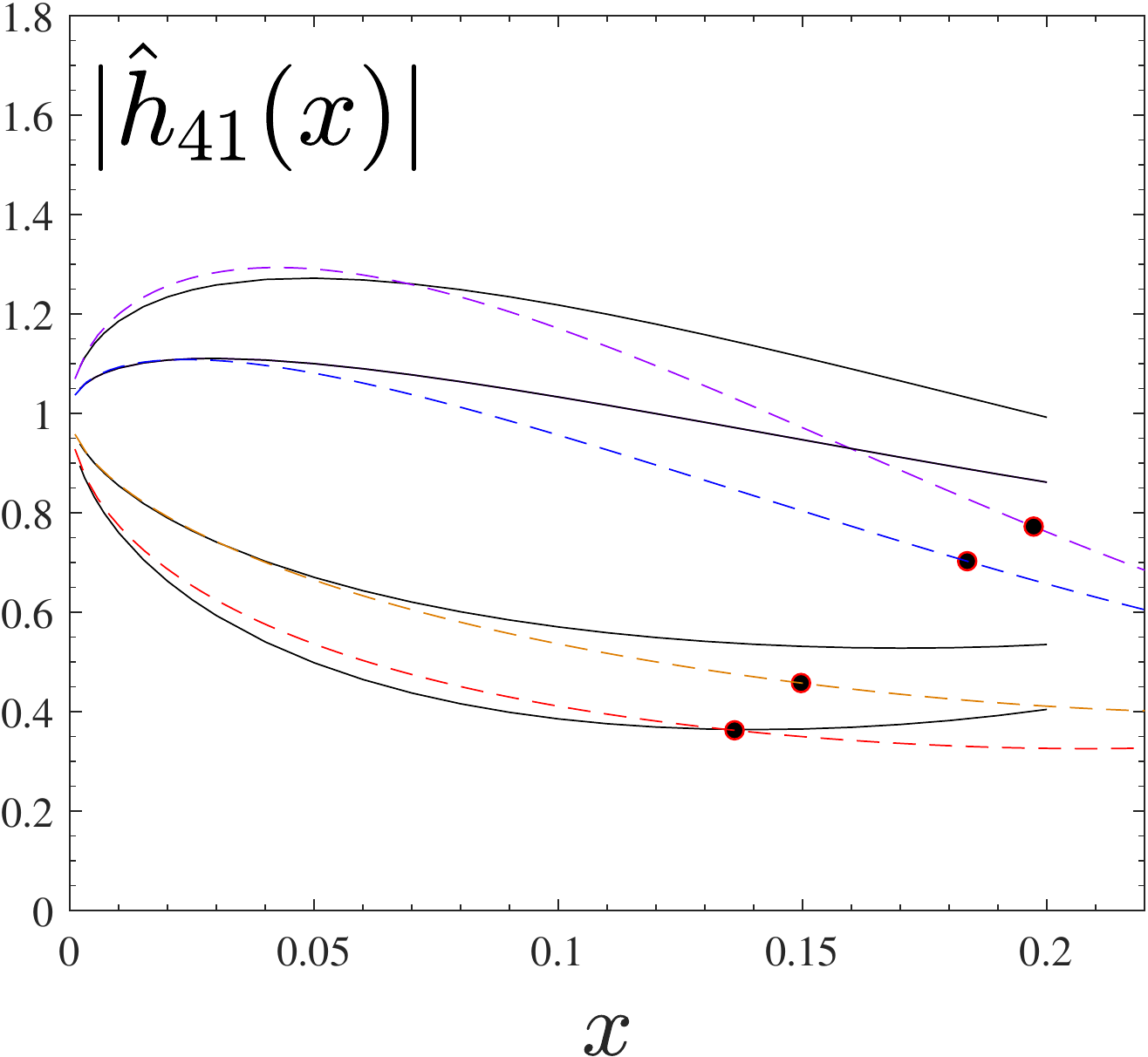}
  \includegraphics[width=0.17\textwidth]{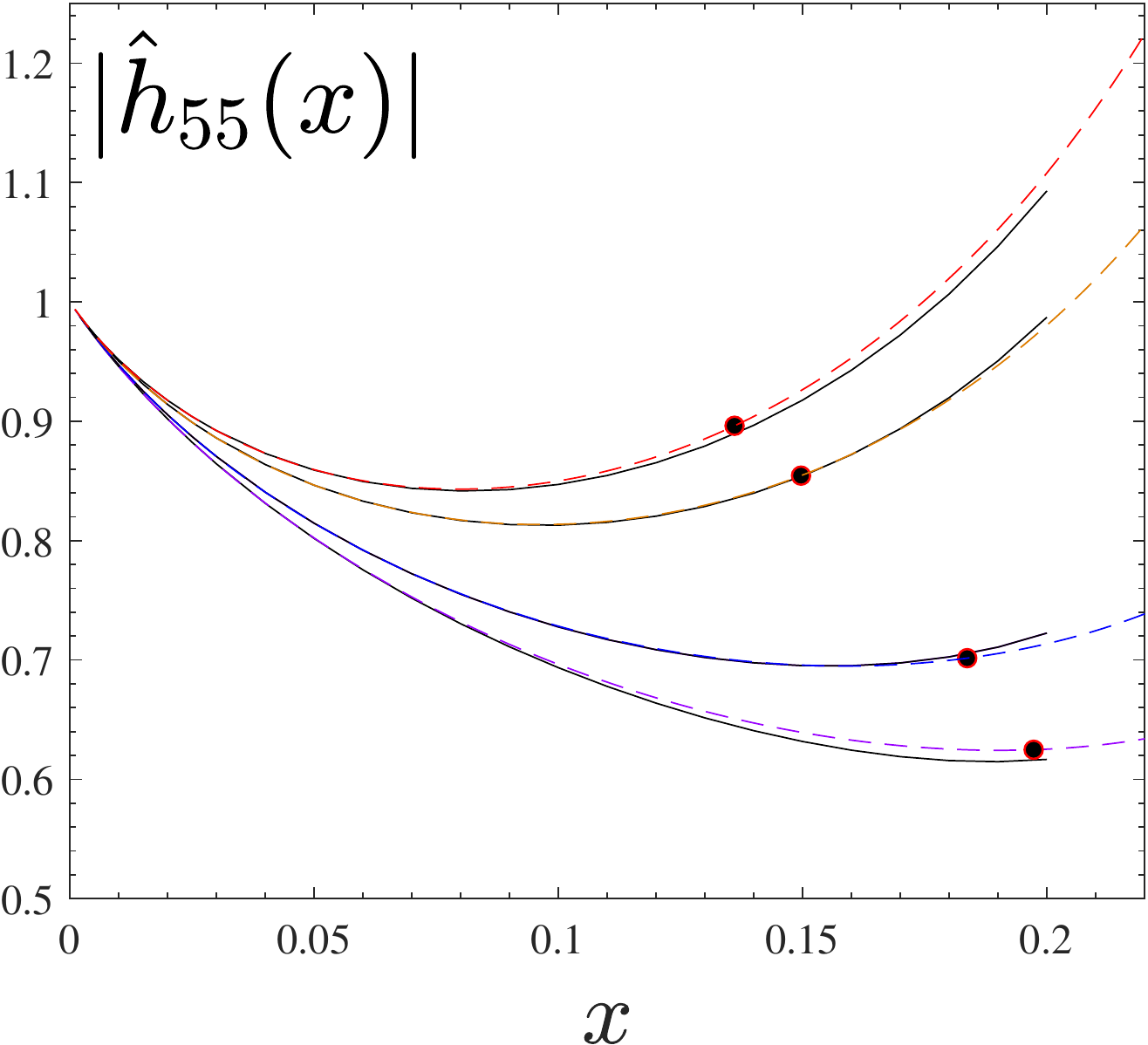}\\
  \includegraphics[width=0.17\textwidth]{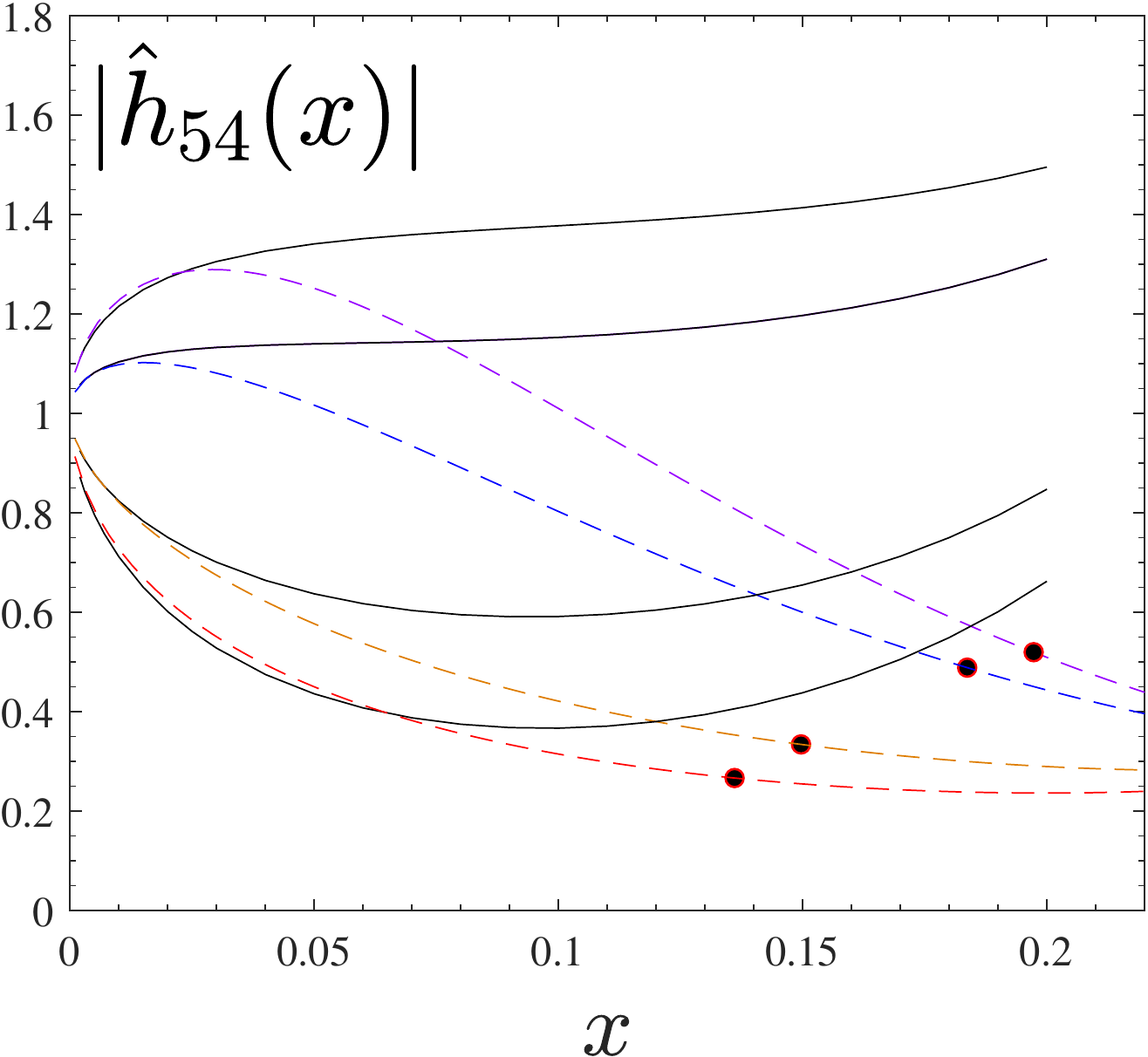}
  \includegraphics[width=0.17\textwidth]{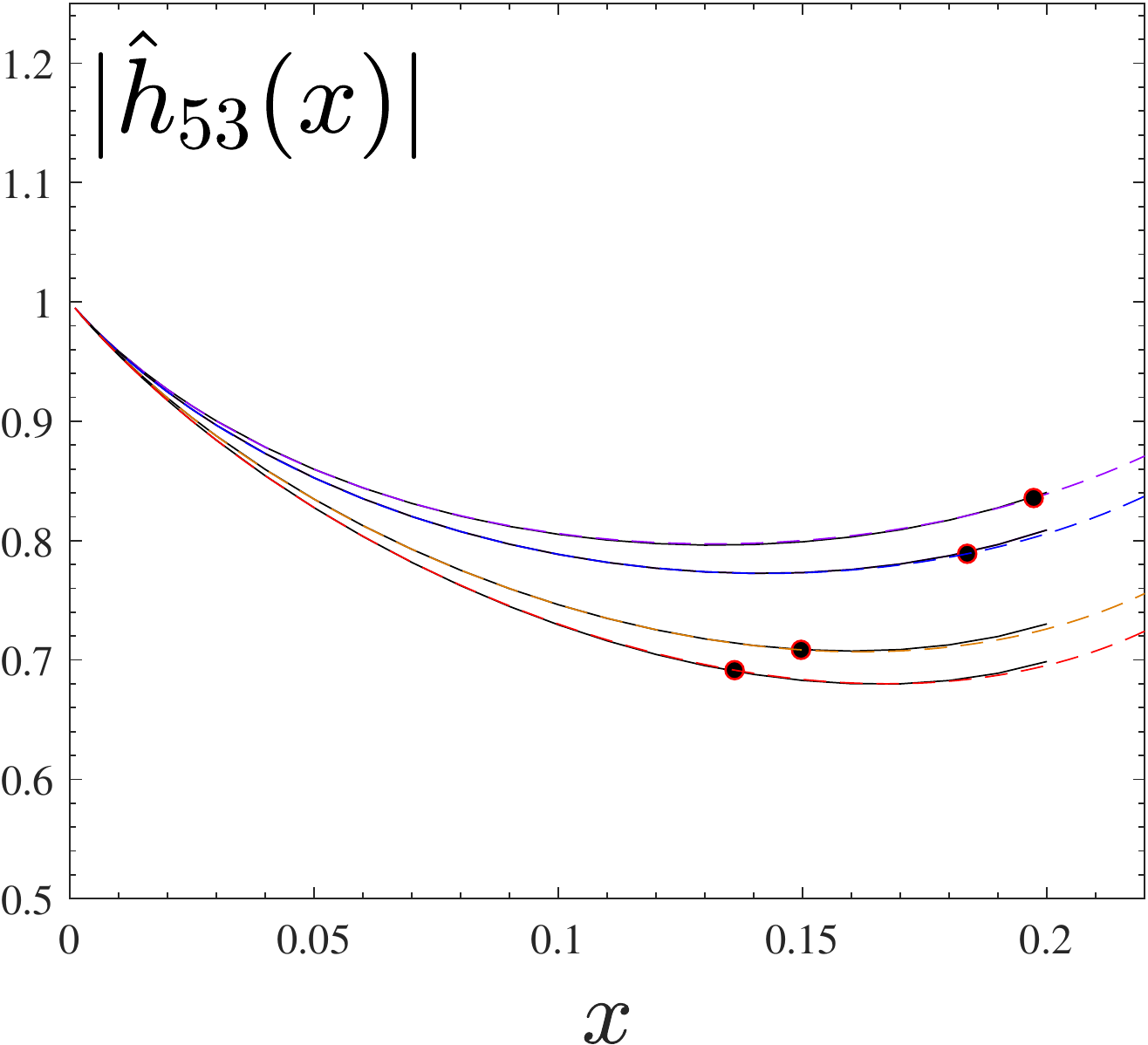}
  \includegraphics[width=0.17\textwidth]{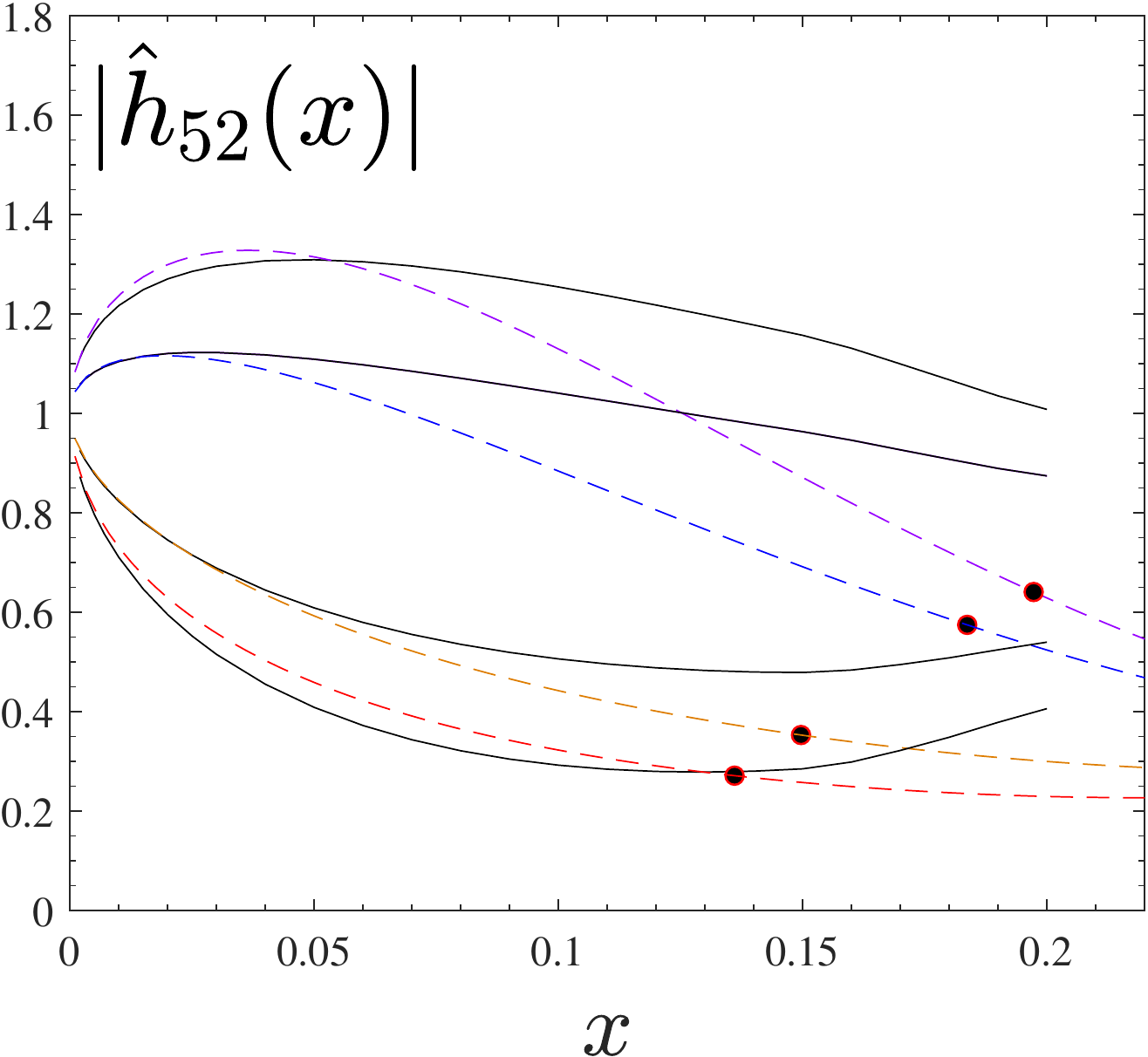}
  \includegraphics[width=0.17\textwidth]{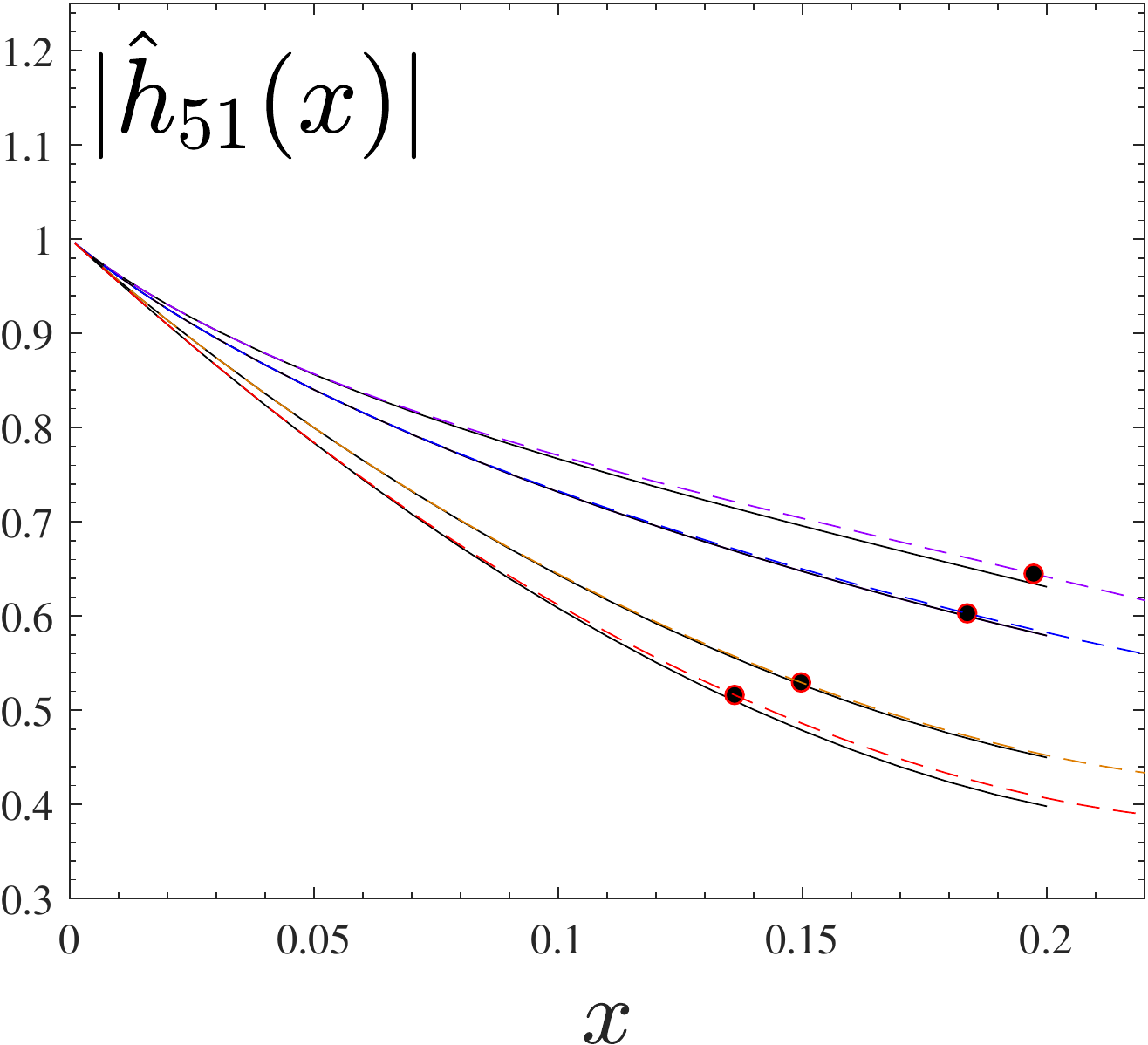}
  \includegraphics[width=0.17\textwidth]{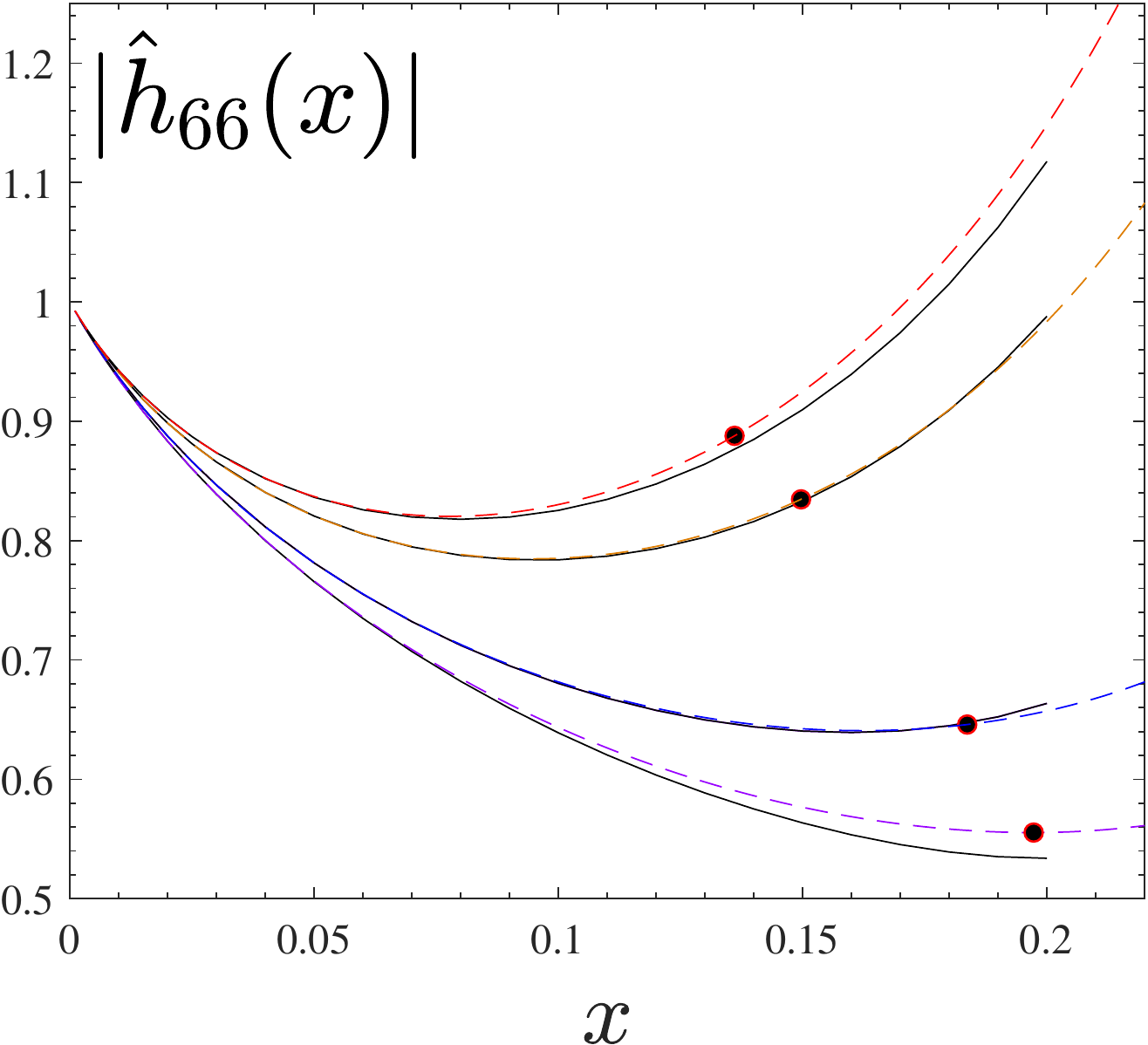}
  \caption{\label{fig:hlmsvs_pan}Comparison between the factorized and resummed,
    Newton-normalized, waveform amplitudes (dotted lines) following the standard procedure of 
    Ref.~\cite{Pan:2011gk} and the corresponding numerical ones (black lines). The values of the 
    particle spin $\sigma=(-0.90, -0.5, +0.5,+0.90)$ are respectively indicated by the colors red,
    orange, blue and purple. The colored markers indicate the location of the LSO, Eq.~\eqref{eq:x_LSO}.
    The analytical approximation is, in general, rather inaccurate for $\ell+m=\text{odd}$ modes.
  } 
\end{figure*}

\subsection{Standard resummation: factorized and resummed amplitudes}
\label{subsec:pan_et_al}
We start by investigating whether the standard factorization and resummation
procedure of waveform amplitudes of Refs.~\cite{Damour:2008gu,Pan:2011gk}
can reduce the gap between the PN-expanded and numerical amplitudes
towards the LSO seen in Fig.~\ref{fig:hlmsvsPN}. The PN-expanded circularized 
waveform amplitudes are resummed in the following factorized form
\begin{equation}
\label{eq:def_hlm}
|\hat{h}^{(\epsilon)}_\lm| =\hat{S}^{(\epsilon)}|\hat{h}^{\rm tail}_\lm(x)| (\rho_\lm)^\ell,
\end{equation}
where $\epsilon$ is the parity of $\ell+m$, $\hat{S}^{(\epsilon)}$ is the 
$\mu$-normalized source of the field, $|\hat{h}_\lm(x)|$ the modulus of the tail factor and 
$\rho_\lm$ the residual factorized amplitudes. The squared modulus of the tail 
factor is given by Eq.~(59) of Ref.~\cite{Damour:2008gu}, and reads
\begin{equation}
\label{eq:tailmod}
|\hat{h}_{\ell m}^{\rm tail}(x)|^{2}=\dfrac{4\pi m x^{3/2}\prod_{s=1}^{\ell}\left(s^{2}+ 2 m x^{3/2}\right)^{2}}{(\ell!)^{2}
  \left(1-e^{-4\pi m}\right)}.
\end{equation}
The source $S^{(\epsilon)}$ is either the $\mu$-normalized energy along circular orbits 
($\epsilon=0$), or the Newton-normalized angular 
momentum ($\epsilon=1$).  From the PN-expanded 
Newton-normalized fluxes $\hat{F}_\lm$ we then calculate the PN-expanded 
residual amplitudes $\rho_\lm$. To do so, we need the energy and angular 
momentum of a spinning body along circular orbits of Schwarzschild spacetime  at linear 
order in the spin. These were obtained in Eqs.~(81) and (82) of Ref.~\cite{Harms:2016ctx},
\begin{align}
\label{energy}
\hat{E}(x)&\equiv E/\mu=\hat{E}^{\rm orb}(x)+\hat{E}^\sigma(x)\nonumber\\
&=\frac{1-2x}{\sqrt{1-3x}}-\frac{x^{5/2}\sigma}{\sqrt{1-3x}},\\
\label{angmom}
\hat{j}(x)&=\hat{j}^{\rm orb}(x)+\hat{j}^\sigma(x)\nonumber\\
                &=\frac{1}{\sqrt{1-3x}}-\sqrt{x}\left(1-\frac{1-4x}{\sqrt{1-3x}}\right)\sigma,
\end{align}
where the expression for $\hat{j}$ is obtained multiplying by $\sqrt{x}$ (i.e., the inverse
of the Newtonian angular momentum along circular orbits) Eq.~(81) of Ref.~\cite{Harms:2016ctx}.
The PN-expanded $\rho_\lm$'s are then given by
\begin{equation}
\label{eq:standardfact}
\rho_{\ell m}=T_n\Biggl[\frac{\sqrt{\hat{F}_{\ell m}}}{|\hat{h}^{\rm tail}_{\ell m}(x)|\hat{S}^{(\epsilon)}}\Biggr]^{1/\ell},
\end{equation}
where $T_n$ denotes a Taylor-expansion of order $n$ and $\hat{F}_\lm$ are the Newton-normalized
PN-expanded energy fluxes. The $\rho_{\ell m}$ functions are here written as the sum of
an orbital (spin-independent) and a spin-dependent term
\begin{equation}
\rho_\lm = \rho_\lm^{\rm orb} + \rho_\lm^\sigma.
\end{equation}
The PN knowledge of $\rho_\lm^\sigma$ before this work was limited to global 3.5PN (NNLO)
and the functions are written explicitly in Ref.~\cite{Messina:2018ghh,Cotesta:2018fcv}.
Here the computation of  {\it each} $\rho_\lm^\sigma$ is pushed up to {\it relative} 5.5PN order,
i.e., next-to-next-to-next-to-next-to-leading-order, N$^4$LO in the spin-orbit coupling.
We list the functions explicitly in Appendix~\ref{app:rholms}. One
verifies that the 3.5PN-accurate truncation of our results agrees in full with the 
corresponding formulas of Refs.~\cite{Messina:2018ghh,Cotesta:2018fcv}.

The analytical $|\hat{h}_\lm|$'s from Eq.~\eqref{eq:def_hlm} are compared with the
corresponding numerical ones in Fig.~\ref{fig:hlmsvs_pan}. The effect of the standard resummation can
be summarized as follows. First, the analytically resummed $\ell=m$ modes deliver a rather
good approximation to the numerical functions. One can see that this remains true up to $\ell=m=6$.
However, we see that the procedure gives a rather unreliable result (even for small values of $x$)
for the $\ell+m=\text{odd}$ modes. This illustrates that the standard resummation approach 
introduced in Ref.~\cite{Pan:2011gk} can become highly inaccurate in some corners of the black-hole
binary parameter space. As such, it should be replaced by something else that is more robust. 
One possibility is proposed in the next section.

\subsection{Improved resummation}
\label{sec:iResum}
References~\cite{Nagar:2016ayt,Messina:2018ghh} presented an alternative
factorization and resummation procedure based on the idea of first factoring
out the orbital, spin-independent, contribution to the $\rho_\lm$'s and then
independently resumming, in various ways, the orbital and spin factors. The
same procedure was implemented in similar ways for the odd-parity and even-parity
modes. In the case of a {\it nonspinning} particle on circular orbits around
a Kerr black hole, Refs.~\cite{Messina:2018ghh} illustrated that this procedure
yields a remarkable analytical/numerical agreement between the $\rho_\lm$'s
(and fluxes) up to the LSO {\it also} for extremal values of the spin
of the black hole (see e.g. Figs.~1 and~2) of~\cite{Messina:2018ghh}).
In this Section we test this procedure on the even-parity modes.
By contrast, we apply a different factorization, that only concerns
the tail factor, to the odd-parity modes. Some of the resummation
procedures of Refs.~\cite{Messina:2018ghh} applied to some of the odd-parity
modes are instead discussed in Sec.~\ref{sec:iResum_odd} below.

\begin{figure*}[t]
  \center
  \includegraphics[width=0.18\textwidth]{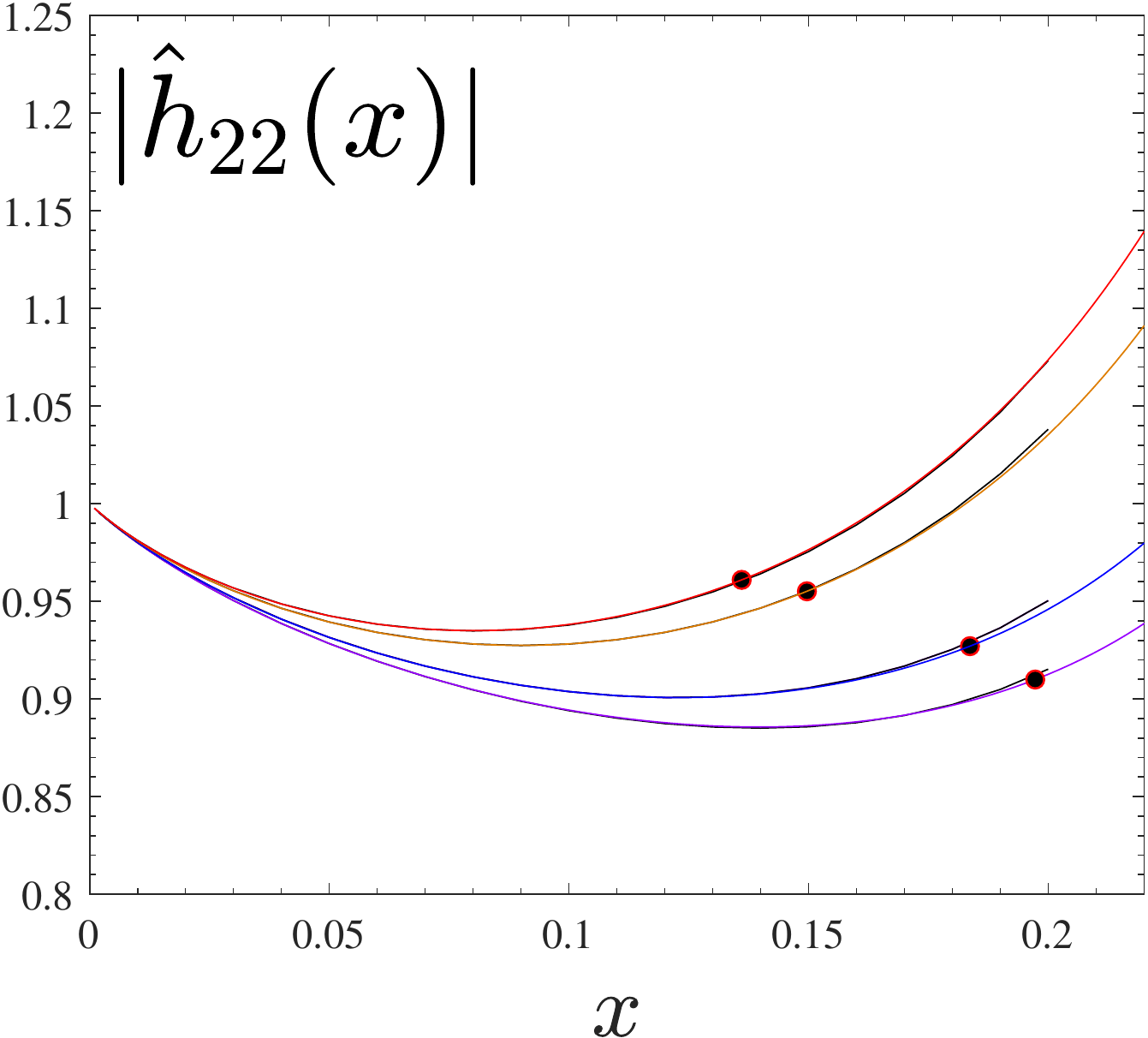}
  \includegraphics[width=0.17\textwidth]{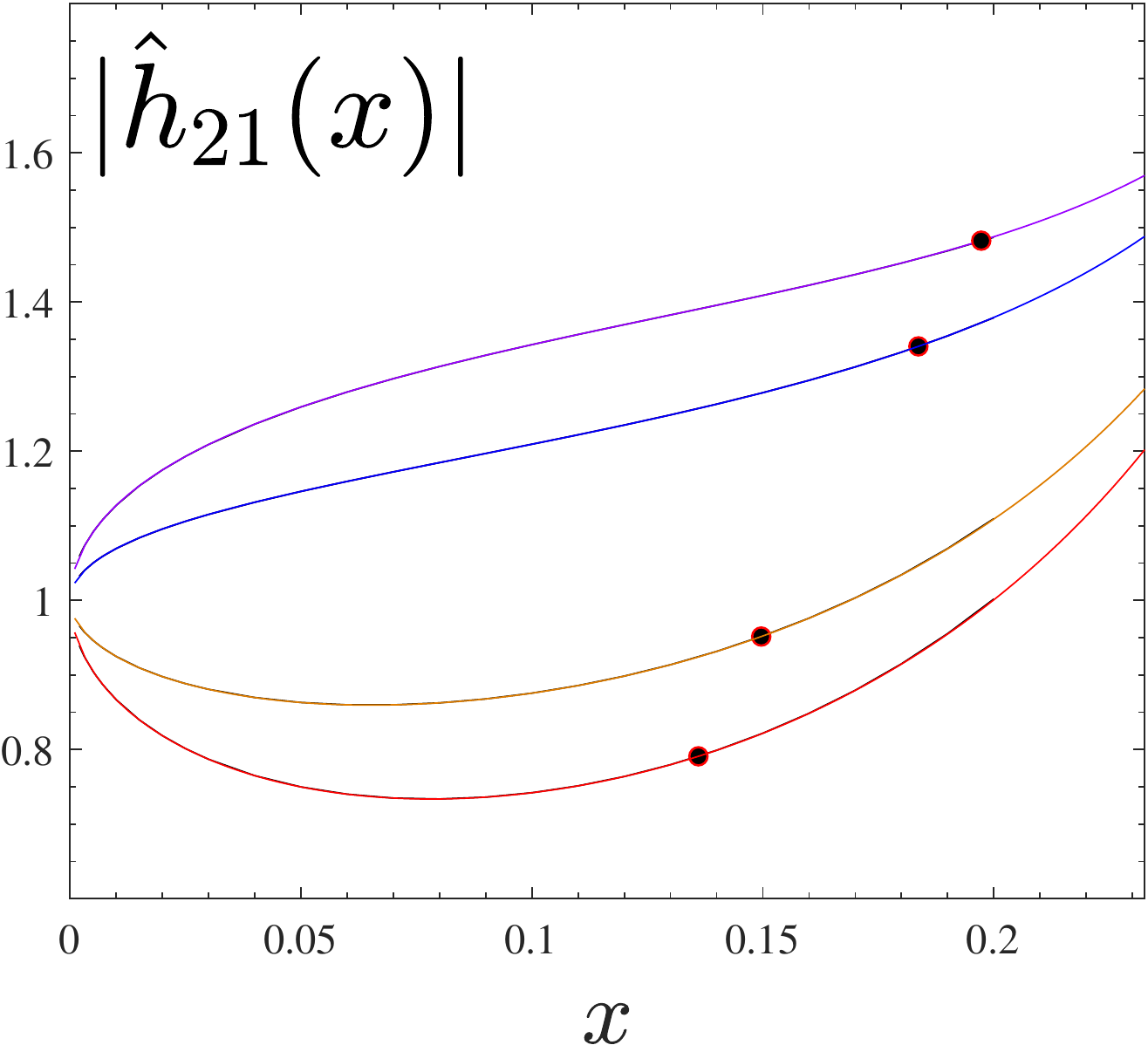}
  \includegraphics[width=0.17\textwidth]{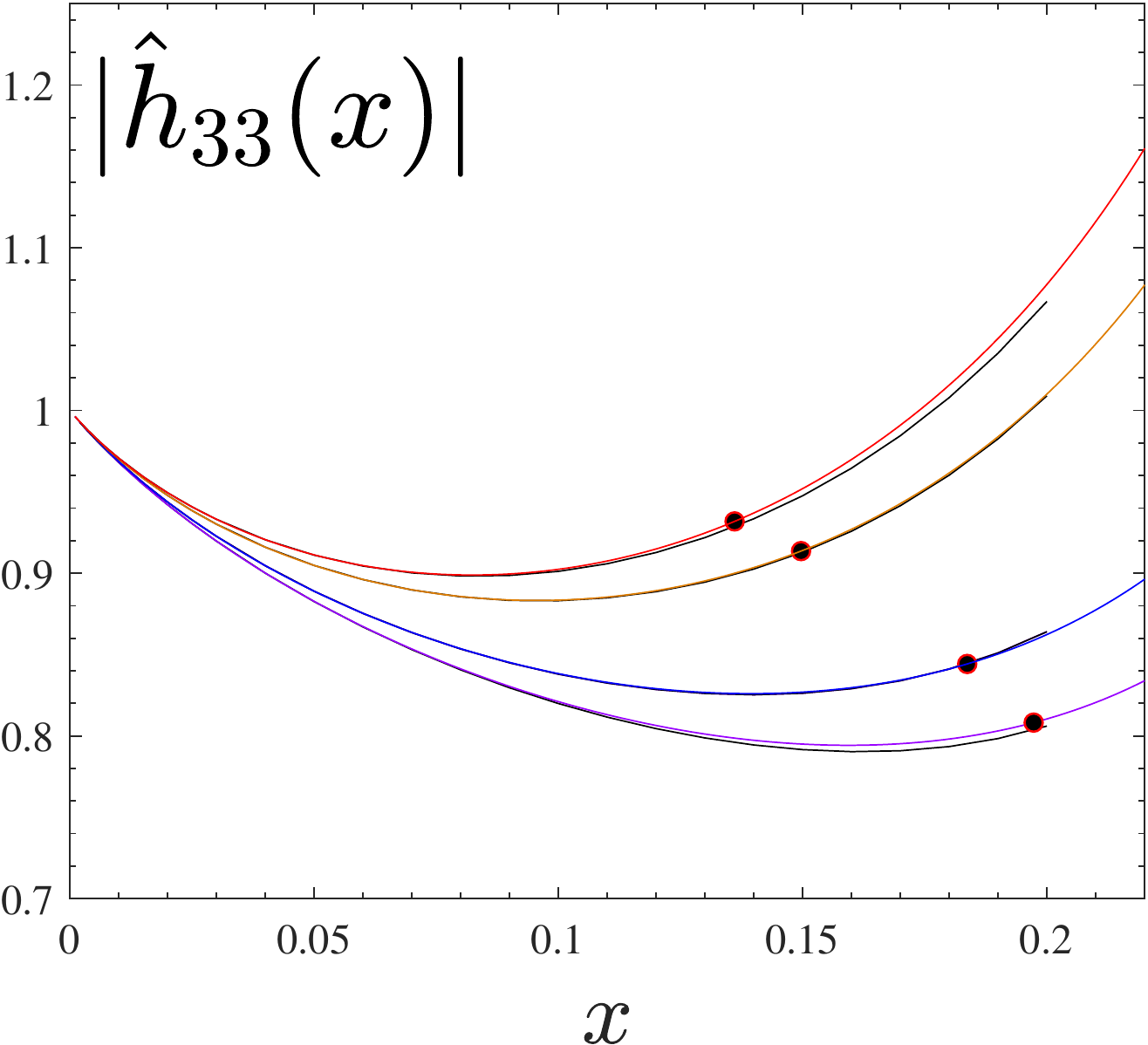}
  \includegraphics[width=0.178\textwidth]{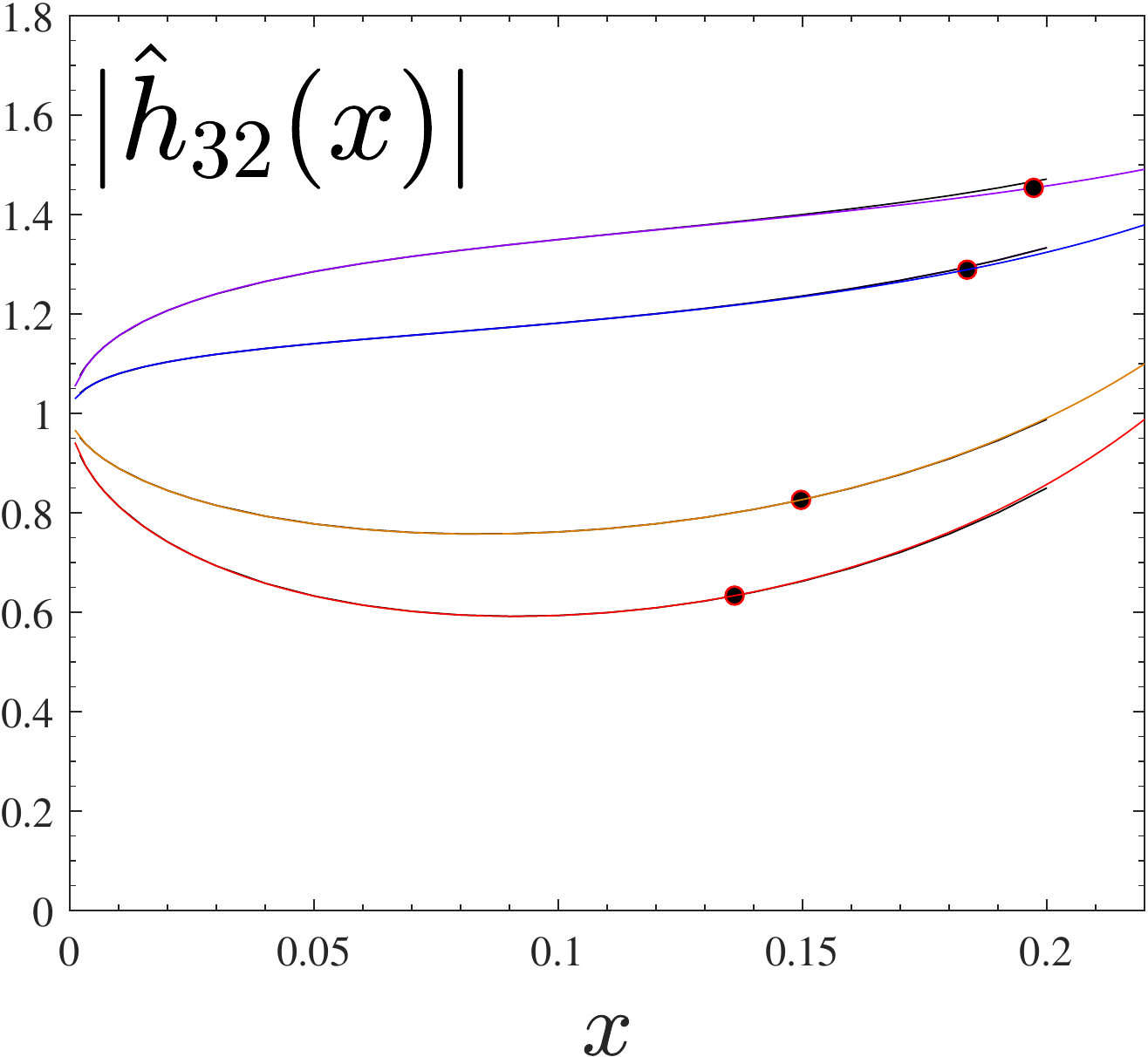}
  \includegraphics[width=0.178\textwidth]{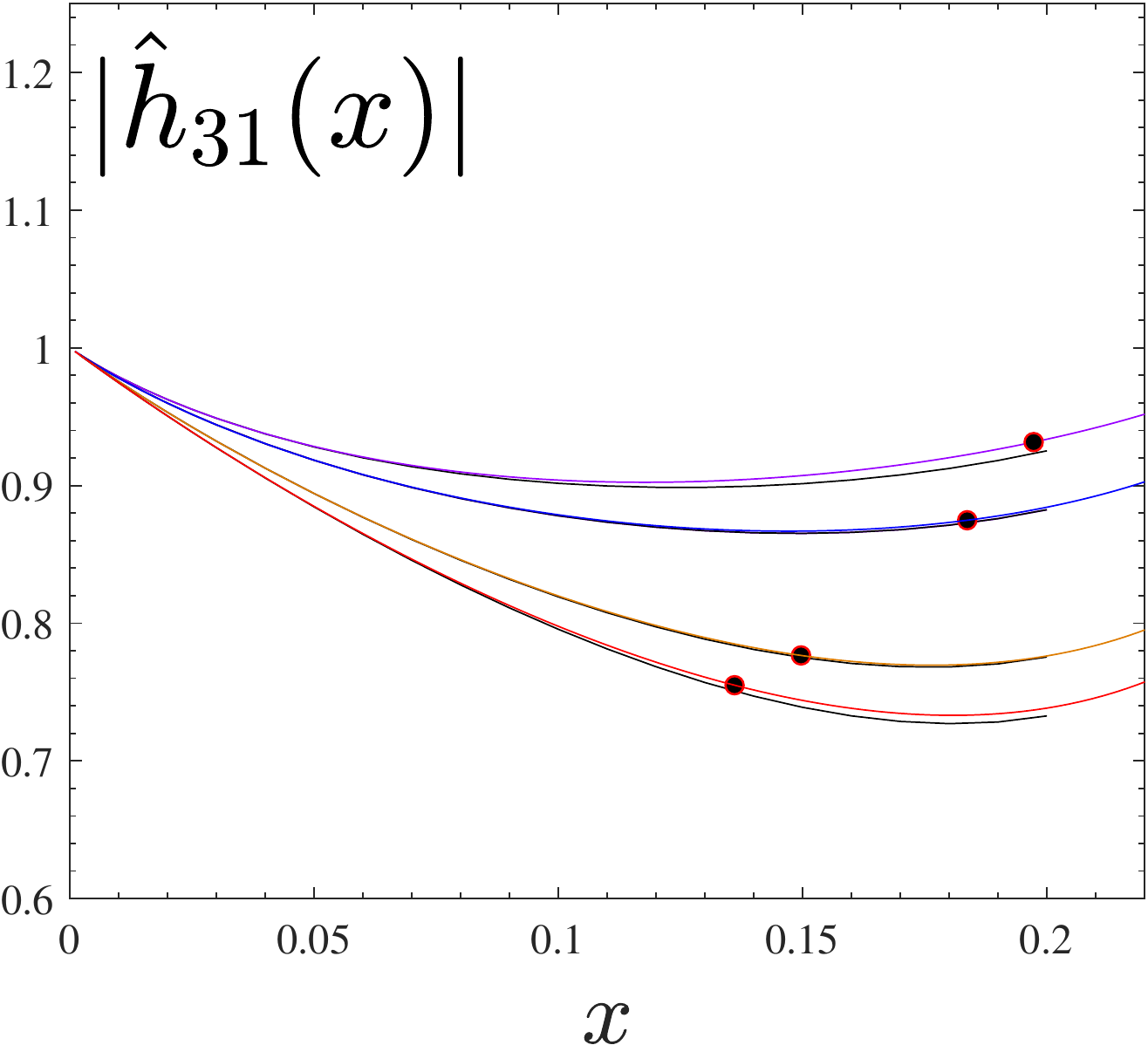}\\
  \includegraphics[width=0.17\textwidth]{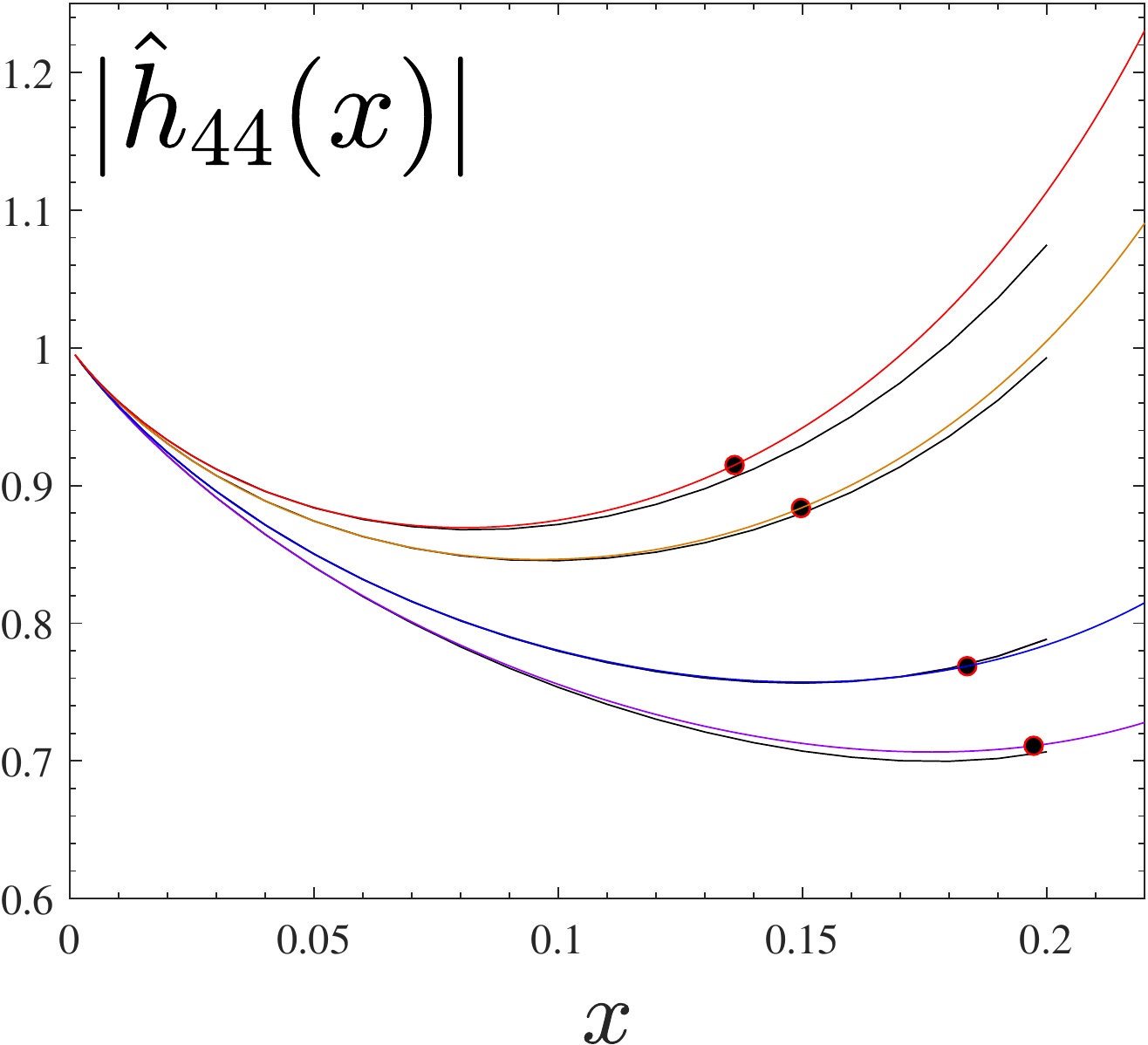}
  \includegraphics[width=0.17\textwidth]{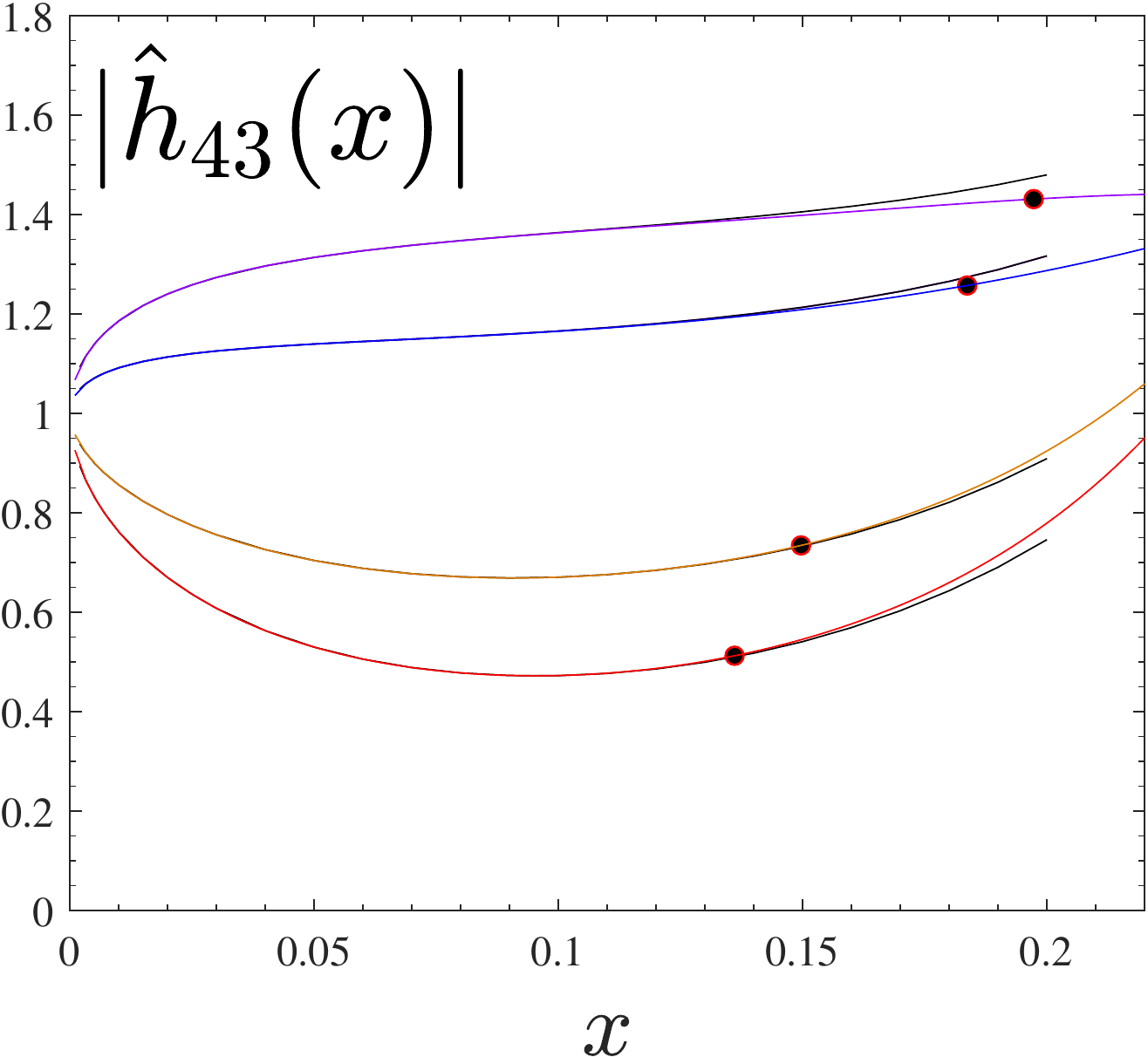}
  \includegraphics[width=0.17\textwidth]{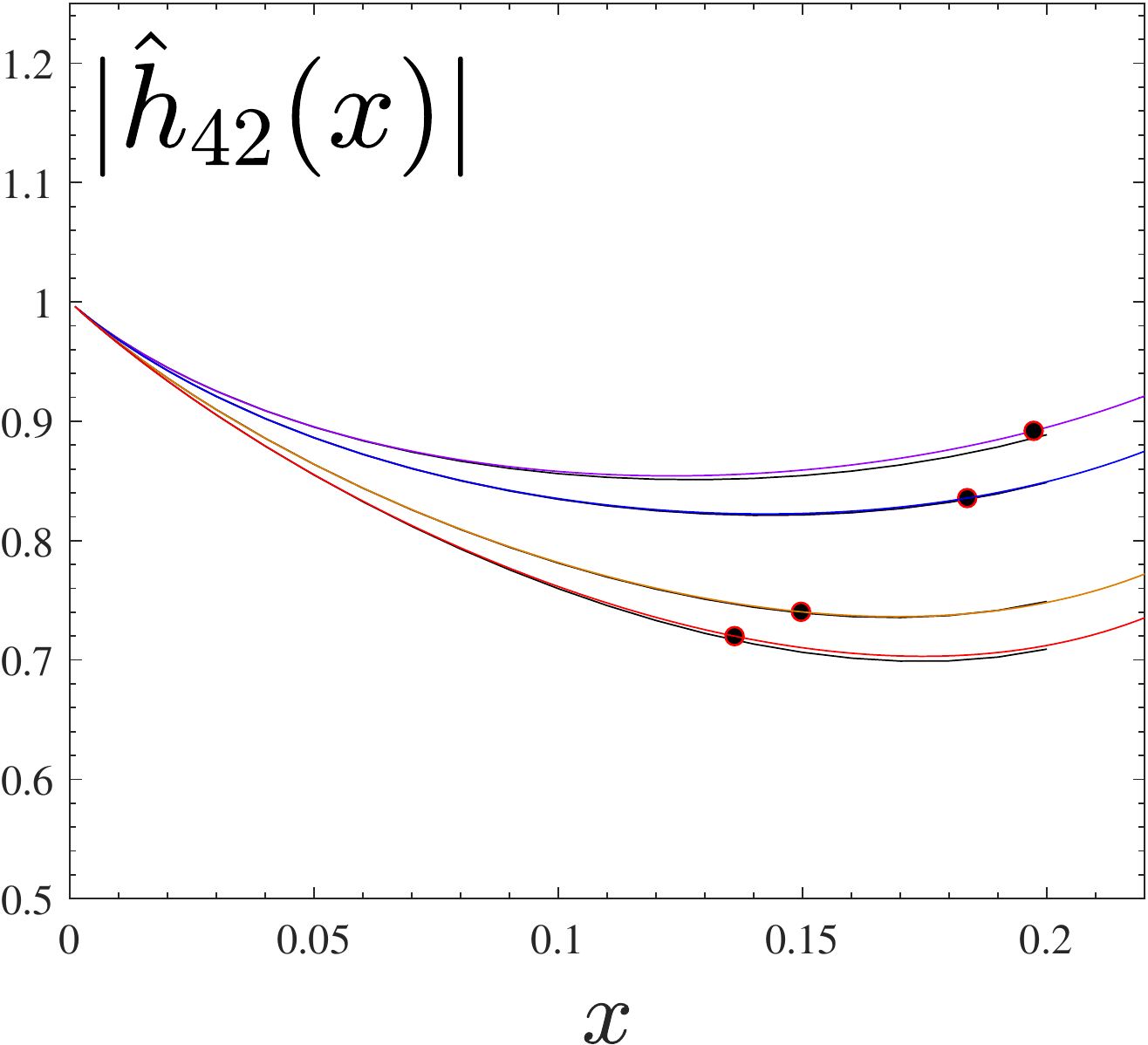}
  \includegraphics[width=0.17\textwidth]{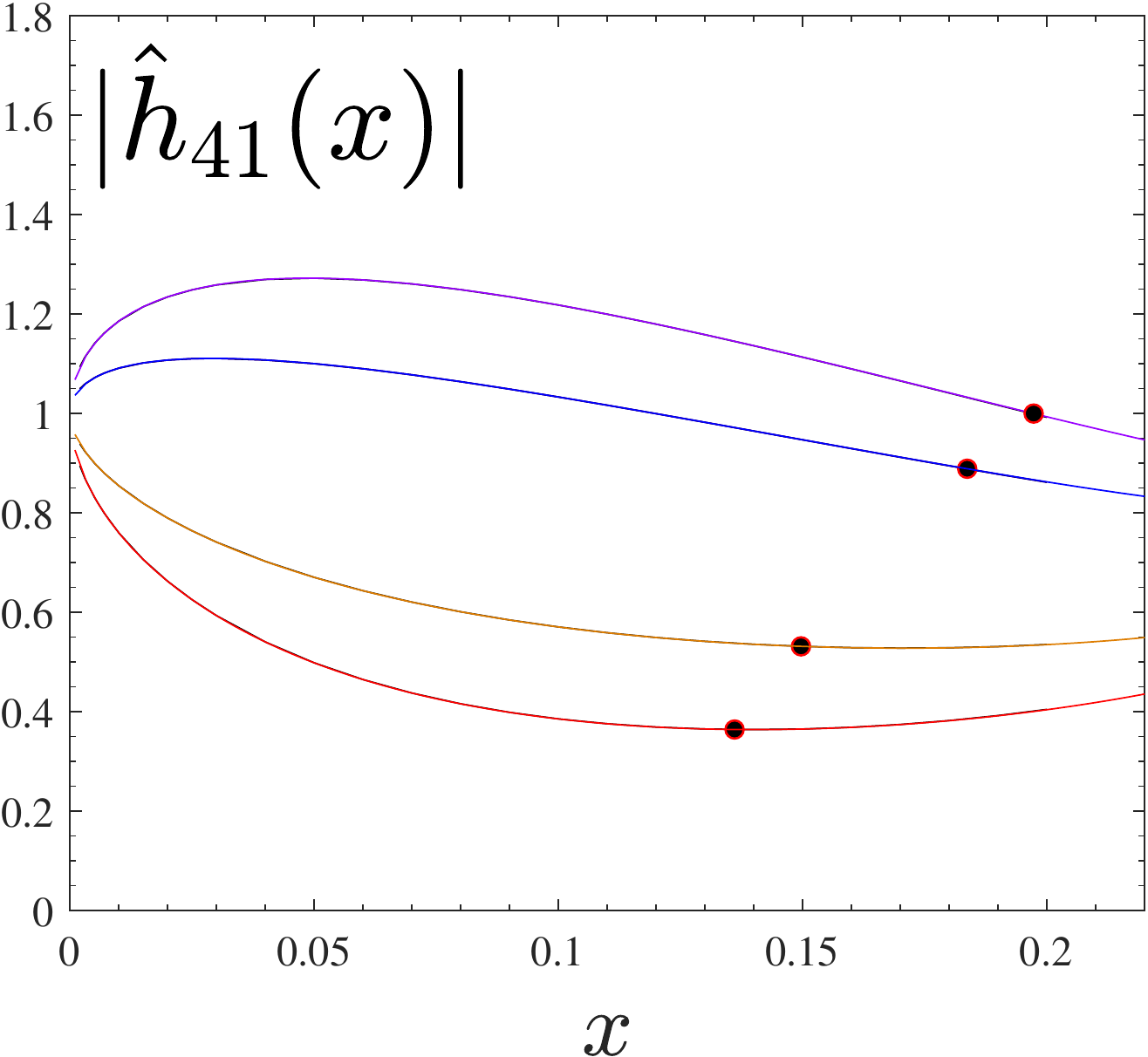}
  \includegraphics[width=0.17\textwidth]{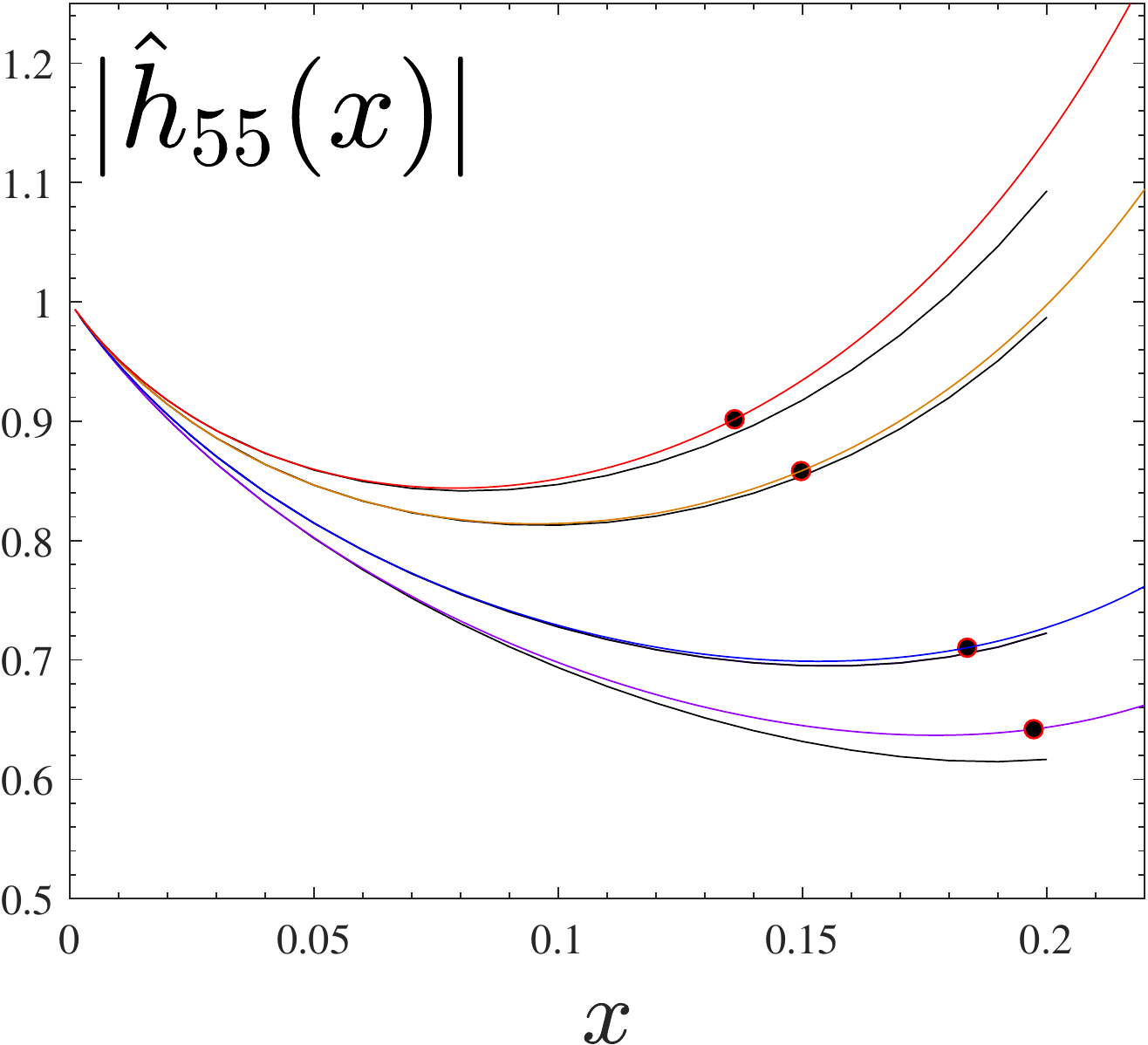}\\
  \includegraphics[width=0.17\textwidth]{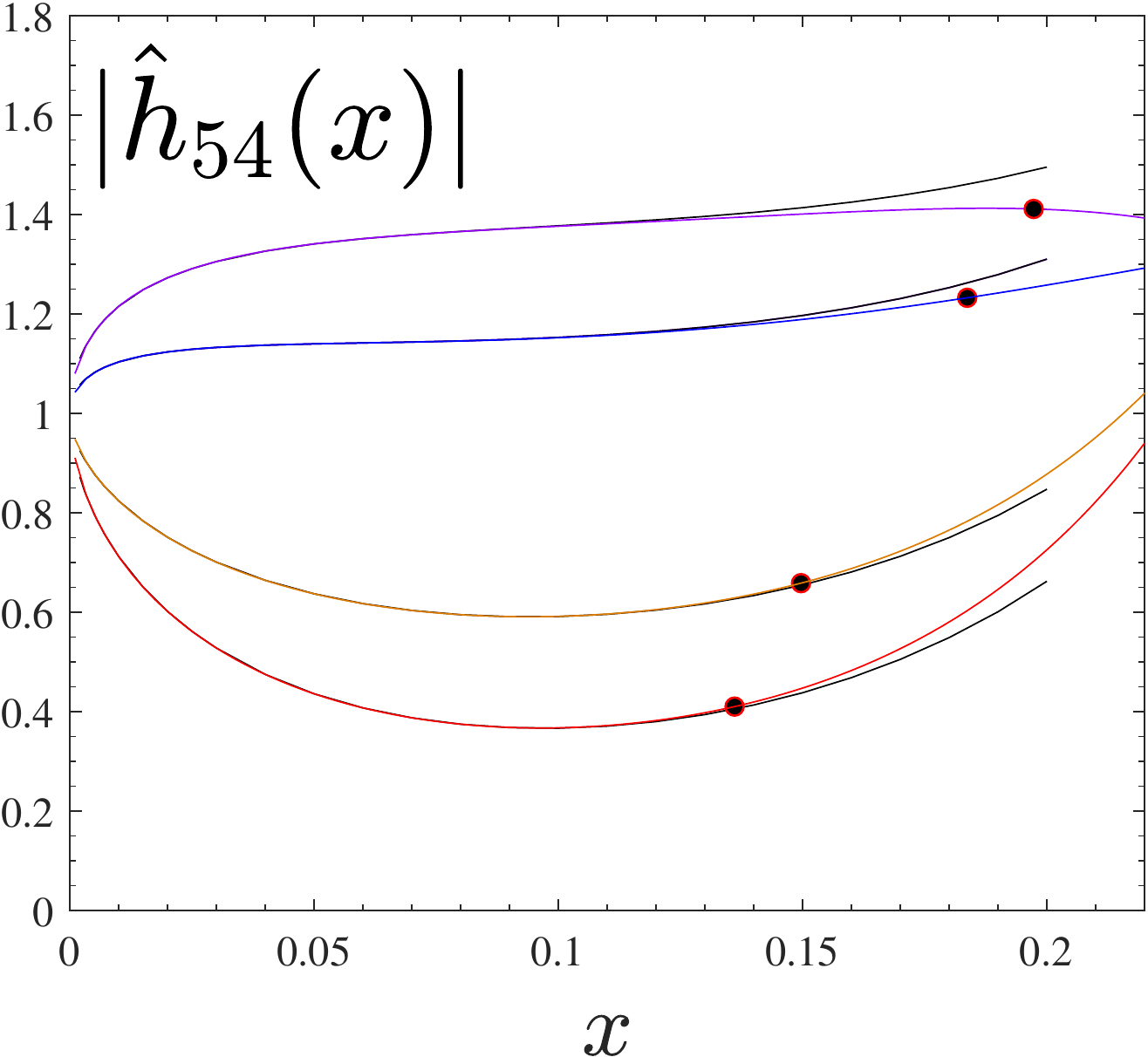}
  \includegraphics[width=0.17\textwidth]{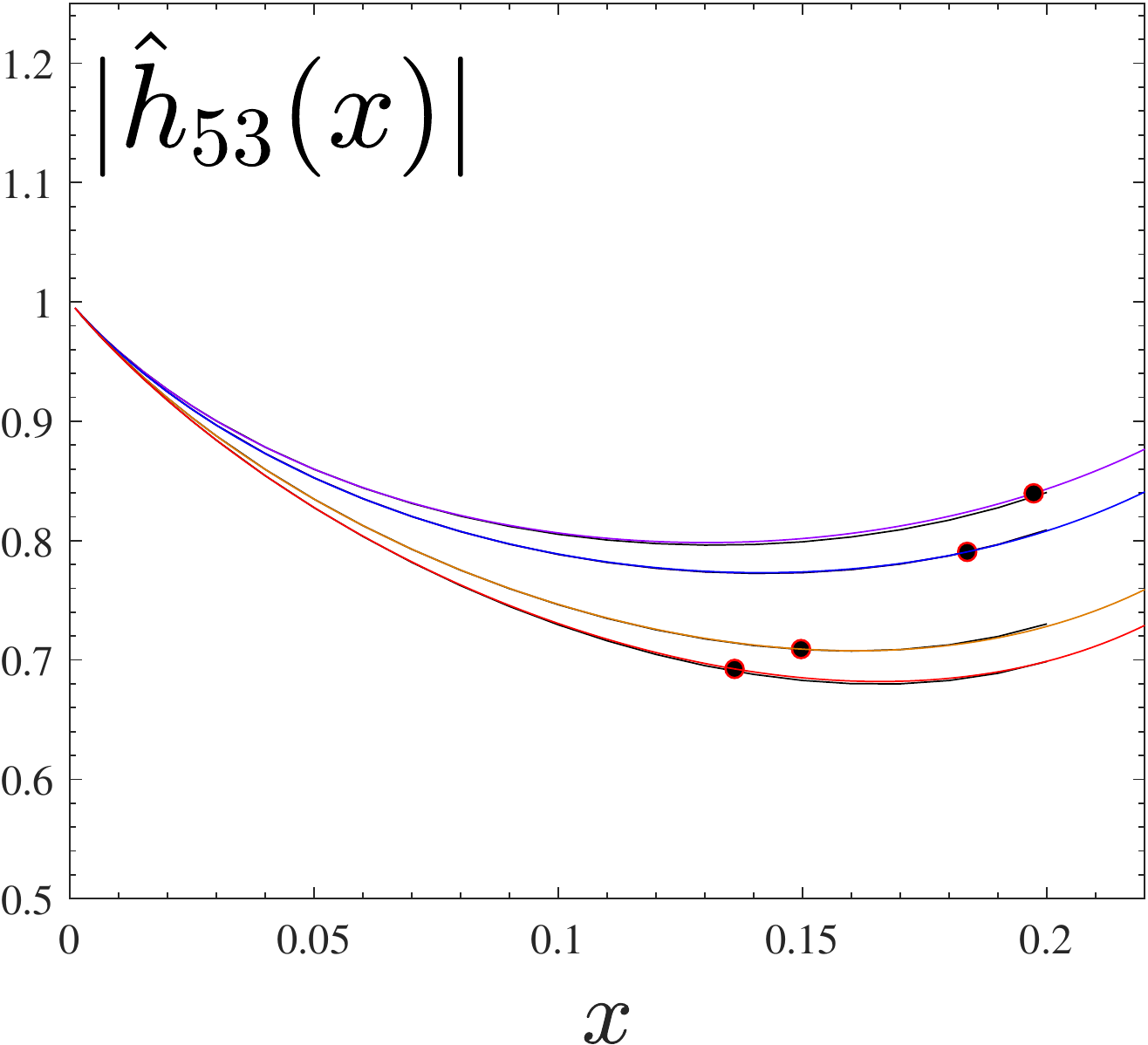}
  \includegraphics[width=0.17\textwidth]{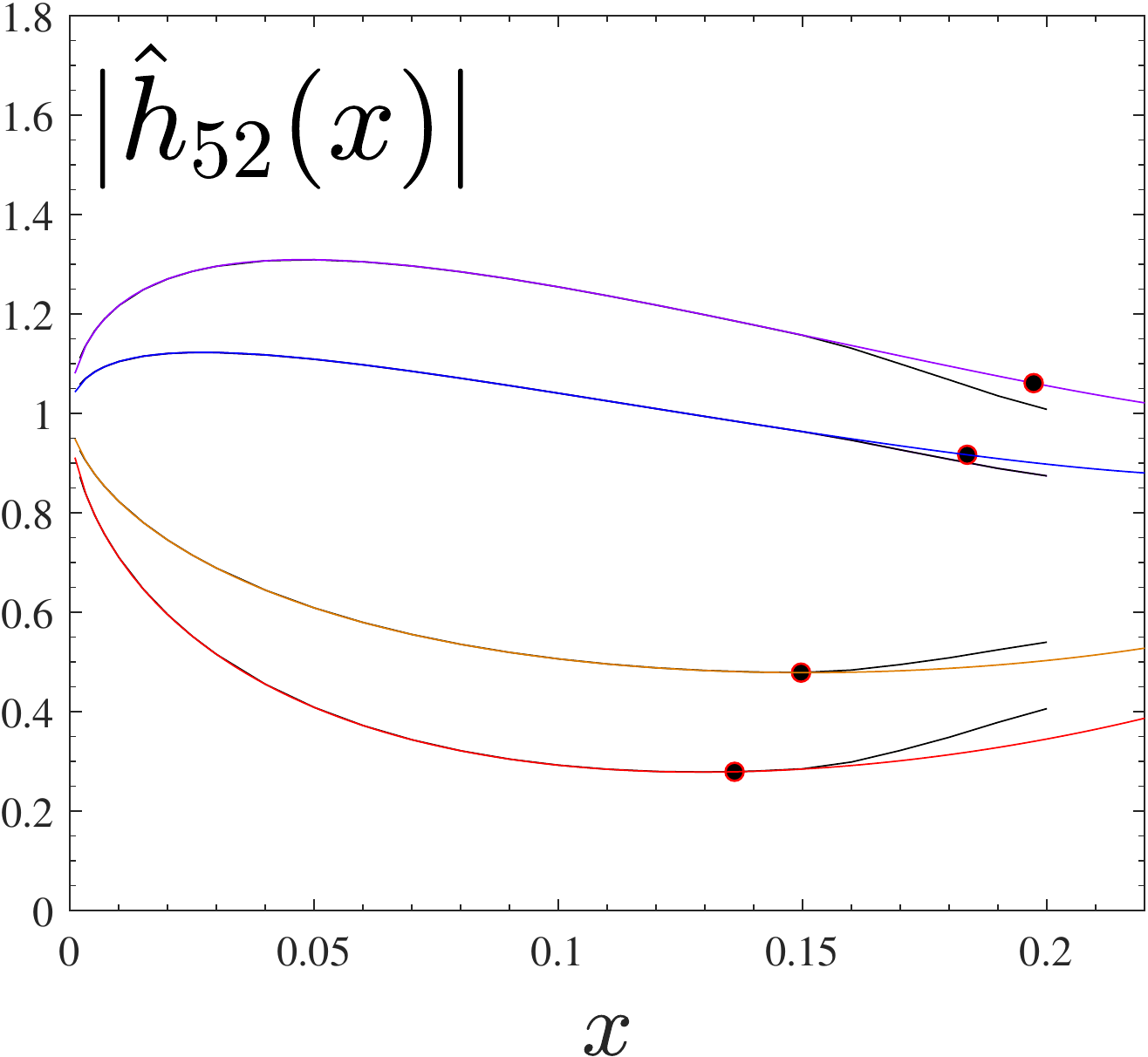}
  \includegraphics[width=0.17\textwidth]{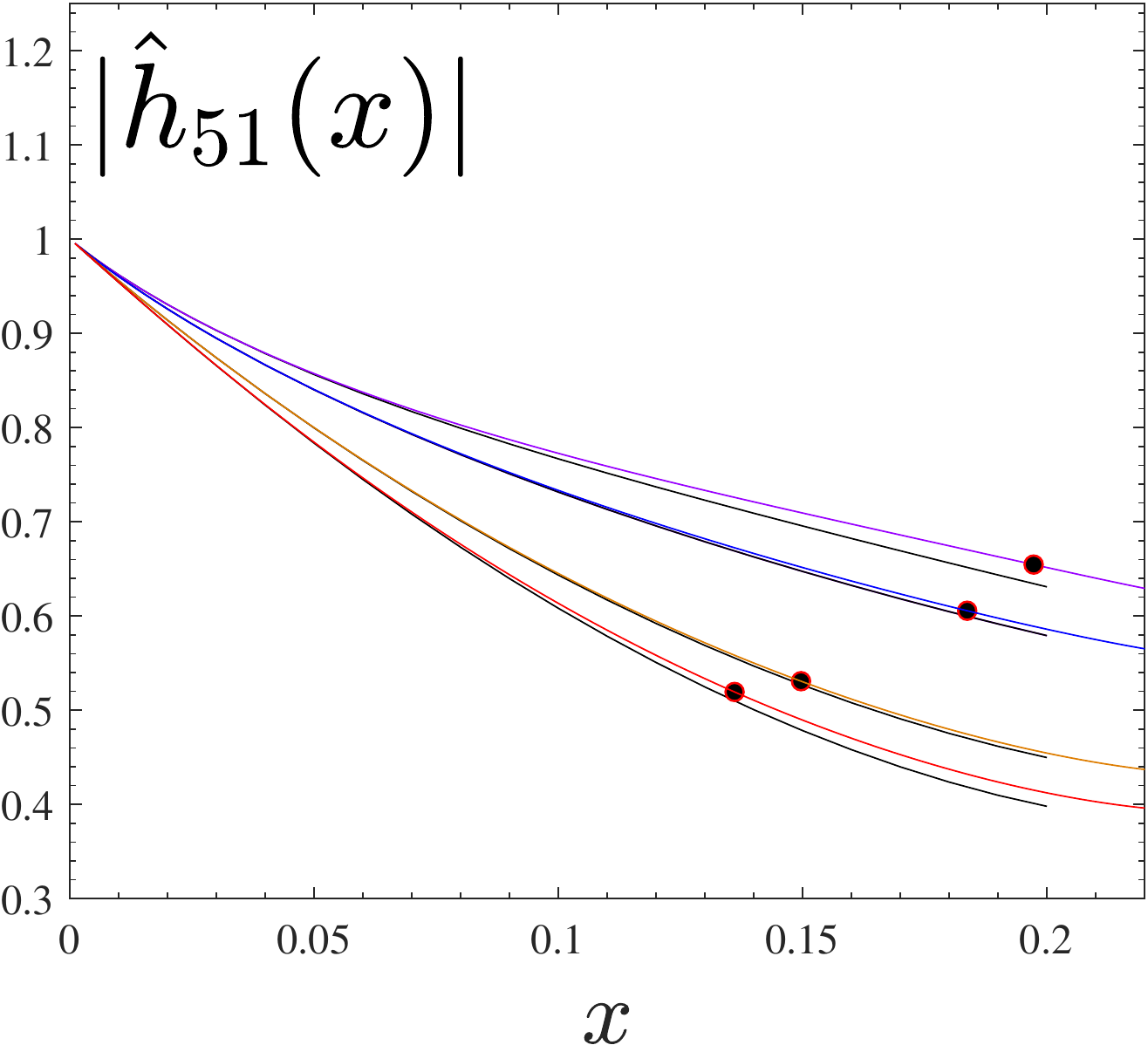}
  \includegraphics[width=0.17\textwidth]{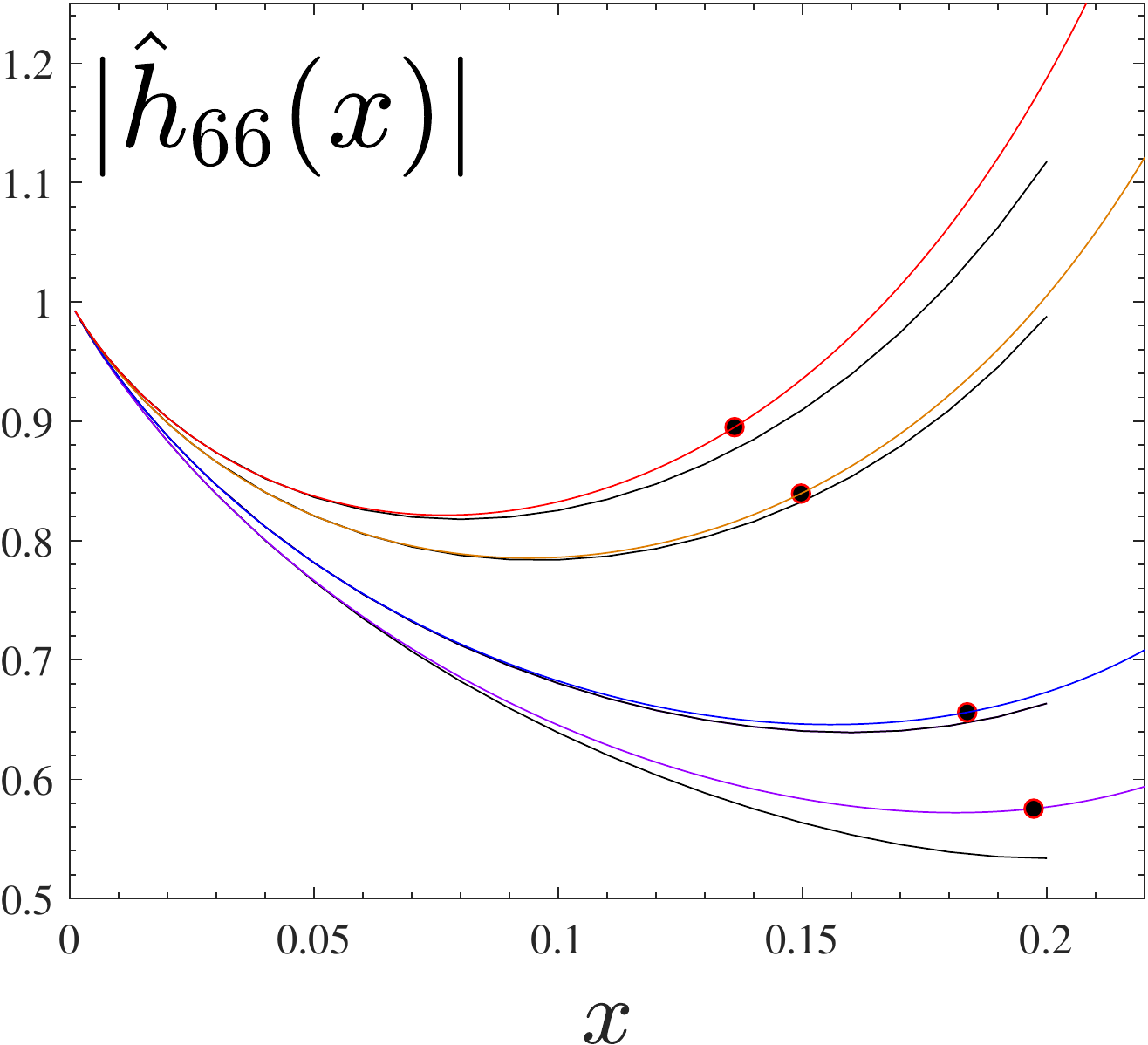}
  \caption{\label{fig:hlm_res} Alternative resummation: the $\ell+m=\text{even}$ modes are resummed
  using Eq.~\eqref{eq:hevenresummed}, while the $\ell+m=\text{odd}$ rely on Eq.~\eqref{eq:hoddresummed},
  where only the tail factor is overall factorized. The values of the particle spin are
  $\sigma=(-0.90, -0.5, +0.5,+0.90)$, and
  are respectively indicated by the colors red, orange, blue and
  purple. The colored markers indicate the location of the LSO, Eq.~\eqref{eq:x_LSO}. 
  The improvement in the $\ell+m=\text{odd}$ modes with respect to
  Fig.~\ref{fig:hlmsvs_pan} is evident. By contrast, the behavior of the $\ell+m=\text{even}$
  modes is similar to those of Fig.~\ref{fig:hlmsvs_pan}, though slightly worse.} 
\end{figure*}

\begin{table*}[t]
  \caption{\label{tab:hhLSO}Fractional differences between the resummed and
    the numerical $\hat{h}_{\ell m}$'s at the LSO shown in Fig.~\ref{fig:hlm_res}.
    The fractional difference is defined as $|\hat{h}^{\rm num}-\hat{h}^{\rm analyt}|/\hat{h}^{\rm num}$
    computed at $x_{\rm LSO}$.}
\begin{center}
\begin{ruledtabular}  
\begin{tabular}{lcccc}
\rule[0,6cm]{0cm}{0cm}
$(\ell,m)$ & \multicolumn{4}{c}{$\Delta(x)\hat{h}(x)/\hat{h}\vert_{x=x_{\rm LSO}}$ for $-0.9 \le \sigma \le +0.9$}\\
\hfill & $-0.9$ & $-0.5$ & $+0.5$ & $+0.9$\\
\hline
\hline
$(2,2)$ & $8\times 10^{-4}$ & $5\times 10^{-5}$ & 0.0023 & 0.0024\\
$(2,1)$ & $4\times 10^{-5}$ & $7\times 10^{-5}$ & $5\times 10^{-5}$ & $4\times 10^{-4}$\\
$(3,3)$ & 0.0035 & 0.0012 & $2\times 10^{-4}$ & 0.0056\\
$(3,2)$ & $9\times 10^{-4}$ & $5\times 10^{-4}$ & 0.0041 & 0.0086\\
$(3,1)$ & 0.0057 & 0.0018 & 0.0023 & 0.0091\\
$(4,4)$ & 0.0098 & 0.0046 & 0.0017 & 0.0085\\
$(4,3)$ & 0.0049 & 0.0029 & 0.013 & 0.029\\
$(4,2)$ & 0.0047 & 0.0013 & 0.0013 & 0.007\\
$(4,1)$ & $7\times 10^{-5}$ & $3\times 10^{-5}$ & $4\times 10^{-4}$ & 0.001\\
$(5,5)$ & 0.014 & 0.0055 & 0.0072 & 0.042\\
$(5,4)$ & 0.012 & 0.0068 & 0.023 & 0.05\\
$(5,3)$ & 0.0029 & $6\times 10^{-4}$ & $10^{-4}$ & 0.0036\\ 
$(5,2)$ & $8\times 10^{-4}$ & $5\times 10^{-4}$ & 0.018 & 0.046\\
$(5,1)$ & 0.019 & 0.0067 & 0.0097 & 0.032\\
$(6,6)$ & 0.02 & 0.0087 & 0.013 & 0.078
\end{tabular}
\end{ruledtabular}
\end{center}
\end{table*}

\subsubsection{$\ell+m$=even: orbital factorization}
\label{sec:resum_even}
We implement the orbital factorization of the $\rho_\lm$'s of~\cite{Messina:2018ghh}
to all even-parity multipoles up to $\ell=6$ included. Consistently with~\cite{Messina:2018ghh}, 
the orbital factors are taken at 6PN order\footnote{Though the $(3,1)$ mode is taken at 5PN only.}
and Pad\'e resummed according to Table~I of~\cite{Messina:2018ghh}. More precisely, we use the standard
$P_2^4$ Padé for all the multipoles except the $(3,1)$ one, that is kept at 5PN with a $P_2^3$ Padé (the bold
values in the table are not considerated). The only exception is the $(4,2)$ multipole, that in this paper 
is resummed with a $P_2^4$ Padé instead of a $P_0^6$ Taylor Series, since the difference between the two choices in
this case is minimal. 
By resumming the spin-dependent factors by taking their inverse-Taylor representation,
the resummed residual amplitudes finally read
\begin{equation}
\label{eq:iRfull}
\rho_{\ell m}(x,\sigma)=P_d^n[\rho_{\ell m}^{\rm orb}]\overline{\hat{\rho}_{\ell m}^{\sigma}},
\end{equation}
where $\hat{\rho}_{\ell m}^{\sigma}=T_n\left[1+\rho_{\ell m}^{\sigma}/\rho_{\ell m}^{\rm orb}\right]$ and
we defined $\overline{\hat{\rho}_\lm^\sigma}\equiv(T_n[(\hat{\rho}^\sigma_\lm)^{-1}])^{-1}$.
Finally, the even-parity waveform amplitudes read
\begin{equation}
\label{eq:hevenresummed}
|\hat{h}_{\ell m}^{(0)}(x,\sigma)|=\hat{E}\,|\hat{h}^{\rm tail}_{\ell m}|\,\left[\rho_{\ell m}(x,\sigma)\right]^\ell.
\end{equation}
The analytical/numerical agreement, displayed in Fig.~\ref{fig:hlm_res}, is essentially 
comparable to the  standard approach shown in Fig.~\ref{fig:hlmsvs_pan} above.

\subsubsection{$\ell+m=$odd: factorizing the tail factor only}
For the odd-parity modes we suggest here to follow a new route:
factor out  {\it only} the tail factor spin part of the modulus,  $|h_{\ell m}^\sigma|$, 
while keeping the orbital part $|h_{\ell m}^{\rm orb}|$ factorized as usual~\eqref{eq:standardfact},
i.e., with the orbital angular momentum $\hat{j}^{\rm orb}$ factored out. Note that for this
particular calculation, we keep $\rho_\lm^{\rm orb}$ at 5PN accuracy and in Taylor-expanded form.
The rationale behind the choice of not factorizing the orbital angular momentum is that,
in the presence of a spinning body, the source of the field is given by the sum of two separate
pieces, one proportional to $\hat{j}^{\rm orb}$ and another one to $\sigma$.
This is in particular the structure of the source of the Regge-Wheeler-Zerilli equation for a 
spinning test-body. It seems then less sound to factor out $\hat{j}^{\rm orb}$ {\it also} from the
$\sigma$-dependent term. Starting from
\begin{equation}
\label{eq:hlm_fact}
|\hat{h}_\lm^{(1)}(x,\sigma)| \equiv |\hat{h}^{\rm orb,(1)}_\lm| + \sigma |\hat{h}_\lm^{\sigma,(1)}|, 
\end{equation}
we factorize each term separately as
\begin{align}
& \rho_\lm^{\rm orb}=T_n\left[\left(\dfrac{|\hat{h}_\lm^{\rm orb,(1)}|}{|\hat{h}_\lm^{\rm tail}|\hat{j}^{\rm orb}}\right)^{1/\ell}\right]\ ,\\
&|\tilde{h}|^\sigma_\lm= T_n\left[\dfrac{|\hat{h}_{\lm}^{\sigma,(1)}|}{|\hat{h}^{\rm tail}_\lm|}\right] \ .
\end{align}
The resummed odd-parity waveform amplitudes finally read
\begin{equation}
\label{eq:hoddresummed}
|\hat{h}_{\ell m}^{(1)}(x,\sigma)|=|\hat{h}_{\ell m}^{\rm tail}|\left\{\left(\rho_{\ell m}^{\rm orb}\right)^{\ell}\hat{j}^{\rm orb}+\sigma|\tilde{h}_{\ell m}^{\sigma,(1)}|\right\}.
\end{equation}
It is important to stress that, for simplicity, we are here not using Pad\'e
approximants of the orbital $\rho_{\ell m}^{\rm orb}$ in Eq.~\eqref{eq:hoddresummed}.  
The analytical and numerical amplitudes are compared in Fig.~\ref{fig:hlm_res}.
The agreement between the two is remarkable, and way better than the one
obtained with the standard approach shown in Fig.~\ref{fig:hlmsvs_pan} above.
To better quantify the agreement, we list in Table~\ref{tab:hhLSO} the fractional
differences computed at $x_{\rm LSO}$. Although not shown in the table, we have
also verified that an analogous quantitative agreement holds for $\sigma =\pm 0.99$,

\subsection{Resumming the dominant $\mathbf{m}$=\text{odd} modes
  consistently with the comparable-mass case.}
\label{sec:iResum_odd}
\begin{figure*}[t]
  \center
  \includegraphics[width=0.25\textwidth]{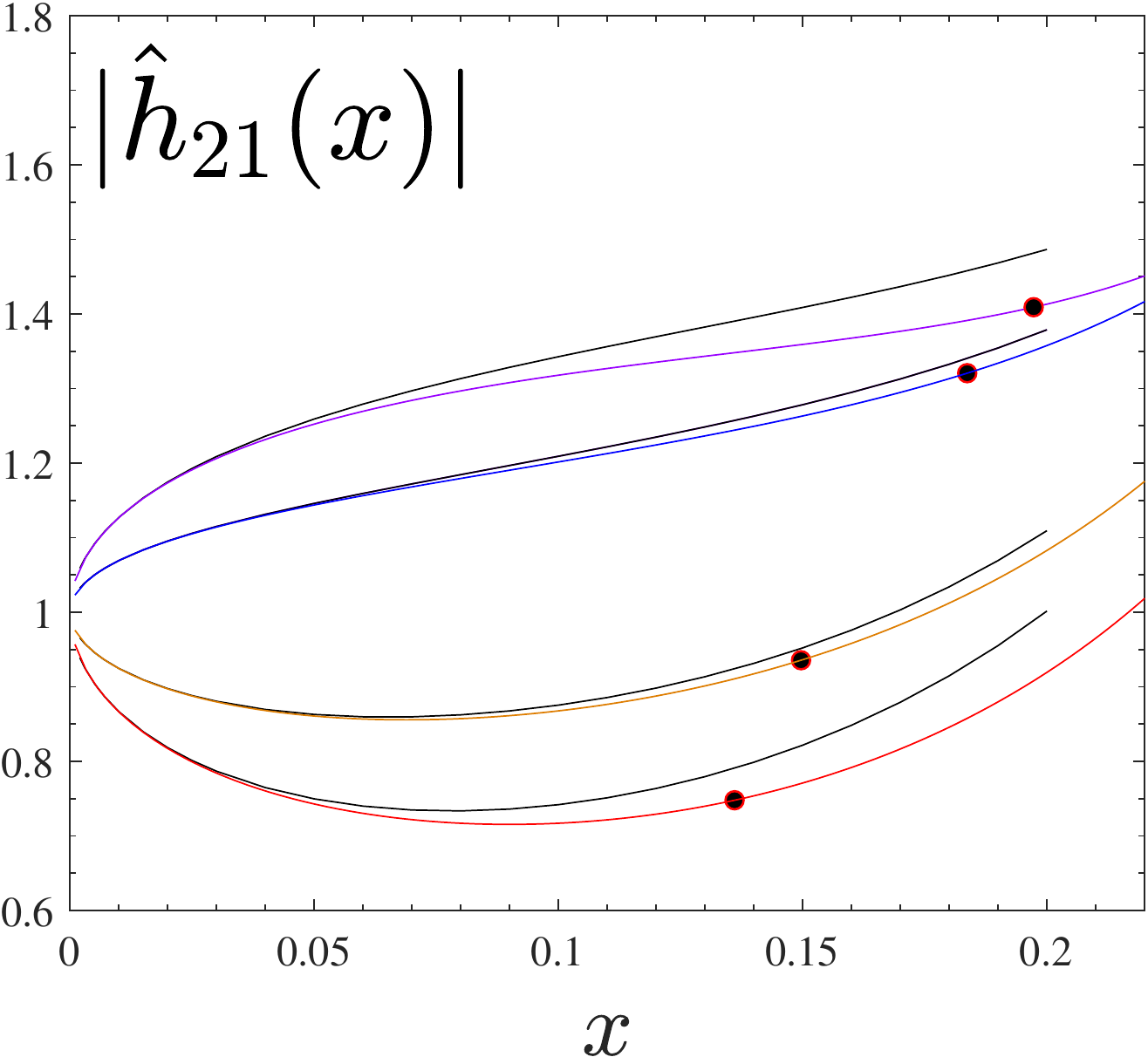}
  \includegraphics[width=0.26\textwidth]{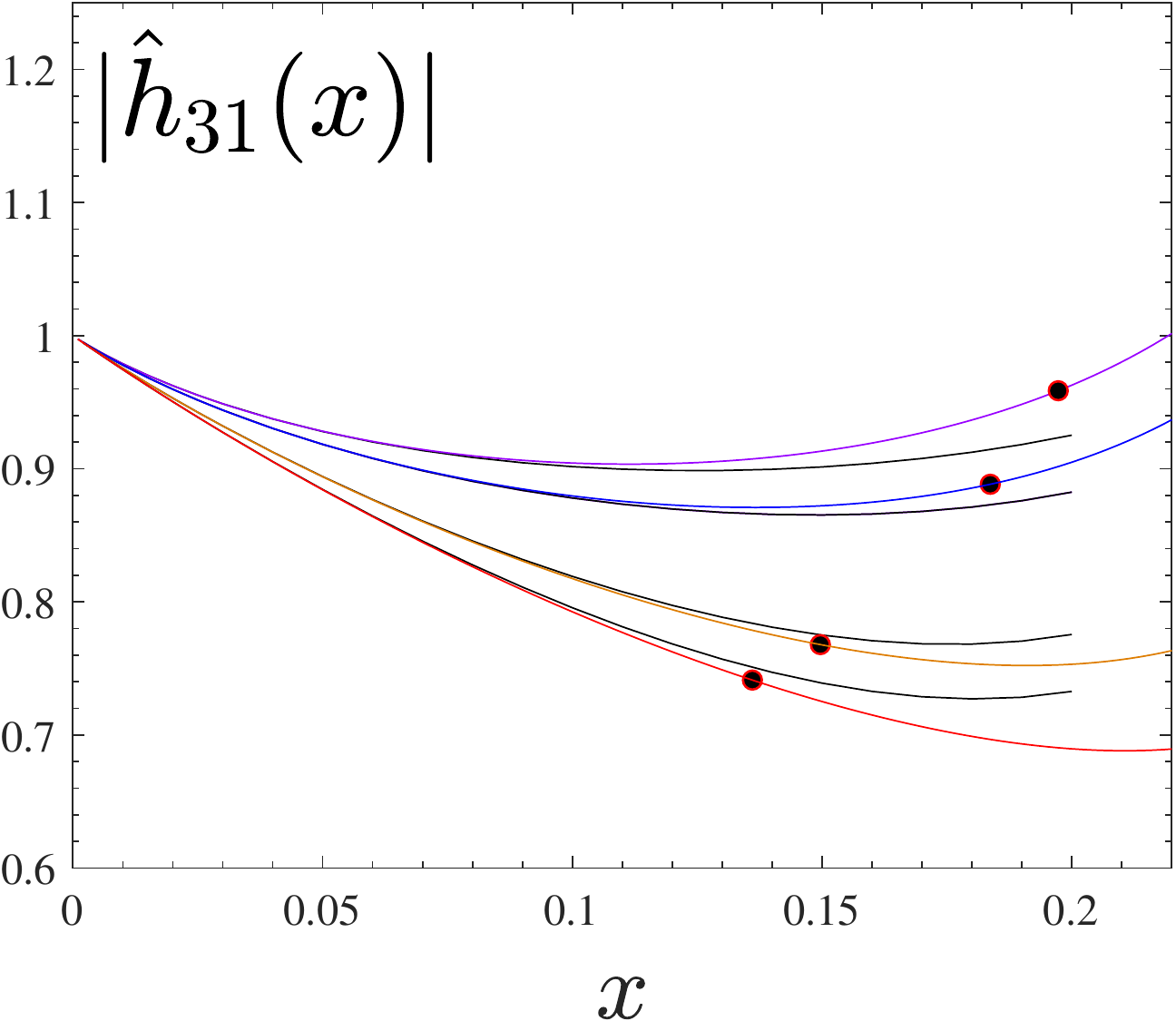}
  \includegraphics[width=0.25\textwidth]{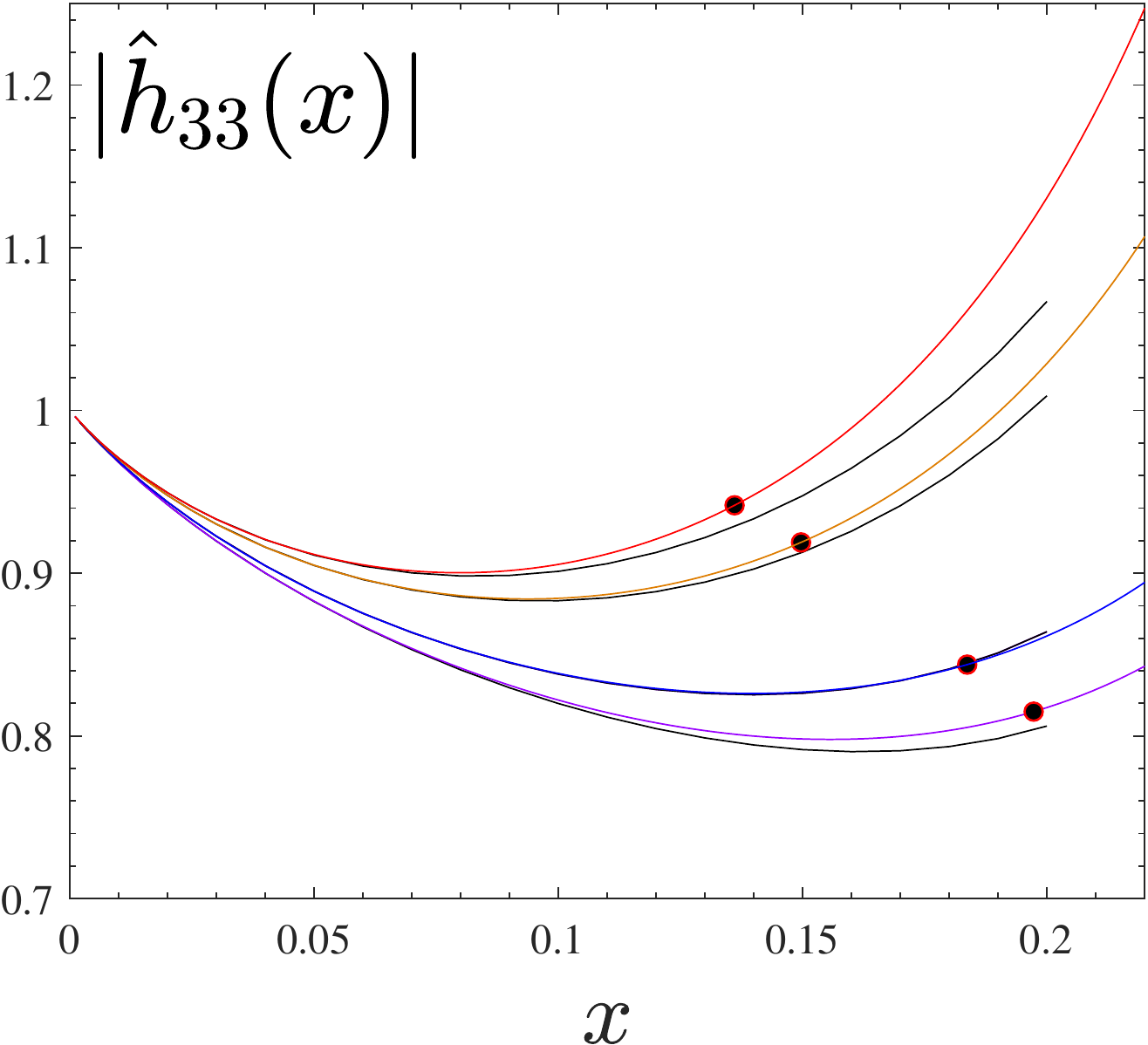}
  \includegraphics[width=0.24\textwidth]{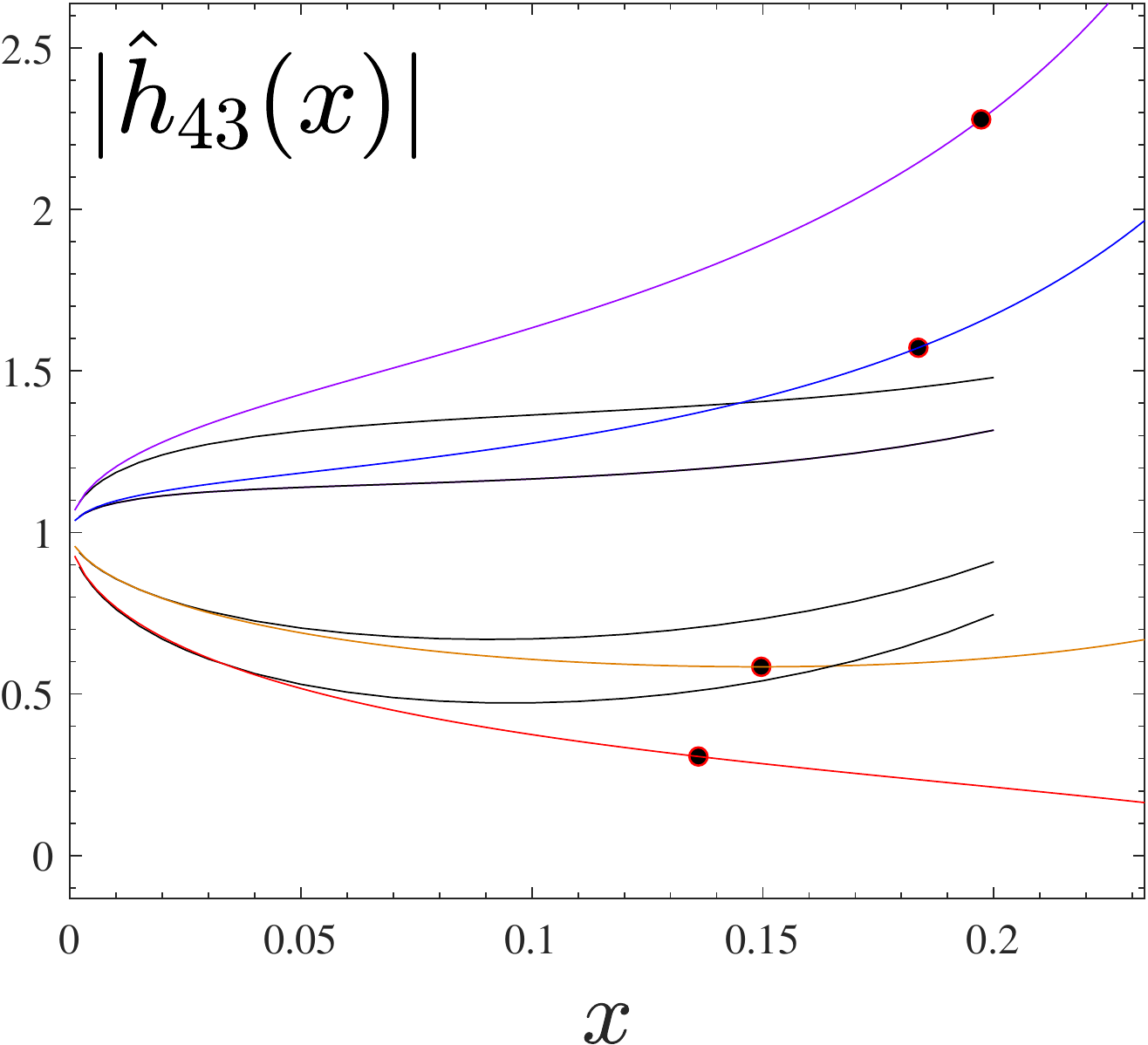}
  \includegraphics[width=0.24\textwidth]{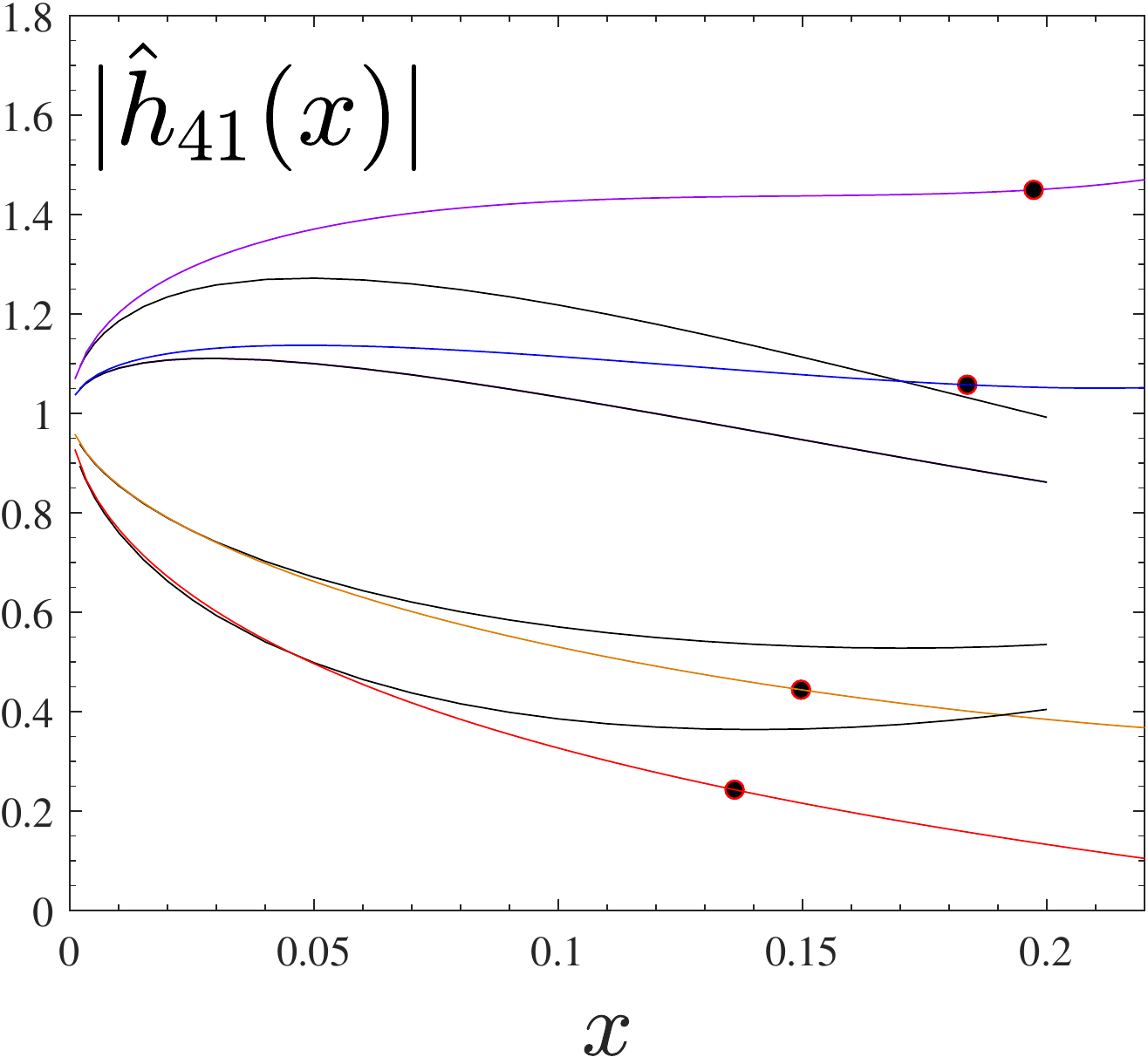}
  \caption{\label{fig:iResumOdd} Testing the resummation of the $m=\text{odd}$ modes suggested
  in the comparable-mass case. The analytical/numerical agreement is rather good for the $(2,1)$ and $(3,3)$ mode, while it is largely inaccurate for the other multipoles.
   } 
\end{figure*}

Up to this point we have seen that there are different procedures for resumming the 
waveform amplitudes depending on the parity of $\ell+m$. Actually, detailed
studies of the waveform amplitudes for two objects
of masses $(M_1,M_2)$~\cite{Nagar:2016ayt,Messina:2018ghh} 
have illustrated that one should also carefully separate the analytic treatment
depending on the parity of $m$ only.
In particular, Refs.~\cite{Nagar:2016ayt,Messina:2018ghh} introduced a special
analytical treatment of the orbital-factorized spin-dependent 
functions $\hat{f}_\lm^S$ when $\nu\equiv M_1 M_2/(M_1+M_2)^2\neq 0$.
More precisely, one shows that, 
when $\nu\neq 0$, these functions are naturally written as the sum of two separate 
Taylor expansions,  one proportional to $X_{12}=X_1-X_2=\sqrt{1-4\nu}$ and the other to
$\tilde{a}_{12}\equiv X1\chi_1-X_2\chi_2$, where $X_i\equiv M_i/(M_1+M_2)$
and $\chi_i\equiv S_i/M_i^2$ are the dimensionless spins of the two objects.
Each Taylor expansion can eventually be resummed taking its inverse-Taylor 
representation.  This resummation of the $m$-odd waveform amplitudes has been 
recently incorporated in a new, multipolar, EOB waveform model. 
One of the remarkable features of this analytical choice is that the zero in the 
$(2,1)$ time-domain amplitude, that exists in certain regions of the parameter space, 
is quantitatively consistent with a similar feature found in state-of-the-art 
NR simulations.  On top of this, Ref.~\cite{Nagar:2016ayt}, pointed out that
the resummation procedure is reliable and accurate  also in the special case 
of a nonspinning particle orbiting a spinning black hole. 

\begin{figure}[t]
  \center
  \includegraphics[width=0.45\textwidth]{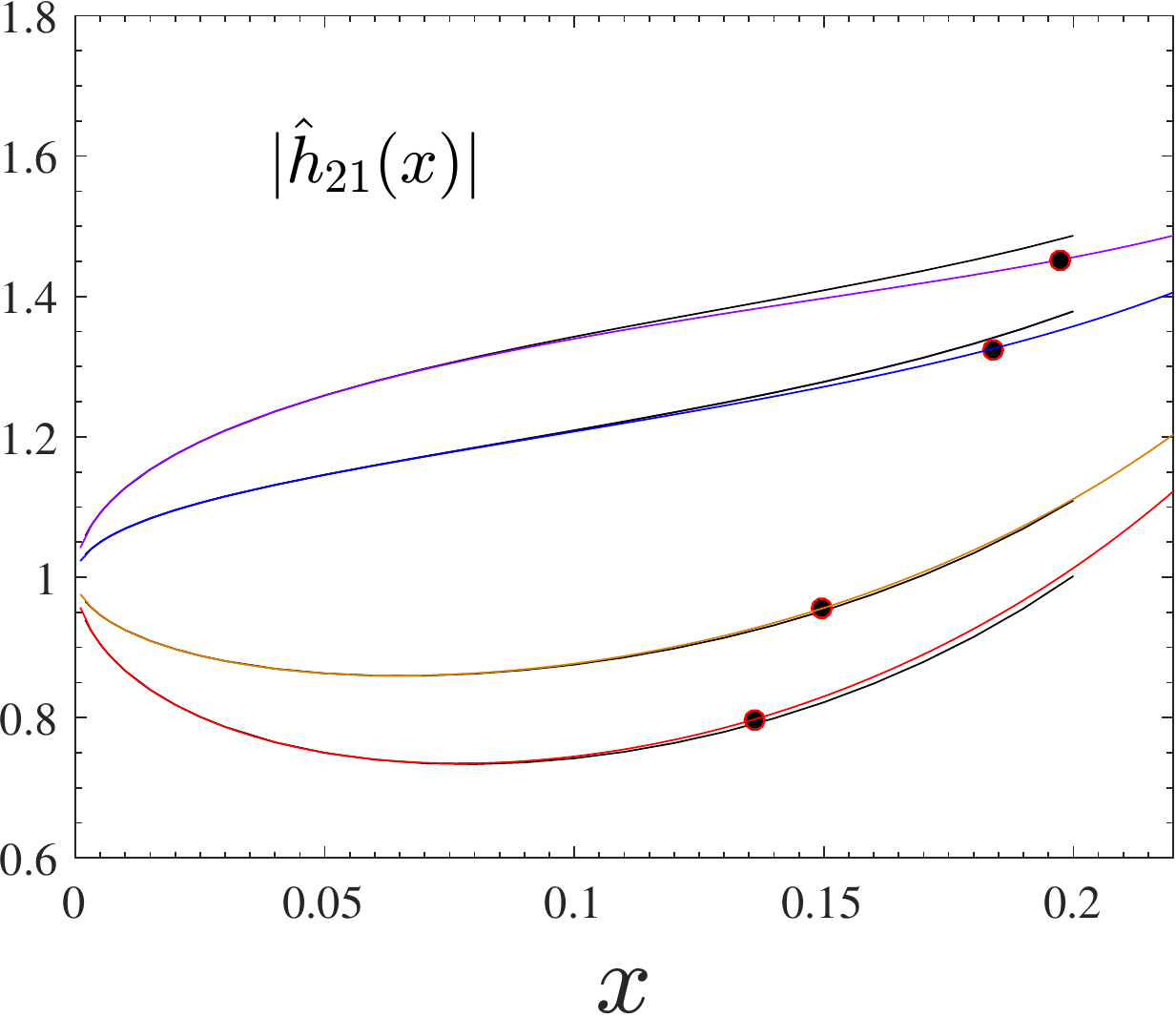}
  \caption{\label{fig:h21_tail_25pn} Analytical/numerical agreement with the tail-factorized $\ell=2$, $m=1$
    amplitude truncated at 2.5PN in the spin sector, consistently with Eq.~\eqref{eq:f21}. Despite the reduced
    amount of spin-dependent information
  reduced amount of PN information the tail-factorized analytical amplitude does not look especially better 
  than the orbital-factorized and resummed one displayed in Fig.~\ref{fig:iResumOdd}.} 
\end{figure}

In this section we investigate to which extent the $m$-odd multipoles resummation
approach remains robust and stable also in the case of a spinning test-body on a Schwarzschild
black hole. To do so, we focus on the same $m$-odd modes considered in the 
$\nu\neq 0$ case,  $(2,1)$, $(3,3)$, $(3,1)$, $(4,3)$, $(4,1)$. 
In addition, each mode is truncated at the largest PN order that carries known
$\nu$-dependent corrections~\footnote{Even if we have more spinning-particle terms, we cannot
use them since we cannot consistently split each PN order in the two separate series.}.
Consistently with Ref.~\cite{Riemenschneider:2019}, the $m$-odd waveform amplitudes 
are written as 
\begin{equation}
|\hat{h}_{\ell m}| =\hat{j}^{\rm orb}|\hat{h}_{\ell m}^{\rm tail}| (P^m_n[\rho_{\ell m}^{\rm orb}])^\ell (1 + \hat{f}_{\lm}^\sigma).
\end{equation}
The Pad\'e approximants of the orbital part $P^m_n[\rho_{\ell m}^{\rm orb}]$ are the same,
adopted in Ref.~\cite{Messina:2018ghh}, that is $P^5_1$ for $(2,1)$,  $P^3_2$ for $(3,1)$,
$P^6_0$ for $(5,5)$ and $P^4_2$ for $(3,2)$, $(4,3)$ and $(4,1)$.
On the other hand, the spin-dependent terms $\hat{f}_\lm^\sigma$ are given by
\begin{align}
  \label{eq:f21}
\hat{f}_{21}^{\sigma}&=\overline{f_{21}^{\sigma_{(0)}}} + \dfrac{3}{2}\sigma x^{1/2} \overline{f_{21}^{\sigma_{(1)}}},\\
\hat{f}_{33}^{\sigma}&=\overline{f_{33}^{\sigma_{(0)}}} + \dfrac{1}{4}\sigma x^{3/2} f_{33}^{\sigma_{(1)}},\\
\hat{f}_{31}^{\sigma}&=\overline{f_{31}^{\sigma_{(0)}}} + \dfrac{9}{4}\sigma x^{3/2} f_{31}^{\sigma_{(1)}},\\
\hat{f}_{43}^{\sigma}&=f_{43}^{\sigma_{(0)}} + \dfrac{5}{4}\sigma x^{1/2} f_{43}^{\sigma_{(1)}},\\
\hat{f}_{41}^{\sigma}&=f_{41}^{\sigma_{(0)}} + \dfrac{5}{4}\sigma x^{1/2} f_{41}^{\sigma_{(1)}}.
\end{align}  
Following Ref.~\cite{Messina:2018ghh}, the overbar indicates that each term is resummed using its inverse-Taylor
representation. The function written above explicitly read
\begin{align}
\overline{f_{21}^{\sigma_{(0)}}} &= \left(1+\frac{13}{84}x^{3/2}\sigma+\frac{14705}{7056}x^{5/2}\sigma\right)^{-1},\\
\overline{f_{21}^{\sigma_{(1)}}} &=\left(1+\frac{349}{252}x+\frac{65969}{31752}x^2\right)^{-1},\\
\overline{f_{33}^{\sigma_{(0)}}} &=\left(1+\frac{7}{4}x^{3/2}\sigma+\frac{211}{60}x^{5/2}\sigma\right)^{-1} ,\\
f_{33}^{\sigma_{(1)}} &= 1+\frac{169}{15}x,\\
\overline{f_{31}^{\sigma_{(0)}}} &= \left(1-\frac{1}{4}x^{3/2}\sigma+\frac{13}{36}x^{5/2}\sigma\right)^{-1},\\
f_{31}^{\sigma_{(1)}} &=1-\frac{1}{9}x,\\
f_{43}^{\sigma_{(0)}} &=\left(1-\frac{5}{4}\right)x^{1/2}\sigma, \\
f_{43}^{\sigma_{(1)}} &=1,\\
f_{41}^{\sigma_{(0)}} &= \left(1-\frac{5}{4}\right)x^{1/2}\sigma, \\
f_{41}^{\sigma_{(1)}} &= 1 \ .
\end{align}
The  analytical $|\hat{h}_\lm|$ are compared to the numerical ones in Fig.~\ref{fig:iResumOdd}.
The most interesting result displayed in the figure is that both the $(2,1)$ and $(3,3)$ analytical
amplitudes deliver a reasonably accurate representation of the numerical data up to the LSO,
although this is not as good as the tail-factorized case discussed in Fig.~\ref{fig:hlm_res}.
Note however that in that case we were using spin information truncated at 5.5PN accuracy.
Thus, to produce a meaningful comparison we need to redo the calculation of Fig.~\ref{fig:hlm_res}
truncating the $\ell=2$, $m=1$ mode at 2.5PN. By contrast, the orbital $\rho_{21}^{\rm orb}$ function
is kept at 5PN accuracy and in Taylor-expanded form. This new comparison is displayed in
Fig.~\ref{fig:h21_tail_25pn}. Interestingly, despite the reduce PN information,
the analytical/numerical agreement is visibly {\it better} than the one displayed
in the top-left panel of Fig.~\ref{fig:iResumOdd}.
We also performed the same analysis for the $\ell=4$ modes. Although we found an
improvement, the truncation at 1.5PN of the spin information is not sufficient
to provide a good agreement up to the LSO location.
In conclusion, the result of Fig.~\ref{fig:h21_tail_25pn} gives further support
to the need of exploring the performance of the tail-only factorization also for
comparable mass binaries.

\section{Conclusions}
\label{sec:conclusions}
In this paper we have collected several new results concerning the gravitational
waveform fluxes and amplitudes emitted by a spinning particle (i.e., a spinning test-black hole) 
on circular orbits around a Schwarzschild black hole at linear order in the particle spin.
Our main findings are summarized as follows:
\begin{itemize}
\item[(i)] We have analytically computed the PN-expanded multipolar energy fluxes up to $\ell=7$. 
Each multipole is obtained at 5.5~PN order beyond the leading, Newtonian, contribution. 
This improves our current analytical knowledge of the fluxes in this corner of the parameter
space, that was previously known only up to {\it global} 3.5PN order.
\item[(ii)] We have computed numerically the multipolar energy fluxes (and waveform amplitudes) up to
$\ell=6$ using two different, and independent, numerical codes. One is a frequency-domain code;
the other uses a time-domain approach. We demonstrated the excellent mutual consistency between the two
numerical approaches once the numerical data are suitably linearized in the spin of the particle. 
This allows us to provide accurate, circularized, waveform amplitudes up to the LSO for different 
values of the particle spin.
%

\item[(iii)]  We compared to these numerical data, considered as  exact, the PN-expanded 
  analytical amplitudes as well as different flavors of their resummation. In particular,
  we could show that the standard resummation approach of Ref.~\cite{Pan:2011gk}, that is
  implemented in state-of-the-art waveform models for coalescing black-hole binaries,
  such as {\tt SEOBNRv4}~\cite{Bohe:2016gbl} or {\tt TEOBResumS}~\cite{Nagar:2018zoe},
  is {\it inaccurate} for $\ell+m$~odd modes and provides nonnegligible differences
  with the numerical data already in the early inspiral. This indicates that the procedure
  of Ref.~\cite{Pan:2011gk} should be improved to construct a waveform model robust
  and accurate all over the BBH parameter space.

\item[(iv)] The most important result of this work is  that the factorization of {\it only} the tail factor 
   in the $\ell+m=\text{odd}$ modes allows one to obtain an excellent analytical/numerical agreement 
   up to the LSO (and even below), for $-0.9\leq \sigma \leq +0.9$, and for {\it all} odd-parity modes 
   analyzed up to $\ell=5$, see Fig.~\ref{fig:hlm_res}.  
   This result opens at least two questions. First of all, one wonders whether the only-tail factorization 
   would be helpful also to resumming the fluxes emitted by a nonspinning particle orbiting
   a Kerr black hole. Resumming that PN series has always been a challenge~\cite{Pan:2011gk}
   that was solved, to a certain extent, only through the orbital factorization (and additional resummation) 
   procedure of Refs.~\cite{Nagar:2016ayt,Messina:2018ghh}. Hopefully the only-tail factorization approach 
   might be useful in that case {\it at least} for what concerns the terms linear in the black-hole spin.
   On the other hand, our findings also suggest that it might be worth investigating the performance 
   of the only-tail factorization procedure {\it also} in the $\nu\neq 0$ case. The accuracy of such odd-parity 
   waveform amplitudes should then be carefully evaluated by performing extensive comparisons 
   between a so-constructed EOB waveform model and state-of-the-art NR simulations. 
       
 \item[(v)] We also explored the accuracy  of the orbital-factorization/resummation
   approach of Ref.~\cite{Nagar:2016ayt,Messina:2018ghh}, notably in the form 
   developed for the odd-$m$ modes in the $\nu\neq 0$ case. We do so by truncating
   the PN information at the same PN order where $\nu$-dependent corrections are known.
   This approach has been recently used to improve the behavior of the $\ell=2$, $m=1$ 
   waveform amplitude in a new, EOB-based, multipolar waveform model~\cite{Riemenschneider:2019}.
   For this mode, we found (see top-left panel of Fig.~\ref{fig:iResumOdd}) that the 
   numerical/analytical agreement is rather good, though not at the level
   of the only-tail factorized case mentioned above. The situation is even worse
   for the subdominant modes (except the $\ell=m=3$ one). Our results seem to suggest
   that the resummation of Ref.~\cite{Nagar:2016ayt,Messina:2018ghh} should be replaced
   by the only-tail factorization one, since it is simpler and more accurate. 
   More investigations, notably when both objects are spinning, will be needed 
   to confirm this preliminary conclusion.
   
\end{itemize}
    
\acknowledgments
FM thanks IHES for hospitality at various stages during the development of this work.
GLG is supported by Grant No. GA\v{C}R-17-06962Y of the Czech Science Foundation that is gratefully acknowledged.
NW gratefully acknowledges support from a Royal Society - Science Foundation Ireland University Research Fellowship.
SB acknowledges support by the EU H2020 under ERC Starting Grant, no.~BinGraSp-714626. 
We acknowledge the networking support received by the COST Action CA16104.

\appendix

\section{Interpreting the spinning-body limit using the EOB dynamics}
\label{sec:Spint}
In the main text we have considered  $-1\leq\sigma\leq 1$. Due to the definition
of $\sigma$ in Eq.~\eqref{eq:def_sigma} we can also write
\begin{equation}
\label{eq:sigma_lim_euristic}
\sigma = \dfrac{\mu}{M}\dfrac{s}{\mu^2}.
\end{equation}
This parameter, per se, can be thus meaningful also for the equal-mass case, with
the understanding that the MPD approach (or the Hamiltonian approach) plus the solution
of the Teukolsky equations for the fluxes can only deliver some of the contributions
at leading-order in the mass ratio. More precisely, the value $\sigma=\pm 1$ can be
interpreted as the case of an extremally spinning black hole, $s/\mu^2=\pm 1$ in an
equal-mass binary, $\mu/M=1$. Although in this case the perturbative approach we
are using is expected to be meaningless, in practice it is useful to compute, and
test numerically, some of the terms that enter the complete PN expansion.

The euristic comparable-mass limit suggested by Eq.~\eqref{eq:sigma_lim_euristic}
can be put on a more solid ground starting from the complete EOB Hamiltonian for
a two-body system with masses $(m_1,m_2)$, with the convention that $m_1\geq m_2$.
We also refer the reader to Sec.~III of Ref.~\cite{Harms:2016ctx} for complementary
information. The complete EOB Hamiltonian reads
\begin{equation}
\hat{H}_{\rm EOB} \equiv \dfrac{H_{\rm EOB}}{\mu}=\frac{1}{\nu}\sqrt{1+2\nu(\hat{H}_{\rm eff}-1)},
\end{equation}
where now it is $\mu\equiv m_1 m_2/(m_1+m_2)$, and $\nu=\mu/M$ is the symmetric mass ratio and
\begin{equation}
\hat{H}_{\rm eff}=\hat{H}^{\rm eff}_{\rm orb}+\hat{H}^{\rm eff}_{\rm SO}
\end{equation}
is the effective Hamiltonian. The spin-orbit sector is given by
\begin{equation}
  \hat{H}^{\rm eff}_{\rm SO}\equiv p_{\varphi}(G_{S^*}\hat{S}_{*}+G_S\hat{S}),
  \end{equation}
where $p_\varphi =P_\varphi/\mu$ is the dimensionless orbital
angular momentum, we introduced the symmetric spin combinations
\begin{align}
\hat{S}_{*}&=\left(\frac{m_2}{m_1}S_1+\frac{m_1}{m_2}S_2\right)\frac{1}{(m_1+m_2)^2}\\
\hat{S}    &=\frac{S_1+S_2}{(m_1+m_2)^2}, 
\end{align}
and $(G_S,G_{S_*})$ are the gyro-gravitomagnetic functions.
For the case where only the secondary, $m_2$ is spinning ($S_1=0$),
$\hat{H}_{\rm SO}^{\rm eff}$ becomes
\begin{equation}
\hat{H}^{\rm eff}_{\rm SO}=p_{\varphi}\biggl\lbrace G_{S^*}\frac{1}{(m_1+m_2)^2}\frac{m_1}{m_2}S_2+G_S\frac{S_2}{(m_1+m_2)^2}\biggr\rbrace .
\end{equation}
After introducing powers of $m_2$ so to explicitly have a dimensionless spin variable $\chi_2\equiv S_2/m_2^2$ one obtains
\begin{equation}
\hat{H}_{\rm SO}=p_{\varphi}\biggl\lbrace G_{S_*}\nu \frac{S_2}{m_2^2}+G_S\frac{1}{4}(1-\sqrt{1-4\nu})^2 \frac{S_2}{m_2^2}\biggr\rbrace .
\end{equation} 
The extreme-mass-ratio limit is now defined by the condition $\nu\to 0$. In this limit, one sees that the term
proportionalt to $G_S$ is suppressed with respect to the one proportional to $G_{S_*}$ because of the different
$\nu$ dependence. In the $\nu\to 0$ limit one finally has that the spin-orbit sector of the {\it real}
spin-orbit Hamiltonian describing a spinning test-body orbiting a Schwarzschild black hole is given by
\begin{equation}
\hat{H}_{\rm SO}=p_{\phi} G_{S_*}\nu \frac{S_2}{m_2^2},
\end{equation}
where $G_{S_*}$ is known analytically in closed, non-PN expanded,
form~\cite{Barausse:2009aa,Barausse:2009xi,Bini:2015xua}.
One then defines the dimensionless spin $\sigma$ as
\begin{equation}
\label{eq:simgaEOBint}
\sigma\equiv \nu \frac{S_2}{m_2^2}=\nu\chi_2.
\end{equation}
Thus, since $-1\leq \chi_2\leq +1$, and $0\leq \nu \leq 1/4$,
one concludes, from the EOB (and thus PN) perspective,
that $-1/4\leq \sigma \leq 1/4$.
The usual limit $-1\leq \sigma \leq 1$ is obtained writing
$\sigma$ as
\begin{equation}
\sigma\equiv\frac{m_1 m_2}{(m_1+m_2)^2}\frac{S_2}{m_2^2}
\end{equation}
and then addtionally expanding the symmetric mass ratio
in powers of the mass ratio $m_2/m_1\approx \mu/M \ll 1$.


\section{Spinning particle on Schwarzschild background: PN-expanded multipolar energy fluxes}
\label{sec:SPfluxes}
\subsection{Newtonian prefactors}
\label{sec:Newt}
Each multipolar flux is written as the product of a Newtonian (or leading-order)
prefactor and of a PN correction, i.e.
\begin{equation}
F_\lm= F_\lm^{(N,\epsilon)}\hat{F}_\lm,
\end{equation}
where $\epsilon=0,1$ indicates the parity of $\ell+m$. For each multipole, the Newtonian
prefactor can be written in closed form as
\begin{equation}
\label{eq:FlmNewt}
F^{(N,\epsilon)}_{\ell m} = \dfrac{1}{8\pi} x^3 m^2(-)^{\ell +\epsilon}\left\vert {\cal R} h_{\ell m}^{(N,\epsilon)}\right\vert^2\ ,
\end{equation}
where the Newtonian waveform ${\cal R}h_\lm^{(N,\epsilon)}$ is explicitly given as~\cite{Damour:2008gu}
\begin{equation}
\label{eq:hlmNexpl}
{\cal R} h_\lm^{(N,\epsilon)} = M\nu \,n_\lm^{(\epsilon)}c_{\ell+\epsilon}(\nu)x^{(\ell + \epsilon)/2}Y_{\ell - \epsilon,-m}(\pi/2,\phi)\ ,
\end{equation}
with
\begin{equation}
c_{\ell+\epsilon}(\nu)=X_2^{\ell+\epsilon-1}+(-)^{\ell+\epsilon}X_1^{\ell + \epsilon-1}
\end{equation}
and
\begin{align}
n_\lm^{(0)}&=({\rm i} m)^\ell\dfrac{8\pi}{(2\ell + 1)!!}\sqrt{\dfrac{(\ell+1)(\ell+2)}{\ell(\ell-1)}}\ ,\\
n_\lm^{(1)}&=-({\rm i}m)^\ell\dfrac{16\pi{\rm i}}{(2\ell+1)!!}\sqrt{\dfrac{(2\ell+1)(\ell+2)(\ell^2-m^2)}{(2\ell-1)(\ell+1)\ell(\ell-1)}}\ .
\end{align}
\subsection{Newton-normalized PN fluxes}
\label{app:fluxes}
The PN-expanded fluxes up to $\ell=7$ are presented below. For completeness and
future reference we also keep all the nonspinning terms. All of the series given below can be found digitally in the \texttt{PostNewtonianSelfForce} package in the Black Hole Perturbation Toolkit \cite{BHPToolkit}.
\begin{widetext}
\begin{align}\label{eq:PN_flux22_normalized}
\hat{F}_{22}&=1-\frac{107}{21}x+\frac{4784}{1323}x^2+\left(4\pi-\frac{4\sigma}{3}\right)x^{3/2}+\left(-\frac{428\pi}{21}+\frac{208\sigma}{63}\right)x^{5/2}\nonumber\\
&+\left(-\frac{16 \pi  \sigma }{3}-\frac{856 \log (x)}{105}+\frac{16 \pi ^2}{3}-\frac{1712 \gamma
   }{105}+\frac{99210071}{1091475}-\frac{3424 \log (2)}{105}\right)x^3\nonumber\\
&+\left(\frac{19136\pi}{1323}+\frac{49844\sigma}{3969}\right)x^{7/2}\nonumber\\
&+\left(\frac{832 \pi  \sigma }{63}+\frac{91592 \log (x)}{2205}-\frac{1712 \pi ^2}{63}+\frac{183184 \gamma
   }{2205}-\frac{27956920577}{81265275}+\frac{366368 \log (2)}{2205}\right)x^4\nonumber\\
&+\biggl[\sigma  \left(\frac{3424 \log (x)}{315}-\frac{64 \pi ^2}{9}+\frac{6848 \gamma }{315}-\frac{46815716}{363825}+\frac{13696 \log
   (2)}{315}\right)\nonumber\\
   &+\pi  \left(-\frac{3424 \log (x)}{105}-\frac{6848 \gamma }{105}+\frac{396840284}{1091475}-\frac{13696 \log
   (2)}{105}\right)\biggr]x^{9/2}\nonumber\\
   &\left(\frac{199376 \pi  \sigma }{3969}-\frac{4095104 \log (x)}{138915}+\frac{76544 \pi ^2}{3969}-\frac{8190208 \gamma
   }{138915}+\frac{187037845924}{6257426175}-\frac{16380416 \log (2)}{138915}\right)x^5\nonumber\\
   &+\biggl[\sigma  \left(-\frac{178048 \log (x)}{6615}+\frac{3328 \pi ^2}{189}-\frac{356096 \gamma
   }{6615}+\frac{123621054016}{893918025}-\frac{712192 \log (2)}{6615}\right)\nonumber\\
   &+\pi  \left(\frac{366368 \log
   (x)}{2205}+\frac{732736 \gamma }{2205}-\frac{111827682308}{81265275}+\frac{1465472 \log (2)}{2205}\right)\biggr]x^{11/2}   
\end{align}

\begin{align}
\hat{F}_{21}&=1+3\sigma x^{1/2}-\frac{17}{14}x+\left(2\pi-\frac{367\sigma}{28}\right)x^{3/2}+\left(6 \pi  \sigma -\frac{2215}{7056}\right)x^2+\left(-\frac{17\pi}{7}+\frac{14383\sigma}{1764}\right)x^{5/2}\nonumber\\
&\left(-\frac{367 \pi  \sigma }{14}-\frac{214 \log (x)}{105}+\frac{4 \pi ^2}{3}-\frac{428 \gamma
   }{105}+\frac{15707221}{727650}-\frac{428 \log (2)}{105}\right)x^3\nonumber\\
   &+\biggl[\sigma  \left(-\frac{214 \log (x)}{35}+4 \pi ^2-\frac{428 \gamma }{35}+\frac{3602793}{53900}-\frac{428 \log
   (2)}{35}\right)-\frac{2215 \pi }{3528}\biggr]x^{7/2}\nonumber\\
   &\left(\frac{14383 \pi  \sigma }{882}+\frac{1819 \log (x)}{735}-\frac{34 \pi ^2}{21}+\frac{3638 \gamma
   }{735}-\frac{6435768121}{1589187600}+\frac{3638 \log (2)}{735}\right)x^4\nonumber\\
   &+\biggl[\sigma  \left(\frac{39269 \log (x)}{1470}-\frac{367 \pi ^2}{21}+\frac{39269 \gamma
   }{735}-\frac{344524801603}{1589187600}+\frac{39269 \log (2)}{735}\right)+\nonumber\\
   &\pi  \left(-\frac{428 \log (x)}{105}-\frac{856 \gamma
   }{105}+\frac{15707221}{363825}-\frac{856 \log (2)}{105}\right)\biggr]x^{9/2}\nonumber\\
   &+\biggl[\pi  \sigma  \left(-\frac{428 \log (x)}{35}-\frac{856 \gamma }{35}+\frac{3602793}{26950}-\frac{856 \log
   (2)}{35}\right)\nonumber\\
   &+\frac{47401 \log (x)}{74088}-\frac{2215 \pi ^2}{5292}+\frac{47401 \gamma
   }{37044}+\frac{113031526373}{11124313200}+\frac{47401 \log (2)}{37044}\biggr]x^5\nonumber\\
   &+\biggl[\sigma  \left(-\frac{1538981 \log (x)}{92610}+\frac{14383 \pi ^2}{1323}-\frac{1538981 \gamma
   }{46305}+\frac{19389238559}{1112431320}-\frac{1538981 \log (2)}{46305}\right)\nonumber\\
   &+\pi  \left(\frac{3638 \log (x)}{735}+\frac{7276
   \gamma }{735}-\frac{6435768121}{794593800}+\frac{7276 \log (2)}{735}\right)\biggr]x^{11/2}
\end{align}

\begin{align}
\hat{F}_{31}&=1-\frac{16}{3}x+(2\pi+5\sigma)x^{3/2}+\frac{437}{33}x^2+\left(-\frac{32\pi}{3}-\frac{269\sigma}{9}\right)x^{5/2}\nonumber\\
&+\left(10 \pi  \sigma -\frac{26 \log (x)}{21}+\frac{4 \pi ^2}{3}-\frac{52 \gamma }{21}-\frac{1137077}{6306300}-\frac{52 \log (2)}{21}\right)x^3\nonumber\\
&+\left(\frac{35839 \sigma }{594}+\frac{874 \pi }{33}\right)x^{7/2}\nonumber\\
&+\left(-\frac{538 \pi  \sigma }{9}+\frac{416 \log (x)}{63}-\frac{64 \pi ^2}{9}+\frac{832 \gamma
   }{63}-\frac{38943317051}{624323700}+\frac{832 \log (2)}{63}\right)x^4\nonumber\\
   &+\biggl[\sigma  \left(-\frac{130 \log (x)}{21}+\frac{20 \pi ^2}{3}-\frac{260 \gamma }{21}+\frac{358636417}{11351340}-\frac{260 \log
   (2)}{21}\right)\nonumber\\
   &+\pi  \left(-\frac{52 \log (x)}{21}-\frac{104 \gamma }{21}-\frac{1137077}{3153150}-\frac{104 \log
   (2)}{21}\right)\biggr]x^{9/2}\nonumber\\
   &+\left(\frac{35839 \pi  \sigma }{297}-\frac{11362 \log (x)}{693}+\frac{1748 \pi ^2}{99}-\frac{22724 \gamma
   }{693}+\frac{806092611133}{4548644100}-\frac{22724 \log (2)}{693}\right)x^5\nonumber\\
   &+\biggl[\sigma  \left(\frac{6994 \log (x)}{189}-\frac{1076 \pi ^2}{27}+\frac{13988 \gamma }{189}-\frac{21000333479}{48024900}+\frac{13988
   \log (2)}{189}\right)\nonumber\\
   &+\pi  \left(\frac{832 \log (x)}{63}+\frac{1664 \gamma }{63}-\frac{38943317051}{312161850}+\frac{1664 \log
   (2)}{63}\right)\biggr]x^{11/2}
\end{align}

\begin{align}
\hat{F}_{32}&=1+4\sigma x^{1/2}-\frac{193}{45}x+\left(4 \pi -\frac{1316 \sigma }{45}\right)x^{3/2}+\left(16 \pi  \sigma +\frac{86111}{22275}\right)x^2+\left(\frac{92498 \sigma }{1485}-\frac{772 \pi }{45}\right)x^{5/2}\nonumber\\
&+\left(-\frac{5264 \pi  \sigma }{45}-\frac{104 \log (x)}{21}+\frac{16 \pi ^2}{3}-\frac{208 \gamma
   }{21}+\frac{960188809}{14189175}-\frac{416 \log (2)}{21}\right)x^3\nonumber\\
   &+\biggl[\sigma  \left(-\frac{416 \log (x)}{21}+\frac{64 \pi ^2}{3}-\frac{832 \gamma }{21}+\frac{3402129202}{14189175}-\frac{1664 \log
   (2)}{21}\right)+\frac{344444 \pi }{22275}\biggr]x^{7/2}\nonumber\\
   &+\left(\frac{369992 \pi  \sigma }{1485}+\frac{20072 \log (x)}{945}-\frac{3088 \pi ^2}{135}+\frac{40144 \gamma
   }{945}-\frac{185796488947}{780404625}+\frac{80288 \log (2)}{945}\right)x^4\nonumber\\
   &+\biggl[\sigma  \left(\frac{19552 \log (x)}{135}-\frac{21056 \pi ^2}{135}+\frac{39104 \gamma
   }{135}-\frac{65791391422}{37162125}+\frac{78208 \log (2)}{135}\right)\nonumber\\
   &+\pi  \left(-\frac{416 \log (x)}{21}-\frac{832 \gamma
   }{21}+\frac{3840755236}{14189175}-\frac{1664 \log (2)}{21}\right)\biggr]x^{9/2}\nonumber\\
   &+\biggl[\pi  \sigma  \left(-\frac{1664 \log (x)}{21}-\frac{3328 \gamma }{21}+\frac{13608516808}{14189175}-\frac{6656 \log
   (2)}{21}\right)\nonumber\\
   &-\frac{8955544 \log (x)}{467775}+\frac{1377776 \pi ^2}{66825}-\frac{17911088 \gamma
   }{467775}+\frac{3746014222817}{27554286375}-\frac{35822176 \log (2)}{467775}\biggr]x^5\nonumber\\
   &+\biggl[\sigma  \left(-\frac{1374256 \log (x)}{4455}+\frac{1479968 \pi ^2}{4455}-\frac{2748512 \gamma
   }{4455}+\frac{367939750486582}{119401907625}-\frac{5497024 \log (2)}{4455}\right)\nonumber\\
   &+\pi  \left(\frac{80288 \log
   (x)}{945}+\frac{160576 \gamma }{945}-\frac{743185955788}{780404625}+\frac{321152 \log (2)}{945}\right)\biggr]x^{11/2}
\end{align}

\begin{align}
\hat{F}_{33}&=1-8x+(6\pi-3\sigma)x^{3/2}+\frac{1003}{55}x^2+\left(-48\pi+\frac{103\sigma}{5}\right)x^{5/2}\nonumber\\
&+\left(-18 \pi  \sigma -\frac{78 \log (x)}{7}+12 \pi ^2-\frac{156 \gamma }{7}+\frac{103033003}{700700}-\frac{156 \log (3)}{7}-\frac{156
   \log (2)}{7}\right)x^3\nonumber\\
   &+\left(\frac{6018 \pi }{55}-\frac{587 \sigma }{22}\right)x^{7/2}\nonumber\\
   &+\left(\frac{618 \pi  \sigma }{5}+\frac{624 \log (x)}{7}-96 \pi ^2+\frac{1248 \gamma }{7}-\frac{8538500227}{7707700}+\frac{1248 \log
   (3)}{7}+\frac{1248 \log (2)}{7}\right)x^4\nonumber\\
   &+\biggl[\sigma  \left(\frac{234 \log (x)}{7}-36 \pi ^2+\frac{468 \gamma }{7}-\frac{350580603}{700700}+\frac{468 \log (3)}{7}+\frac{468
   \log (2)}{7}\right)\nonumber\\
   &+\pi  \left(-\frac{468 \log (x)}{7}-\frac{936 \gamma }{7}+\frac{309099009}{350350}-\frac{936 \log
   (3)}{7}-\frac{936 \log (2)}{7}\right)\biggr]x^{9/2}\nonumber\\
   &+\biggl(-\frac{1761 \pi  \sigma }{11}-\frac{78234 \log (x)}{385}+\frac{12036 \pi ^2}{55}\nonumber\\
   &-\frac{156468 \gamma
   }{385}+\frac{1337469126533}{655154500}-\frac{156468 \log (3)}{385}-\frac{156468 \log (2)}{385}\biggr)x^5\nonumber\\
   &+\biggl[\sigma  \left(-\frac{8034 \log (x)}{35}+\frac{1236 \pi ^2}{5}-\frac{16068 \gamma }{35}+\frac{107744864609}{38538500}-\frac{16068
   \log (3)}{35}-\frac{16068 \log (2)}{35}\right)\nonumber\\
   &+\pi  \left(\frac{3744 \log (x)}{7}+\frac{7488 \gamma
   }{7}-\frac{25615500681}{3853850}+\frac{7488 \log (3)}{7}+\frac{7488 \log (2)}{7}\right)\biggr]x^{11/2}
\end{align}

\begin{align}
\hat{F}_{41}&=1+5\sigma x^{1/2}-\frac{202}{33}x+\left(2\pi-\frac{1432\sigma}{33}\right)x^{3/2}+\left(10 \pi  \sigma +\frac{7478267}{495495}\right)x^2+\left(\frac{24367582 \sigma }{165165}-\frac{404 \pi }{33}\right)x^{5/2}\nonumber\\
&+\left(-\frac{2864 \pi  \sigma }{33}-\frac{3142 \log (x)}{3465}+\frac{4 \pi ^2}{3}-\frac{6284 \gamma
   }{3465}-\frac{1847674847}{624323700}-\frac{6284 \log (2)}{3465}\right)x^3\nonumber\\
   &+\biggl[\sigma  \left(-\frac{3142 \log (x)}{693}+\frac{20 \pi ^2}{3}-\frac{6284 \gamma }{693}-\frac{22314389969}{124864740}-\frac{6284
   \log (2)}{693}\right)+\frac{14956534 \pi }{495495}\biggr]x^{7/2}\nonumber\\
   &+\biggl(\frac{48735164 \pi  \sigma }{165165}+\frac{634684 \log (x)}{114345}-\frac{808 \pi ^2}{99}+\frac{1269368 \gamma
   }{114345}-\frac{6373046696371}{91063854882}+\frac{1269368 \log (2)}{114345}\biggr)x^4\nonumber\\
   &+\biggl[\sigma  \left(\frac{4499344 \log (x)}{114345}-\frac{5728 \pi ^2}{99}+\frac{8998688 \gamma
   }{114345}-\frac{831081952398961}{2276596372050}+\frac{8998688 \log (2)}{114345}\right)\nonumber\\
   &+\pi  \left(-\frac{6284 \log
   (x)}{3465}-\frac{12568 \gamma }{3465}-\frac{1847674847}{312161850}-\frac{12568 \log (2)}{3465}\right)\biggr]x^{9/2}\nonumber\\
   &+\biggl[\pi  \sigma  \left(-\frac{6284 \log (x)}{693}-\frac{12568 \gamma }{693}-\frac{22314389969}{62432370}-\frac{12568 \log
   (2)}{693}\right)\nonumber\\
   &-\frac{23496714914 \log (x)}{1716890175}+\frac{29913068 \pi ^2}{1486485}-\frac{46993429828 \gamma
   }{1716890175}+\frac{348496299110928742}{1921534899409125}-\frac{46993429828 \log (2)}{1716890175}\biggr]x^5\nonumber\\
   &+\biggl[\sigma  \biggl(-\frac{76562942644 \log (x)}{572296725}+\frac{97470328 \pi ^2}{495495}-\frac{153125885288 \gamma
   }{572296725}+\frac{116422245347531790853}{66613209846183000}\nonumber\\
   &-\frac{153125885288 \log (2)}{572296725}\biggr)
   +\pi 
   \left(\frac{1269368 \log (x)}{114345}+\frac{2538736 \gamma }{114345}-\frac{6373046696371}{45531927441}+\frac{2538736 \log
   (2)}{114345}\right)\biggr]x^{11/2}
\end{align}

\begin{align}
\hat{F}_{42}&=1-\frac{437}{55}x+\left(4\pi+\frac{24\sigma}{5}\right)x^{3/2}+\frac{7199152}{275275}x^2+\left(-\frac{11504 \sigma }{275}-\frac{1748 \pi }{55}\right)x^{5/2}\nonumber\\
&+\left(\frac{96 \pi  \sigma }{5}-\frac{12568 \log (x)}{3465}+\frac{16 \pi ^2}{3}-\frac{25136 \gamma
   }{3465}+\frac{9729776708}{780404625}-\frac{50272 \log (2)}{3465}\right)x^3\nonumber\\
   &+\left(\frac{187271932 \sigma }{1376375}+\frac{28796608 \pi }{275275}\right)x^{7/2}\nonumber\\
   &+\left(-\frac{46016 \pi  \sigma }{275}+\frac{5492216 \log (x)}{190575}-\frac{6992 \pi ^2}{165}+\frac{10984432 \gamma
   }{190575}-\frac{3621954717305192}{9485818216875}+\frac{21968864 \log (2)}{190575}\right)x^4\nonumber\\
   &+\biggl[\sigma  \left(-\frac{100544 \log (x)}{5775}+\frac{128 \pi ^2}{5}-\frac{201088 \gamma
   }{5775}+\frac{18538641488}{260134875}-\frac{402176 \log (2)}{5775}\right)\nonumber\\
   &+\pi  \left(-\frac{50272 \log (x)}{3465}-\frac{100544
   \gamma }{3465}+\frac{38919106832}{780404625}-\frac{201088 \log (2)}{3465}\right)\biggr]x^{9/2}\nonumber\\
   &+\biggl[\frac{749087728 \pi  \sigma }{1376375}-\frac{90478942336 \log (x)}{953827875}+\frac{115186432 \pi ^2}{825825}-\frac{180957884672
   \gamma }{953827875}\nonumber\\
   &+\frac{17759046991305753284}{13877752051288125}-\frac{361915769344 \log (2)}{953827875}\biggr]x^5\nonumber\\
   &+\biggl[\sigma  \left(\frac{144582272 \log (x)}{952875}-\frac{184064 \pi ^2}{825}+\frac{289164544 \gamma
   }{952875}-\frac{20087730595257316}{9485818216875}+\frac{578329088 \log (2)}{952875}\right)\nonumber\\
   &+\pi  \left(\frac{21968864 \log
   (x)}{190575}+\frac{43937728 \gamma }{190575}-\frac{14487818869220768}{9485818216875}+\frac{87875456 \log (2)}{190575}\right)\biggr]x^{11/2}
\end{align}

\begin{align}
\hat{F}_{43}&=1+5\sigma x^{1/2}-\frac{78}{11}x+\left(6\pi-\frac{556\sigma}{11}\right)x^{3/2}+\left(30 \pi  \sigma +\frac{850587}{55055}\right)x^2+\left(\frac{9743466 \sigma }{55055}-\frac{468 \pi }{11}\right)x^{5/2}\nonumber\\
&+\left(-\frac{3336 \pi  \sigma }{11}-\frac{3142 \log (x)}{385}+12 \pi ^2-\frac{6284 \gamma }{385}+\frac{8495997793}{69369300}-\frac{6284
   \log (3)}{385}-\frac{6284 \log (2)}{385}\right)x^3\nonumber\\
   &+\biggl[\sigma  \left(-\frac{3142 \log (x)}{77}+60 \pi ^2-\frac{6284 \gamma }{77}+\frac{5881588639}{13873860}-\frac{6284 \log
   (3)}{77}-\frac{6284 \log (2)}{77}\right)+\frac{5103522 \pi }{55055}\biggr]x^{7/2}\nonumber\\
   &+\biggl(\frac{58460796 \pi  \sigma }{55055}+\frac{245076 \log (x)}{4235}-\frac{936 \pi ^2}{11}+\frac{490152 \gamma
   }{4235}\nonumber\\
   &-\frac{23942096774363}{28106128050}+\frac{490152 \log (3)}{4235}+\frac{490152 \log (2)}{4235}\biggr)x^4\nonumber\\
   &+\biggl[\sigma  \left(\frac{1746952 \log (x)}{4235}-\frac{6672 \pi ^2}{11}+\frac{3493904 \gamma
   }{4235}-\frac{520642287162973}{84318384150}+\frac{3493904 \log (3)}{4235}+\frac{3493904 \log (2)}{4235}\right)\nonumber\\
   &+\pi 
   \left(-\frac{18852 \log (x)}{385}-\frac{37704 \gamma }{385}+\frac{8495997793}{11561550}-\frac{37704 \log (3)}{385}-\frac{37704
   \log (2)}{385}\right)\biggr]x^{9/2}\nonumber\\
   &+\biggl[\pi  \sigma  \left(-\frac{18852 \log (x)}{77}-\frac{37704 \gamma }{77}+\frac{5881588639}{2312310}-\frac{37704 \log
   (3)}{77}-\frac{37704 \log (2)}{77}\right)\nonumber\\
   &-\frac{2672544354 \log (x)}{21196175}+\frac{10207044 \pi ^2}{55055}-\frac{5345088708
   \gamma }{21196175}+\frac{164198437359891442}{102798163342875}\nonumber\\
   &-\frac{5345088708 \log (3)}{21196175}-\frac{5345088708 \log
   (2)}{21196175}\biggr]x^5\nonumber\\
   &+\biggl[\sigma  \biggl(-\frac{30613970172 \log (x)}{21196175}+\frac{116921592 \pi ^2}{55055}-\frac{61227940344 \gamma
   }{21196175}+\frac{16347298755161569813}{822385306743000}\nonumber\\
   &-\frac{61227940344 \log (3)}{21196175}-\frac{61227940344 \log
   (2)}{21196175}\biggr)\nonumber\\
   &+\pi  \left(\frac{1470456 \log (x)}{4235}+\frac{2940912 \gamma
   }{4235}-\frac{23942096774363}{4684354675}+\frac{2940912 \log (3)}{4235}+\frac{2940912 \log (2)}{4235}\right)\biggr]x^{11/2}
\end{align}

\begin{align}
\hat{F}_{44}&=1-\frac{593}{55}x+\left(8\pi-\frac{24\sigma}{5}\right)x^{3/2}+\frac{2187772}{55055}x^2+\left(\frac{13236 \sigma }{275}-\frac{4744 \pi }{55}\right)x^{5/2}\nonumber\\
&+\left(-\frac{192 \pi  \sigma }{5}-\frac{50272 \log (x)}{3465}+\frac{64 \pi ^2}{3}-\frac{100544 \gamma
   }{3465}+\frac{143850002468}{780404625}-\frac{100544 \log (2)}{1155}\right)x^3\nonumber\\
   &+\left(\frac{17502176 \pi }{55055}-\frac{205853488 \sigma }{1376375}\right)x^{7/2}\nonumber\\
   &+\left(\frac{105888 \pi  \sigma }{275}+\frac{29811296 \log (x)}{190575}-\frac{37952 \pi ^2}{165}+\frac{59622592 \gamma
   }{190575}-\frac{461493531002948}{193588126875}+\frac{59622592 \log (2)}{63525}\right)x^4\nonumber\\
   &+\biggl[\sigma  \left(\frac{402176 \log (x)}{5775}-\frac{512 \pi ^2}{5}+\frac{804352 \gamma
   }{5775}-\frac{269452289384}{260134875}+\frac{804352 \log (2)}{1925}\right)\nonumber\\
   &+\pi  \left(-\frac{402176 \log
   (x)}{3465}-\frac{804352 \gamma }{3465}+\frac{1150800019744}{780404625}-\frac{804352 \log (2)}{1155}\right)\biggr]x^{9/2}\nonumber\\
   &+\biggl(-\frac{1646827904 \pi  \sigma }{1376375}-\frac{109983673984 \log (x)}{190765575}+\frac{140017408 \pi
   ^2}{165165}-\frac{219967347968 \gamma }{190765575}\nonumber\\
   &+\frac{110876948362954366588}{13877752051288125}-\frac{219967347968 \log
   (2)}{63588525}\biggr)x^5\nonumber\\
   &+\biggl[\sigma  \left(-\frac{221800064 \log (x)}{317625}+\frac{282368 \pi ^2}{275}-\frac{443600128 \gamma
   }{317625}+\frac{33963566444720392}{3161939405625}-\frac{443600128 \log (2)}{105875}\right)\nonumber\\
   &+\pi  \left(\frac{238490368 \log
   (x)}{190575}+\frac{476980736 \gamma }{190575}-\frac{3691948248023584}{193588126875}+\frac{476980736 \log (2)}{63525}\right)\biggr]x^{11/2}
\end{align}

\begin{align}
\hat{F}_{51}&=1-\frac{358}{39}x+\left(2\pi+\frac{28\sigma}{3}\right)x^{3/2}+\frac{290803}{7605}x^2+\left(-\frac{54892 \sigma }{585}-\frac{716 \pi }{39}\right)x^{5/2}\nonumber\\
&+\biggl(\frac{56 \pi  \sigma }{3}-\frac{1546 \log (x)}{2145}+\frac{4 \pi ^2}{3}-\frac{3092 \gamma
   }{2145}-\frac{70678556867}{884458575}-\frac{3092 \log (2)}{2145}\biggr)x^3\nonumber\\
   &+\left(\frac{1818512 \sigma }{4563}+\frac{581606 \pi }{7605}\right)x^{7/2}\nonumber\\
   &+\left(-\frac{109784 \pi  \sigma }{585}+\frac{553468 \log (x)}{83655}-\frac{1432 \pi ^2}{117}+\frac{1106936 \gamma
   }{83655}+\frac{30751133534746}{946665494775}+\frac{1106936 \log (2)}{83655}\right)x^4\nonumber\\
   &+\biggl[\sigma  \left(-\frac{43288 \log (x)}{6435}+\frac{112 \pi ^2}{9}-\frac{86576 \gamma
   }{6435}-\frac{4023624605818}{4927697775}-\frac{86576 \log (2)}{6435}\right)\nonumber\\
   &+\pi  \left(-\frac{3092 \log (x)}{2145}-\frac{6184
   \gamma }{2145}-\frac{141357113734}{884458575}-\frac{6184 \log (2)}{2145}\right)\biggr]x^{9/2}\nonumber\\
   &+\biggl(\frac{3637024 \pi  \sigma }{4563}-\frac{449581438 \log (x)}{16312725}+\frac{1163212 \pi ^2}{22815}-\frac{899162876 \gamma
   }{16312725}+\frac{71365115368720132}{237342563332875}-\frac{899162876 \log (2)}{16312725}\biggr)x^5\nonumber\\
   &+\biggl[\sigma  \left(\frac{84863032 \log (x)}{1254825}-\frac{219568 \pi ^2}{1755}+\frac{169726064 \gamma
   }{1254825}+\frac{21412695417260944}{127799841794625}+\frac{169726064 \log (2)}{1254825}\right)\nonumber\\
   &+\pi  \left(\frac{1106936 \log
   (x)}{83655}+\frac{2213872 \gamma }{83655}+\frac{61502267069492}{946665494775}+\frac{2213872 \log (2)}{83655}\right)\biggr]x^{11/2}	
\end{align}

\begin{align}
\hat{F}_{52}&=1+6\sigma x^{1/2}-\frac{3911}{455}x+\left(4\pi-\frac{30458\sigma}{455}\right)x^{3/2}+\left(24 \pi  \sigma +\frac{18688127}{621075}\right)x^2+\left(\frac{5403626 \sigma }{17745}-\frac{15644 \pi }{455}\right)x^{5/2}\nonumber\\
&+\left(-\frac{121832 \pi  \sigma }{455}-\frac{6184 \log (x)}{2145}+\frac{16 \pi ^2}{3}-\frac{12368 \gamma
   }{2145}-\frac{3565375256}{2299592295}-\frac{24736 \log (2)}{2145}\right)x^3\nonumber\\
   &+\biggl[\sigma  \left(-\frac{12368 \log (x)}{715}+32 \pi ^2-\frac{24736 \gamma }{715}-\frac{225656637512}{547521975}-\frac{49472 \log
   (2)}{715}\right)+\frac{74752508 \pi }{621075}\biggr]x^{7/2}\nonumber\\
   &+\biggl(\frac{21614504 \pi  \sigma }{17745}+\frac{24185624 \log (x)}{975975}-\frac{62576 \pi ^2}{1365}+\frac{48371248 \gamma
   }{975975}-\frac{37120039834250764}{99399876951375}+\frac{96742496 \log (2)}{975975}\biggr)x^4\nonumber\\
   &+\biggl[\sigma  \left(\frac{188352272 \log (x)}{975975}-\frac{487328 \pi ^2}{1365}+\frac{376704544 \gamma
   }{975975}-\frac{237272216817464212}{99399876951375}+\frac{753409088 \log (2)}{975975}\right)\nonumber\\
   &+\pi  \left(-\frac{24736 \log
   (x)}{2145}-\frac{49472 \gamma }{2145}-\frac{14261501024}{2299592295}-\frac{98944 \log (2)}{2145}\right)\biggr]x^{9/2}\nonumber\\
   &+\biggl[\pi  \sigma  \left(-\frac{49472 \log (x)}{715}-\frac{98944 \gamma }{715}-\frac{902626550048}{547521975}-\frac{197888 \log
   (2)}{715}\right)-\frac{115567377368 \log (x)}{1332205875}\nonumber\\
   &+\frac{299010032 \pi ^2}{1863225}-\frac{231134754736 \gamma
   }{1332205875}+\frac{2962795414725501668}{2153664000613125}-\frac{462269509472 \log (2)}{1332205875}\biggr]x^5\nonumber\\
   &+\biggl[\sigma  \left(-\frac{33416023184 \log (x)}{38063025}+\frac{86458016 \pi ^2}{53235}-\frac{66832046368 \gamma
   }{38063025}+\frac{209759979290346544}{15202334121975}-\frac{133664092736 \log (2)}{38063025}\right)\nonumber\\
   &+\pi  \left(\frac{96742496
   \log (x)}{975975}+\frac{193484992 \gamma }{975975}-\frac{148480159337003056}{99399876951375}+\frac{386969984 \log
   (2)}{975975}\right)\biggr]x^{11/2}
\end{align}

\begin{align}
\hat{F}_{53}&=1-\frac{138}{13}x+(6\pi+ 4\sigma)x^{3/2}+\frac{823943}{17745}x^2+\left(-\frac{20844 \sigma }{455}-\frac{828 \pi }{13}\right)x^{5/2}\nonumber\\
&+\biggl(24 \pi  \sigma -\frac{4638 \log (x)}{715}+12 \pi ^2-\frac{9276 \gamma }{715}+\frac{4687046283}{425850425}-\frac{9276 \log
   (3)}{715}-\frac{9276 \log (2)}{715}\biggr)x^3\nonumber\\
   &+\left(\frac{3663136 \sigma }{17745}+\frac{1647886 \pi }{5915}\right)x^{7/2}\nonumber\\
   &+\biggl(-\frac{125064 \pi  \sigma }{455}+\frac{640044 \log (x)}{9295}-\frac{1656 \pi ^2}{13}+\frac{1280088 \gamma
   }{9295}-\frac{28614963763754}{27047585565}+\frac{1280088 \log (3)}{9295}+\frac{1280088 \log (2)}{9295}\biggr)x^4\nonumber\\
   &+\biggl[\sigma  \left(-\frac{18552 \log (x)}{715}+48 \pi ^2-\frac{37104 \gamma }{715}+\frac{4975310434}{425850425}-\frac{37104 \log
   (3)}{715}-\frac{37104 \log (2)}{715}\right)\nonumber\\
   &+\pi  \left(-\frac{27828 \log (x)}{715}-\frac{55656 \gamma
   }{715}+\frac{28122277698}{425850425}-\frac{55656 \log (3)}{715}-\frac{55656 \log (2)}{715}\right)\biggr]x^{9/2}\nonumber\\
   &+\biggl(\frac{7326272 \pi  \sigma }{5915}-\frac{1273815878 \log (x)}{4229225}+\frac{3295772 \pi ^2}{5915}-\frac{2547631756 \gamma
   }{4229225}\nonumber\\
   &+\frac{2096233487612758084}{430732800122625}-\frac{2547631756 \log (3)}{4229225}-\frac{2547631756 \log (2)}{4229225}\biggr)x^5\nonumber\\
   &+\biggl[\sigma  \left(\frac{96674472 \log (x)}{325325}-\frac{250128 \pi ^2}{455}+\frac{193348944 \gamma
   }{325325}-\frac{51479627364210592}{11044430772375}+\frac{193348944 \log (3)}{325325}+\frac{193348944 \log
   (2)}{325325}\right)\nonumber\\
   &+\pi  \left(\frac{3840264 \log (x)}{9295}+\frac{7680528 \gamma
   }{9295}-\frac{57229927527508}{9015861855}+\frac{7680528 \log (3)}{9295}+\frac{7680528 \log (2)}{9295}\right)\biggr]x^{11/2}
\end{align}

\begin{align}
\hat{F}_{54}&=1+6\sigma x^{1/2}-\frac{4451}{455}x+\left(8\pi-\frac{35018\sigma}{455}\right)x^{3/2}+\left(48 \pi  \sigma +\frac{20952707}{621075}\right)x^2+\left(\frac{6596234 \sigma }{17745}-\frac{35608 \pi }{455}\right)x^{5/2}\nonumber\\
&+\left(-\frac{280144 \pi  \sigma }{455}-\frac{24736 \log (x)}{2145}+\frac{64 \pi ^2}{3}-\frac{49472 \gamma
   }{2145}+\frac{1919077079981}{11497961475}-\frac{49472 \log (2)}{715}\right)x^3\nonumber\\
   &+\biggl[\sigma  \left(-\frac{49472 \log (x)}{715}+128 \pi ^2-\frac{98944 \gamma }{715}+\frac{1797855496834}{3832653825}-\frac{296832 \log
   (2)}{715}\right)+\frac{167621656 \pi }{621075}\biggr]x^{7/2}\nonumber\\
   &+\left(\frac{52769872 \pi  \sigma }{17745}+\frac{110099936 \log (x)}{975975}-\frac{284864 \pi ^2}{1365}+\frac{220199872 \gamma
   }{975975}-\frac{193747438701841261}{99399876951375}+\frac{220199872 \log (2)}{325325}\right)x^4\nonumber\\
   &+\biggl[\sigma  \left(\frac{866205248 \log (x)}{975975}-\frac{2241152 \pi ^2}{1365}+\frac{1732410496 \gamma
   }{975975}-\frac{1489526042743253458}{99399876951375}+\frac{1732410496 \log (2)}{325325}\right)\nonumber\\
   &+\pi  \left(-\frac{197888 \log
   (x)}{2145}-\frac{395776 \gamma }{2145}+\frac{15352616639848}{11497961475}-\frac{395776 \log (2)}{715}\right)\biggr]x^{9/2}\nonumber\\
   &+\biggl[\pi  \sigma  \left(-\frac{395776 \log (x)}{715}-\frac{791552 \gamma }{715}+\frac{14382843974672}{3832653825}-\frac{2374656 \log
   (2)}{715}\right)\nonumber\\
   &-\frac{518286160352 \log (x)}{1332205875}+\frac{1340973248 \pi ^2}{1863225}-\frac{1036572320704 \gamma
   }{1332205875}+\frac{13465702596429393907}{2153664000613125}-\frac{1036572320704 \log (2)}{444068625}\biggr]x^5\nonumber\\
   &+\biggl[\sigma  \left(-\frac{163164444224 \log (x)}{38063025}+\frac{422158976 \pi ^2}{53235}-\frac{326328888448 \gamma
   }{38063025}+\frac{92163273485725245586}{1292198400367875}-\frac{326328888448 \log (2)}{12687675}\right)\nonumber\\
   &+\pi 
   \left(\frac{880799488 \log (x)}{975975}+\frac{1761598976 \gamma
   }{975975}-\frac{1549979509614730088}{99399876951375}+\frac{1761598976 \log (2)}{325325}\right)\biggr]x^{11/2}
\end{align}

\begin{align}
\hat{F}_{55}&=1-\frac{526}{39}x+\left(10\pi-\frac{20\sigma}{3}\right)x^{3/2}+\frac{722993}{10647}x^2+\left(\frac{70324 \sigma }{819}-\frac{5260 \pi }{39}\right)x^{5/2}\nonumber\\
&+\left(-\frac{200 \pi  \sigma }{3}-\frac{7730 \log (x)}{429}+\frac{100 \pi ^2}{3}-\frac{15460 \gamma
   }{429}+\frac{6552129589}{35378343}-\frac{15460 \log (5)}{429}-\frac{15460 \log (2)}{429}\right)x^3\nonumber\\
   &+\left(\frac{7229930 \pi }{10647}-\frac{12610048 \sigma }{31941}\right)x^{7/2}\nonumber\\
   &+\biggl(\frac{703240 \pi  \sigma }{819}+\frac{4065980 \log (x)}{16731}-\frac{52600 \pi ^2}{117}+\frac{8131960 \gamma
   }{16731}-\frac{477890910720122}{113599859373}+\frac{8131960 \log (5)}{16731}+\frac{8131960 \log (2)}{16731})x^4\nonumber\\
   &+\biggl[\sigma  \left(\frac{154600 \log (x)}{1287}-\frac{2000 \pi ^2}{9}+\frac{309200 \gamma
   }{1287}-\frac{302007543634}{197107911}+\frac{309200 \log (5)}{1287}+\frac{309200 \log (2)}{1287}\right)\nonumber\\
   &+\pi 
   \left(-\frac{77300 \log (x)}{429}-\frac{154600 \gamma }{429}+\frac{65521295890}{35378343}-\frac{154600 \log
   (5)}{429}-\frac{154600 \log (2)}{429}\right)\biggr]x^{9/2}\nonumber\\
   &+\biggl(-\frac{126100480 \pi  \sigma }{31941}-\frac{5588735890 \log (x)}{4567563}+\frac{72299300 \pi ^2}{31941}-\frac{11177471780 \gamma
   }{4567563}\nonumber\\
   &+\frac{1907636757769320236}{93038284826487}-\frac{11177471780 \log (5)}{4567563}-\frac{11177471780 \log
   (2)}{4567563}\biggr)x^5\nonumber\\
   &+\biggl[\sigma  \biggl(-\frac{543604520 \log (x)}{351351}+\frac{7032400 \pi ^2}{2457}-\frac{1087209040 \gamma
   }{351351}\nonumber\\
   &+\frac{195345738616319360}{7156791140499}-\frac{1087209040 \log (5)}{351351}-\frac{1087209040 \log
   (2)}{351351}\biggr)\nonumber\\
   &+\pi  \left(\frac{40659800 \log (x)}{16731}+\frac{81319600 \gamma
   }{16731}-\frac{4778909107201220}{113599859373}+\frac{81319600 \log (5)}{16731}+\frac{81319600 \log (2)}{16731}\right)\biggr]x^{11/2}
\end{align}

\begin{align}
\hat{F}_{61}&=1+7\sigma x^{1/2}-\frac{125}{12}x+\left(2\pi-\frac{2147\sigma}{24}\right)x^{3/2}+\left(14 \pi  \sigma +\frac{180021689}{3769920}\right)x^2+\left(\frac{78074681 \sigma }{157080}-\frac{125 \pi }{6}\right)x^{5/2}\nonumber\\
&+\left(-\frac{2147 \pi  \sigma }{12}-\frac{1802 \log (x)}{3003}+\frac{4 \pi ^2}{3}-\frac{3604 \gamma
   }{3003}-\frac{24598440919576}{218461268025}-\frac{3604 \log (2)}{3003}\right)x^3\nonumber\\
   &+\biggl[\sigma  \left(-\frac{1802 \log (x)}{429}+\frac{28 \pi ^2}{3}-\frac{3604 \gamma
   }{429}-\frac{186251710369819}{124835010300}-\frac{3604 \log (2)}{429}\right)+\frac{180021689 \pi }{1884960}\biggr]x^{7/2}\nonumber\\
   &+\left(\frac{78074681 \pi  \sigma }{78540}+\frac{112625 \log (x)}{18018}-\frac{125 \pi ^2}{9}+\frac{112625 \gamma
   }{9009}+\frac{707729869694513}{8643122167680}+\frac{112625 \log (2)}{9009}\right)x^4\nonumber\\
   &+\biggl[\sigma  \left(\frac{1934447 \log (x)}{36036}-\frac{2147 \pi ^2}{18}+\frac{1934447 \gamma
   }{18018}+\frac{39883212208000103}{19014868768896}+\frac{1934447 \log (2)}{18018}\right)\nonumber\\
   &+\pi  \left(-\frac{3604 \log
   (x)}{3003}-\frac{7208 \gamma }{3003}-\frac{49196881839152}{218461268025}-\frac{7208 \log (2)}{3003}\right)\biggr]x^{9/2}\nonumber\\
   &+\biggl[\pi  \sigma  \left(-\frac{3604 \log (x)}{429}-\frac{7208 \gamma }{429}-\frac{186251710369819}{62417505150}-\frac{7208 \log
   (2)}{429}\right)\nonumber\\
   &-\frac{9541149517 \log (x)}{332972640}+\frac{180021689 \pi ^2}{2827440}-\frac{9541149517 \gamma
   }{166486320}+\frac{1443709506661055325611}{4457029313344608000}-\frac{9541149517 \log (2)}{166486320}\biggr]x^5\nonumber\\
   &+\biggl[\sigma  \left(-\frac{4137958093 \log (x)}{13873860}+\frac{78074681 \pi ^2}{117810}-\frac{4137958093 \gamma
   }{6936930}+\frac{4849567930889021794613}{3157062430285764000}-\frac{4137958093 \log (2)}{6936930}\right)\nonumber\\
   &+\pi 
   \left(\frac{112625 \log (x)}{9009}+\frac{225250 \gamma }{9009}+\frac{707729869694513}{4321561083840}+\frac{225250 \log
   (2)}{9009}\right)\biggr]x^{11/2}
\end{align}

\begin{align}
\hat{F}_{62}&=1-\frac{81}{7}x+\left(4\pi+\frac{68\sigma}{7}\right)x^{3/2}+\frac{8221522}{137445}x^2+\left(-\frac{88714 \sigma }{735}-\frac{324 \pi }{7}\right)x^{5/2}\nonumber\\
&+\left(\frac{272 \pi  \sigma }{7}-\frac{7208 \log (x)}{3003}+\frac{16 \pi ^2}{3}-\frac{14416 \gamma
   }{3003}-\frac{161484749374217}{1223383100940}-\frac{28832 \log (2)}{3003}\right)x^3\nonumber\\
   &+\left(\frac{7295384 \sigma }{11319}+\frac{32886088 \pi }{137445}\right)x^{7/2}\nonumber\\
   &+\left(-\frac{354856 \pi  \sigma }{735}+\frac{194616 \log (x)}{7007}-\frac{432 \pi ^2}{7}+\frac{389232 \gamma
   }{7007}-\frac{138951395311225087}{727912945059300}+\frac{778464 \log (2)}{7007}\right)x^4\nonumber\\
   &+\biggl[\sigma  \left(-\frac{490144 \log (x)}{21021}+\frac{1088 \pi ^2}{21}-\frac{980288 \gamma
   }{21021}-\frac{2244361964428541}{1529228876175}-\frac{1960576 \log (2)}{21021}\right)\nonumber\\
   &+\pi  \left(-\frac{28832 \log
   (x)}{3003}-\frac{57664 \gamma }{3003}-\frac{161484749374217}{305845775235}-\frac{115328 \log (2)}{3003}\right)\biggr]x^{9/2}\nonumber\\
   &+\biggl[\frac{29181536 \pi  \sigma }{11319}-\frac{3485925328 \log (x)}{24279255}+\frac{131544352 \pi ^2}{412335}-\frac{6971850656 \gamma
   }{24279255}+\frac{69009387572205137581}{30693662516667150}-\frac{13943701312 \log (2)}{24279255}\biggr]x^5\nonumber\\
   &+\biggl[\sigma  \left(\frac{639450512 \log (x)}{2207205}-\frac{1419424 \pi ^2}{2205}+\frac{1278901024 \gamma
   }{2207205}-\frac{51564217241617}{25040004990}+\frac{2557802048 \log (2)}{2207205}\right)\nonumber\\
   &+\pi  \left(\frac{778464 \log
   (x)}{7007}+\frac{1556928 \gamma }{7007}-\frac{138951395311225087}{181978236264825}+\frac{3113856 \log (2)}{7007}\right)\biggr]x^{11/2}
\end{align}

\begin{align}
\hat{F}_{63}&=1+7\sigma x^{1/2}-\frac{133}{21}x+\left(6\pi-\frac{2299\sigma}{24}\right)x^{3/2}+\left(42 \pi  \sigma +\frac{191999369}{3769920}\right)x^2+\left(\frac{86132401 \sigma }{157080}-\frac{133 \pi }{2}\right)x^{5/2}\nonumber\\
&+\left(-\frac{2299 \pi  \sigma }{4}-\frac{5406 \log (x)}{1001}+12 \pi ^2-\frac{10812 \gamma
   }{1001}-\frac{8318725110259}{582563381400}-\frac{10812 \log (3)}{1001}-\frac{10812 \log (2)}{1001}\right)x^3\nonumber\\
   &+\biggl[\sigma  \left(-\frac{5406 \log (x)}{143}+84 \pi ^2-\frac{10812 \gamma }{143}-\frac{6435219971227}{6935278350}-\frac{10812 \log
   (3)}{143}-\frac{10812 \log (2)}{143}\right)+\frac{191999369 \pi }{628320}\biggr]x^{7/2}\nonumber\\
   &\left(\frac{86132401 \pi  \sigma }{26180}+\frac{17119 \log (x)}{286}-133 \pi ^2+\frac{17119 \gamma
   }{143}-\frac{158600290311609547}{158457239740800}+\frac{17119 \log (3)}{143}+\frac{17119 \log (2)}{143}\right)x^4\nonumber\\
   &+\biggl[\sigma  \left(\frac{188309 \log (x)}{364}-\frac{2299 \pi ^2}{2}+\frac{188309 \gamma
   }{182}-\frac{59571165214052209}{8339854723200}+\frac{188309 \log (3)}{182}+\frac{188309 \log (2)}{182}\right)\nonumber\\
   &+\pi 
   \left(-\frac{32436 \log (x)}{1001}-\frac{64872 \gamma }{1001}-\frac{8318725110259}{97093896900}-\frac{64872 \log
   (3)}{1001}-\frac{64872 \log (2)}{1001}\right)\biggr]x^{9/2}\nonumber\\
   &+\biggl[\pi  \sigma  \left(-\frac{32436 \log (x)}{143}-\frac{64872 \gamma }{143}-\frac{6435219971227}{1155879725}-\frac{64872 \log
   (3)}{143}-\frac{64872 \log (2)}{143}\right)\nonumber\\
   &-\frac{10175966557 \log (x)}{36996960}+\frac{191999369 \pi
   ^2}{314160}-\frac{10175966557 \gamma }{18498480}+\frac{42300157953961348695307}{8418833147428704000}\nonumber\\
   &-\frac{10175966557 \log
   (3)}{18498480}-\frac{10175966557 \log (2)}{18498480}\biggr]x^5\nonumber\\
   &+\biggl[\sigma  \biggl(-\frac{4565017253 \log (x)}{1541540}+\frac{86132401 \pi ^2}{13090}-\frac{4565017253 \gamma
   }{770770}+\frac{6208405960475571319531}{116928238158732000}\nonumber\\
   &-\frac{4565017253 \log (3)}{770770}-\frac{4565017253 \log
   (2)}{770770}\biggr)\nonumber\\
   &+\pi  \left(\frac{51357 \log (x)}{143}+\frac{102714 \gamma
   }{143}-\frac{158600290311609547}{26409539956800}+\frac{102714 \log (3)}{143}+\frac{102714 \log (2)}{143}\right)\biggr]x^{11/2}
\end{align}

\begin{align}
\hat{F}_{64}&=1-\frac{93}{7}x+\left(8\pi+\frac{20\sigma}{7}\right)x^{3/2}+\frac{2028464}{27489}x^2+\left(-\frac{5996 \sigma }{147}-\frac{744 \pi }{7}\right)x^{5/2}\nonumber\\
&+\left(\frac{160 \pi  \sigma }{7}-\frac{28832 \log (x)}{3003}+\frac{64 \pi ^2}{3}-\frac{57664 \gamma
   }{3003}-\frac{7473136770797}{305845775235}-\frac{57664 \log (2)}{1001}\right)x^3\nonumber\\
   &+\left(\frac{45679300 \sigma }{192423}+\frac{16227712 \pi }{27489}\right)x^{7/2}\nonumber\\
   &+\left(-\frac{47968 \pi  \sigma }{147}+\frac{893792 \log (x)}{7007}-\frac{1984 \pi ^2}{7}+\frac{1787584 \gamma
   }{7007}-\frac{389504245584812167}{181978236264825}+\frac{5362752 \log (2)}{7007}\right)x^4\nonumber\\
   &+\biggl[\sigma  \left(-\frac{576640 \log (x)}{21021}+\frac{1280 \pi ^2}{21}-\frac{1153280 \gamma
   }{21021}-\frac{9680228679908}{61169155047}-\frac{1153280 \log (2)}{7007}\right)\nonumber\\
   &+\pi  \left(-\frac{230656 \log
   (x)}{3003}-\frac{461312 \gamma }{3003}-\frac{59785094166376}{305845775235}-\frac{461312 \log (2)}{1001}\right)\biggr]x^{9/2}\nonumber\\
   &+\biggl(\frac{365434400 \pi  \sigma }{192423}-\frac{3440274944 \log (x)}{4855851}+\frac{129821696 \pi ^2}{82467}-\frac{6880549888 \gamma
   }{4855851}\nonumber\\
   &+\frac{200984978531762355824}{15346831258333575}-\frac{6880549888 \log (2)}{1618617}\biggr)x^5\nonumber\\
   &+\biggl[\sigma  \left(\frac{172876672 \log (x)}{441441}-\frac{383744 \pi ^2}{441}+\frac{345753344 \gamma
   }{441441}-\frac{79666062343153756}{12131882417655}+\frac{345753344 \log (2)}{147147}\right)\nonumber\\
   &+\pi  \left(\frac{7150336 \log
   (x)}{7007}+\frac{14300672 \gamma }{7007}-\frac{3116033964678497336}{181978236264825}+\frac{42902016 \log (2)}{7007}\right)\biggr]x^{11/2}
\end{align}

\begin{align}
\hat{F}_{65}&=1+7\sigma x^{1/2}-\frac{149}{12}x+\left(10\pi-\frac{2603\sigma}{24}\right)x^{3/2}+\left(70 \pi  \sigma +\frac{44119381}{753984}\right)x^2+\left(\frac{20910829 \sigma }{31416}-\frac{745 \pi }{6}\right)x^{5/2}\nonumber\\
&+\left(-\frac{13015 \pi  \sigma }{12}-\frac{45050 \log (x)}{3003}+\frac{100 \pi ^2}{3}-\frac{90100 \gamma
   }{3003}+\frac{12846599702749}{69907605768}-\frac{90100 \log (5)}{3003}-\frac{90100 \log (2)}{3003}\right)x^3\nonumber\\
   &+\biggl[\sigma  \left(-\frac{45050 \log (x)}{429}+\frac{700 \pi ^2}{3}-\frac{90100 \gamma
   }{429}+\frac{198692130113}{1248350103}-\frac{90100 \log (5)}{429}-\frac{90100 \log (2)}{429}\right)+\frac{220596905 \pi
   }{376992}\biggr]x^{7/2}\nonumber\\
   &+\left(\frac{104554145 \pi  \sigma }{15708}+\frac{3356225 \log (x)}{18018}-\frac{3725 \pi ^2}{9}+\frac{3356225 \gamma
   }{9009}-\frac{9734336137180063}{2716409824128}+\frac{3356225 \log (5)}{9009}+\frac{3356225 \log (2)}{9009}\right)x^4\nonumber\\
   &+\biggl[\sigma  \left(\frac{58632575 \log (x)}{36036}-\frac{65075 \pi ^2}{18}+\frac{58632575 \gamma
   }{18018}-\frac{562883137821630721}{19014868768896}+\frac{58632575 \log (5)}{18018}+\frac{58632575 \log (2)}{18018}\right)\nonumber\\
   &+\pi 
   \left(-\frac{450500 \log (x)}{3003}-\frac{901000 \gamma }{3003}+\frac{64232998513745}{34953802884}-\frac{901000 \log
   (5)}{3003}-\frac{901000 \log (2)}{3003}\right)\biggr]x^{9/2}\nonumber\\
   &+\biggl[\pi  \sigma  \left(-\frac{450500 \log (x)}{429}-\frac{901000 \gamma }{429}+\frac{1986921301130}{1248350103}-\frac{901000 \log
   (5)}{429}-\frac{901000 \log (2)}{429}\right)\nonumber\\
   &-\frac{58458179825 \log (x)}{66594528}+\frac{1102984525 \pi
   ^2}{565488}-\frac{58458179825 \gamma }{33297264}+\frac{10003750718547616854575}{606155986614866688}\nonumber\\
   &-\frac{58458179825 \log
   (5)}{33297264}-\frac{58458179825 \log (2)}{33297264}\biggr]x^5\nonumber\\
   &+\biggl[\sigma  \biggl(-\frac{27706848425 \log (x)}{2774772}+\frac{522770725 \pi ^2}{23562}-\frac{27706848425 \gamma
   }{1387386}+\frac{4801663674782405880901}{25256499442286112}\nonumber\\
   &-\frac{27706848425 \log (5)}{1387386}-\frac{27706848425 \log
   (2)}{1387386}\biggr)\nonumber\\
   &+\pi  \left(\frac{16781125 \log (x)}{9009}+\frac{33562250 \gamma
   }{9009}-\frac{48671680685900315}{1358204912064}+\frac{33562250 \log (5)}{9009}+\frac{33562250 \log (2)}{9009}\right)\biggr]x^{11/2}
\end{align}

\begin{align}
\hat{F}_{66}&=1-\frac{113}{7}x+\left(12\pi-\frac{60\sigma}{7}\right)x^{3/2}+\frac{134290}{1309}x^2+\left(\frac{6554 \sigma }{49}-\frac{1356 \pi }{7}\right)x^{5/2}\nonumber\\
&+\left(-\frac{720 \pi  \sigma }{7}-\frac{21624 \log (x)}{1001}+48 \pi ^2-\frac{43248 \gamma
   }{1001}+\frac{18044070358903}{135931455660}-\frac{43248 \log (3)}{1001}-\frac{86496 \log (2)}{1001}\right)x^3\nonumber\\
   &+\left(\frac{1611480 \pi }{1309}-\frac{51347864 \sigma }{64141}\right)x^{7/2}\nonumber\\
   &+\left(\frac{78648 \pi  \sigma }{49}+\frac{2443512 \log (x)}{7007}-\frac{5424 \pi ^2}{7}+\frac{4887024 \gamma
   }{7007}-\frac{530645101377839167}{80879216117700}+\frac{4887024 \log (3)}{7007}+\frac{9774048 \log (2)}{7007}\right)x^4\nonumber\\
   &+\biggl[\sigma  \left(\frac{1297440 \log (x)}{7007}-\frac{2880 \pi ^2}{7}+\frac{2594880 \gamma
   }{7007}-\frac{3745245301351}{2265524261}+\frac{2594880 \log (3)}{7007}+\frac{5189760 \log (2)}{7007}\right)\nonumber\\
   &+\pi 
   \left(-\frac{259488 \log (x)}{1001}-\frac{518976 \gamma }{1001}+\frac{18044070358903}{11327621305}-\frac{518976 \log
   (3)}{1001}-\frac{1037952 \log (2)}{1001}\right)\biggr]x^{9/2}\nonumber\\
   &+\biggl[-\frac{616174368 \pi  \sigma }{64141}-\frac{13139760 \log (x)}{5929}+\frac{6445920 \pi ^2}{1309}-\frac{26279520 \gamma
   }{5929}\nonumber\\
   &+\frac{656585911003784857}{15431705639350}-\frac{26279520 \log (3)}{5929}
   -\frac{52559040 \log (2)}{5929}\biggr]x^5\nonumber\\
   &+\biggl[\sigma  \left(-\frac{141723696 \log (x)}{49049}+\frac{314592 \pi ^2}{49}-\frac{283447392 \gamma
   }{49049}+\frac{450585773170908589}{8087921611770}-\frac{283447392 \log (3)}{49049}-\frac{566894784 \log (2)}{49049}\right)\nonumber\\
   &+\pi
    \left(\frac{29322144 \log (x)}{7007}+\frac{58644288 \gamma }{7007}-\frac{530645101377839167}{6739934676475}+\frac{58644288
   \log (3)}{7007}+\frac{117288576 \log (2)}{7007}\right)\biggr]x^{11/2}
\end{align}

\begin{align}
\hat{F}_{71}&=1-\frac{223}{17}x+\left(2\pi+\frac{27\sigma}{2}\right)x^{3/2}+\frac{39565047}{499681}x^2+\left(-\frac{90301 \sigma }{476}-\frac{446 \pi }{17}\right)x^{5/2}\nonumber\\
&+\left(27 \pi  \sigma -\frac{11948 \log (x)}{23205}+\frac{4 \pi ^2}{3}-\frac{23896 \gamma
   }{23205}-\frac{900375181510781}{3241170812880}-\frac{23896 \log (2)}{23205}\right)x^3\nonumber\\
   &+\left(\frac{9523157277 \sigma }{7994896}+\frac{79130094 \pi }{499681}\right)x^{7/2}\nonumber\\
   &+\left(-\frac{90301 \pi  \sigma }{238}+\frac{2664404 \log (x)}{394485}-\frac{892 \pi ^2}{51}+\frac{5328808 \gamma
   }{394485}+\frac{272158571374959377057}{481573159377710400}+\frac{5328808 \log (2)}{394485}\right)x^4\nonumber\\
   &+\biggl[\sigma  \left(-\frac{53766 \log (x)}{7735}+18 \pi ^2-\frac{107532 \gamma }{7735}-\frac{204661562601577}{48017345376}-\frac{107532
   \log (2)}{7735}\right)\nonumber\\
   &+\pi  \left(-\frac{23896 \log (x)}{23205}-\frac{47792 \gamma
   }{23205}-\frac{900375181510781}{1620585406440}-\frac{47792 \log (2)}{23205}\right)\biggr]x^{9/2}\nonumber\\
   &+\biggl(\frac{9523157277 \pi  \sigma }{3997448}-\frac{157574393852 \log (x)}{3865032535}+\frac{52753396 \pi
   ^2}{499681}-\frac{315148787704 \gamma }{3865032535}\nonumber\\
   &-\frac{235741151452316569687157}{620828064631098324000}-\frac{315148787704
   \log (2)}{3865032535}\biggr)x^5\nonumber\\
   &+\biggl[\sigma  \left(\frac{269729087 \log (x)}{2761395}-\frac{90301 \pi ^2}{357}+\frac{539458174 \gamma
   }{2761395}+\frac{29113672696708980165059}{3371012115643972800}+\frac{539458174 \log (2)}{2761395}\right)\nonumber\\
   &+\pi 
   \left(\frac{5328808 \log (x)}{394485}+\frac{10657616 \gamma
   }{394485}+\frac{272158571374959377057}{240786579688855200}+\frac{10657616 \log (2)}{394485}\right)\biggr]x^{11/2}
\end{align}

\begin{align}
\hat{F}_{72}&=1+8\sigma x^{1/2}-\frac{13619}{1071}x+\left(4\pi-\frac{129320}{1071}\right)x^{3/2}+\left(32 \pi  \sigma +\frac{20288614207}{283319127}\right)x^2+\left(\frac{3613352872 \sigma }{4497129}-\frac{54476 \pi }{1071}\right)x^{5/2}\nonumber\\
&+\left(-\frac{517280 \pi  \sigma }{1071}-\frac{47792 \log (x)}{23205}+\frac{16 \pi ^2}{3}-\frac{95584 \gamma
   }{23205}-\frac{469120756856}{2500903405}-\frac{191168 \log (2)}{23205}\right)x^3\nonumber\\
   &+\biggl[\sigma  \left(-\frac{382336 \log (x)}{23205}+\frac{128 \pi ^2}{3}-\frac{764672 \gamma
   }{23205}-\frac{110775486737506}{40514635161}-\frac{1529344 \log (2)}{23205}\right)+\frac{81154456828 \pi }{283319127}\biggr]x^{7/2}\nonumber\\
   &+\biggl(\frac{14453411488 \pi  \sigma }{4497129}+\frac{650879248 \log (x)}{24852555}-\frac{217904 \pi ^2}{3213}+\frac{1301758496 \gamma
   }{24852555}-\frac{5565093875687384492}{67721225537490525}+\frac{2603516992 \log (2)}{24852555}\biggr)x^4\nonumber\\
   &+\biggl[\sigma  \left(\frac{1236092288 \log (x)}{4970511}-\frac{2069120 \pi ^2}{3213}+\frac{2472184576 \gamma
   }{4970511}+\frac{153689215630344417094}{67721225537490525}+\frac{4944369152 \log (2)}{4970511}\right)\nonumber\\
   &+\pi  \left(-\frac{191168
   \log (x)}{23205}-\frac{382336 \gamma }{23205}-\frac{1876483027424}{2500903405}-\frac{764672 \log (2)}{23205}\right)\biggr]x^{9/2}\nonumber\\
   &+\biggl[\pi  \sigma  \left(-\frac{1529344 \log (x)}{23205}-\frac{3058688 \gamma
   }{23205}-\frac{443101946950024}{40514635161}-\frac{6117376 \log (2)}{23205}\right)\nonumber\\
   &-\frac{969633450180944 \log
   (x)}{6574420342035}+\frac{324617827312 \pi ^2}{849957381}-\frac{1939266900361888 \gamma
   }{6574420342035}\nonumber\\
   &+\frac{26804007745506705831944402}{11000297270182273428375}-\frac{3878533800723776 \log (2)}{6574420342035}\biggr]x^5\nonumber\\
   &+\biggl[\sigma  \biggl(-\frac{172689360458624 \log (x)}{104355878445}+\frac{57813645952 \pi ^2}{13491387}-\frac{345378720917248 \gamma
   }{104355878445}\nonumber\\
   &+\frac{11711346651047295703868141}{523823679532489210875}-\frac{690757441834496 \log
   (2)}{104355878445}\biggr)\nonumber\\
   &+\pi  \left(\frac{2603516992 \log (x)}{24852555}+\frac{5207033984 \gamma
   }{24852555}-\frac{22260375502749537968}{67721225537490525}+\frac{10414067968 \log (2)}{24852555}\right)\biggr]x^{11/2}
\end{align}

\begin{align}
\hat{F}_{73}&=1-\frac{239}{17}x+\left(6\pi+\frac{19}{2}\right)x^{3/2}+\frac{131958599}{1499043}x^2+\left(-\frac{201767 \sigma }{1428}-\frac{1434 \pi }{17}\right)x^{5/2}\nonumber\\
&+\left(57 \pi  \sigma -\frac{35844 \log (x)}{7735}+12 \pi ^2-\frac{71688 \gamma }{7735}-\frac{78766297655021}{360130090320}-\frac{71688
   \log (3)}{7735}-\frac{71688 \log (2)}{7735}\right)x^3\nonumber\\
   &+\left(\frac{1154491277 \sigma }{1262352}+\frac{263917198 \pi }{499681}\right)x^{7/2}\nonumber\\
   &+\biggl(-\frac{201767 \pi  \sigma }{238}+\frac{8566716 \log (x)}{131495}-\frac{2868 \pi ^2}{17}+\frac{17133432 \gamma
   }{131495}-\frac{79572683467659663}{121333625441600}+\frac{17133432 \log (3)}{131495}+\frac{17133432 \log (2)}{131495}\biggr)x^4\nonumber\\
   &+\biggl[\sigma  \left(-\frac{340518 \log (x)}{7735}+114 \pi ^2-\frac{681036 \gamma
   }{7735}-\frac{344989969285499}{144052036128}-\frac{681036 \log (3)}{7735}-\frac{681036 \log (2)}{7735}\right)\nonumber\\
   &+\pi 
   \left(-\frac{215064 \log (x)}{7735}-\frac{430128 \gamma }{7735}-\frac{78766297655021}{60021681720}-\frac{430128 \log
   (3)}{7735}-\frac{430128 \log (2)}{7735}\right)\biggr]x^{9/2}\nonumber\\
   &+\biggl(\frac{1154491277 \pi  \sigma }{210392}-\frac{1576641340852 \log (x)}{3865032535}+\frac{527834396 \pi
   ^2}{499681}-\frac{3153282681704 \gamma
   }{3865032535}\nonumber\\
   &+\frac{4609993098496362230956667}{620828064631098324000}-\frac{3153282681704 \log
   (3)}{3865032535}-\frac{3153282681704 \log (2)}{3865032535}\biggr)x^5\nonumber\\
   &+\biggl[\sigma  \left(\frac{602678029 \log (x)}{920465}-\frac{201767 \pi ^2}{119}+\frac{1205356058 \gamma
   }{920465}-\frac{20019219044238975311}{3112661233281600}+\frac{1205356058 \log (3)}{920465}+\frac{1205356058 \log
   (2)}{920465}\right)\nonumber\\
   &+\pi  \left(\frac{51400296 \log (x)}{131495}+\frac{102800592 \gamma
   }{131495}-\frac{238718050402978989}{60666812720800}+\frac{102800592 \log (3)}{131495}+\frac{102800592 \log (2)}{131495}\right)\biggr]x^{11/2}
\end{align}

\begin{align}
\hat{F}_{74}&=1+8\sigma x^{1/2}-\frac{14543}{1071}x+\left(8 \pi -\frac{139064 \sigma }{1071}\right)x^{3/2}+\left(64 \pi  \sigma +\frac{22046072455}{283319127}\right)x^2+\left(\frac{4041717718 \sigma }{4497129}-\frac{116344 \pi }{1071}\right)x^{5/2}\nonumber\\
&+\left(-\frac{1112512 \pi  \sigma }{1071}-\frac{191168 \log (x)}{23205}+\frac{64 \pi ^2}{3}-\frac{382336 \gamma
   }{23205}-\frac{3897310139143}{67524391935}-\frac{382336 \log (2)}{7735}\right)x^3\nonumber\\
   &+\biggl[\sigma  \left(-\frac{1529344 \log (x)}{23205}+\frac{512 \pi ^2}{3}-\frac{3058688 \gamma
   }{23205}-\frac{79565406120562}{40514635161}-\frac{3058688 \log (2)}{7735}\right)+\frac{176368579640 \pi }{283319127}\biggr]x^{7/2}\nonumber\\
   &+\left(\frac{32333741744 \pi  \sigma }{4497129}+\frac{2780156224 \log (x)}{24852555}-\frac{930752 \pi ^2}{3213}+\frac{5560312448 \gamma
   }{24852555}-\frac{19342673411957889689}{9674460791070075}+\frac{5560312448 \log (2)}{8284185}\right)x^4\nonumber\\
   &+\biggl[\sigma  \left(\frac{26584586752 \log (x)}{24852555}-\frac{8900096 \pi ^2}{3213}+\frac{53169173504 \gamma
   }{24852555}-\frac{1052245252615435125134}{67721225537490525}+\frac{53169173504 \log (2)}{8284185}\right)\nonumber\\
   &+\pi 
   \left(-\frac{1529344 \log (x)}{23205}-\frac{3058688 \gamma }{23205}-\frac{31178481113144}{67524391935}-\frac{3058688 \log
   (2)}{7735}\right)\biggr]x^{9/2}\nonumber\\
   &+\biggl[\pi  \sigma  \left(-\frac{12234752 \log (x)}{23205}-\frac{24469504 \gamma
   }{23205}-\frac{636523248964496}{40514635161}-\frac{24469504 \log (2)}{7735}\right)\nonumber\\
   &-\frac{842900715815488 \log
   (x)}{1314884068407}+\frac{1410948637120 \pi ^2}{849957381}-\frac{1685801431630976 \gamma
   }{1314884068407}\nonumber\\
   &+\frac{143862077225308606071034793}{11000297270182273428375}-\frac{1685801431630976 \log (2)}{438294689469}\biggr]x^5\nonumber\\
   &+\biggl[\sigma  \biggl(-\frac{772647092714624 \log (x)}{104355878445}+\frac{258669933952 \pi ^2}{13491387}-\frac{1545294185429248 \gamma
   }{104355878445}\nonumber\\
   &+\frac{76962980238610492704134354}{523823679532489210875}-\frac{1545294185429248 \log
   (2)}{34785292815}\biggr)\nonumber\\
   &+\pi  \biggl(\frac{22241249792 \log (x)}{24852555}+\frac{44482499584 \gamma
   }{24852555}\nonumber\\
   &-\frac{154741387295663117512}{9674460791070075}+\frac{44482499584 \log (2)}{8284185}\biggr)\biggr]x^{11/2}
\end{align}

\begin{align}
\hat{F}_{75}&=1-\frac{271}{17}x+\left(10\pi+\frac{3\sigma}{2}\right)x^{3/2}+\frac{161912939}{1499043}x^2+\left(-\frac{37511 \sigma }{1428}-\frac{2710 \pi }{17}\right)x^{5/2}\nonumber\\
&+\left(15 \pi  \sigma -\frac{59740 \log (x)}{4641}+\frac{100 \pi ^2}{3}-\frac{119480 \gamma
   }{4641}-\frac{72708277771009}{648234162576}-\frac{119480 \log (5)}{4641}-\frac{119480 \log (2)}{4641}\right)x^3\nonumber\\
   &+\left(\frac{4609070647 \sigma }{23984688}+\frac{1619129390 \pi }{1499043}\right)x^{7/2}\nonumber\\
   &+\biggl(-\frac{187555 \pi  \sigma }{714}+\frac{16189540 \log (x)}{78897}-\frac{27100 \pi ^2}{51}+\frac{32379080 \gamma
   }{78897}\nonumber\\
   &-\frac{69798849954372148039}{19262926375108416}+\frac{32379080 \log (5)}{78897}+\frac{32379080 \log (2)}{78897}\biggr)x^4\nonumber\\
   &+\biggl[\sigma  \left(-\frac{29870 \log (x)}{1547}+50 \pi ^2-\frac{59740 \gamma }{1547}-\frac{139128548929393}{432156108384}-\frac{59740
   \log (5)}{1547}-\frac{59740 \log (2)}{1547}\right)\nonumber\\
   &+\pi  \left(-\frac{597400 \log (x)}{4641}-\frac{1194800 \gamma
   }{4641}-\frac{363541388855045}{324117081288}-\frac{1194800 \log (5)}{4641}-\frac{1194800 \log (2)}{4641}\right)\biggr]x^{9/2}\nonumber\\
   &+\biggl[\frac{23045353235 \pi  \sigma }{11992344}-\frac{9672678975860 \log (x)}{6957058563}+\frac{16191293900 \pi
   ^2}{4497129}-\frac{19345357951720 \gamma
   }{6957058563}\nonumber\\
   &+\frac{1275474878427485531926255}{44699620653439079328}-\frac{19345357951720 \log
   (5)}{6957058563}-\frac{19345357951720 \log (2)}{6957058563}\biggr]x^5\nonumber\\
   &+\biggl[\sigma  \biggl(\frac{560226785 \log (x)}{1656837}-\frac{937775 \pi ^2}{1071}+\frac{1120453570 \gamma
   }{1656837}-\frac{210794710578688517173}{36774677625206976}\nonumber\\
   &+\frac{1120453570 \log (5)}{1656837}+\frac{1120453570 \log
   (2)}{1656837}\biggr)\nonumber\\
   &+\pi  \biggl(\frac{161895400 \log (x)}{78897}+\frac{323790800 \gamma
   }{78897}-\frac{348994249771860740195}{9631463187554208}+\frac{323790800 \log (5)}{78897}+\frac{323790800 \log
   (2)}{78897}\biggr)\biggr]x^{11/2}
\end{align}

\begin{align}
\hat{F}_{76}&=1+8\sigma x^{1/2}-\frac{1787}{119}x+\left(12 \pi -\frac{17256 \sigma }{119}\right)x^{3/2}+\left(96 \pi  \sigma +\frac{313730895}{3497767}\right)x^2+\left(\frac{538436792 \sigma }{499681}-\frac{21444 \pi }{119}\right)x^{5/2}\nonumber\\
&+\biggl(-\frac{207072 \pi  \sigma }{119}-\frac{143376 \log (x)}{7735}+48 \pi ^2-\frac{286752 \gamma
   }{7735}+\frac{69080576624}{441335895}-\frac{286752 \log (3)}{7735}-\frac{573504 \log (2)}{7735}\biggr)x^3\nonumber\\
   &+\biggl[\sigma  \left(-\frac{1147008 \log (x)}{7735}+384 \pi ^2-\frac{2294016 \gamma
   }{7735}-\frac{1179430181126}{1500542043}-\frac{2294016 \log (3)}{7735}-\frac{4588032 \log (2)}{7735}\right)+\frac{3764770740
   \pi }{3497767}\biggr]x^{7/2}\nonumber\\
   &+\biggl(\frac{6461241504 \pi  \sigma }{499681}+\frac{256212912 \log (x)}{920465}-\frac{85776 \pi ^2}{119}+\frac{512425824 \gamma
   }{920465}\nonumber\\
   &-\frac{535457980021354132}{92896056978725}+\frac{512425824 \log (3)}{920465}+\frac{1024851648 \log (2)}{920465}\biggr)x^4\nonumber\\
   &+\biggl[\sigma  \biggl(\frac{2474096256 \log (x)}{920465}-\frac{828288 \pi ^2}{119}+\frac{4948192512 \gamma
   }{920465}\nonumber\\
   &-\frac{14231093283367007518}{278688170936175}+\frac{4948192512 \log (3)}{920465}+\frac{9896385024 \log
   (2)}{920465}\biggr)\nonumber\\
   &+\pi  \left(-\frac{1720512 \log (x)}{7735}-\frac{3441024 \gamma
   }{7735}+\frac{276322306496}{147111965}-\frac{3441024 \log (3)}{7735}-\frac{6882048 \log (2)}{7735}\right)\biggr]x^{9/2}\nonumber\\
   &+\biggl[\pi  \sigma  \left(-\frac{13764096 \log (x)}{7735}-\frac{27528192 \gamma }{7735}-\frac{4717720724504}{500180681}-\frac{27528192
   \log (3)}{7735}-\frac{55056384 \log (2)}{7735}\right)\nonumber\\
   &-\frac{8996296160304 \log (x)}{5411045549}+\frac{15059082960 \pi
   ^2}{3497767}-\frac{17992592320608 \gamma
   }{5411045549}+\frac{530642810226274540730842}{15089571015339195375}\nonumber\\
   &-\frac{17992592320608 \log
   (3)}{5411045549}-\frac{35985184641216 \log (2)}{5411045549}\biggr]x^5\nonumber\\
   &+\biggl[\sigma  \biggl(-\frac{77198913489792 \log (x)}{3865032535}+\frac{25844966016 \pi ^2}{499681}-\frac{154397826979584 \gamma
   }{3865032535}\nonumber\\
   &+\frac{303009886036669594783701}{718551000730437875}-\frac{154397826979584 \log
   (3)}{3865032535}-\frac{308795653959168 \log (2)}{3865032535}\biggr)\nonumber\\
   &+\pi  \left(\frac{3074554944 \log
   (x)}{920465}+\frac{6149109888 \gamma }{920465}-\frac{6425495760256249584}{92896056978725}+\frac{6149109888 \log
   (3)}{920465}+\frac{12298219776 \log (2)}{920465}\right)\biggr]x^{11/2}
\end{align}

\begin{align}
\hat{F}_{77}&=1-\frac{319}{17}x+\left(14\pi-\frac{21\sigma}{2}\right)x^{3/2}+\frac{30773287}{214149}x^2+\left(\frac{39119 \sigma }{204}-\frac{4466 \pi }{17}\right)x^{5/2}\nonumber\\
&+\left(-147 \pi  \sigma -\frac{83636 \log (x)}{3315}+\frac{196 \pi ^2}{3}-\frac{167272 \gamma
   }{3315}+\frac{699898468579}{66146343120}-\frac{167272 \log (7)}{3315}-\frac{167272 \log (2)}{3315}\right)x^3\nonumber\\
   &+\left(\frac{430826018 \pi }{214149}-\frac{4813117711 \sigma }{3426384}\right)x^{7/2}\nonumber\\
   &+\biggl(\frac{273833 \pi  \sigma }{102}+\frac{26679884 \log (x)}{56355}-\frac{62524 \pi ^2}{51}+\frac{53359768 \gamma
   }{56355}\nonumber\\
   &-\frac{92269694194596106543}{9828023660769600}+\frac{53359768 \log (7)}{56355}+\frac{53359768 \log (2)}{56355}\biggr)x^4\nonumber\\
   &+\biggl[\sigma  \left(\frac{292726 \log (x)}{1105}-686 \pi ^2+\frac{585452 \gamma }{1105}-\frac{8231923476353}{8819512416}+\frac{585452
   \log (7)}{1105}+\frac{585452 \log (2)}{1105}\right)\nonumber\\
   &+\pi  \left(-\frac{1170904 \log (x)}{3315}-\frac{2341808 \gamma
   }{3315}+\frac{4899289280053}{33073171560}-\frac{2341808 \log (7)}{3315}-\frac{2341808 \log (2)}{3315}\right)\biggr]x^{9/2}\nonumber\\
   &+\biggl(-\frac{33691823977 \pi  \sigma }{1713192}-\frac{2573754631532 \log (x)}{709903935}+\frac{6031564252 \pi
   ^2}{642447}-\frac{5147509263064 \gamma }{709903935}\nonumber\\
   &+\frac{66286435739816234692759}{857365748301348000}-\frac{5147509263064
   \log (7)}{709903935}-\frac{5147509263064 \log (2)}{709903935}\biggr)x^5\nonumber\\
   &+\biggl[\sigma  \biggl(-\frac{817939171 \log (x)}{169065}+\frac{1916831 \pi ^2}{153}-\frac{1635878342 \gamma
   }{169065}\nonumber\\
   &+\frac{2911813721908485052367}{29484070982308800}-\frac{1635878342 \log (7)}{169065}-\frac{1635878342 \log
   (2)}{169065}\biggr)\nonumber\\
   &+\pi  \biggl(\frac{373518376 \log (x)}{56355}+\frac{747036752 \gamma
   }{56355}
   -\frac{645887859362172745801}{4914011830384800}+\frac{747036752 \log (7)}{56355}+\frac{747036752 \log
   (2)}{56355}\biggr)\biggr]x^{11/2}
\end{align}
\end{widetext}

\section{Multipolar 5.5PN EOB relativistic residual amplitudes derived from the new spinning particle on Schwartzschild fluxes results}
\label{app:rholms}

The spin-dependent part of the PN-expanded residual relativistic amplitudes obtained from the fluxes of Sec. \ref{app:fluxes} read

\begin{widetext}
\begin{align}
\rho_{22}^{\sigma}&=-\frac{1}{3} \sigma  x^{3/2}-\frac{2}{63} \sigma  x^{5/2}+\frac{61153\sigma}{31752}x^{7/2}+\sigma  \left(\frac{214 \log (x)}{315}+\frac{428 \gamma }{315}-\frac{14661629}{8731800}+\frac{856 \log (2)}{315}\right)x^{9/2}\nonumber\\
&+\sigma  \left(\frac{428 \log (x)}{6615}+\frac{856 \gamma }{6615}-\frac{90273995723}{88994505600}+\frac{1712 \log
   (2)}{6615}\right)x^{11/2}\\
   \nonumber\\
\rho_{21}^{\sigma}&=\frac{3 \sigma }{4}x^{1/2}+-\frac{61 \sigma }{32}x^{3/2}+-\frac{137839 \sigma }{225792}x^{5/2}+\sigma  \left(-\frac{107 \log (x)}{280}-\frac{107 \gamma }{140}+\frac{8865471731}{3477196800}-\frac{107 \log (2)}{140}\right)x^{7/2}\nonumber\\
&+\sigma  \left(\frac{6527 \log (x)}{6720}+\frac{6527 \gamma }{3360}-\frac{79695561168973}{18226074746880}+\frac{6527 \log
   (2)}{3360}\right)x^{9/2}\nonumber\\
   &+\sigma  \left(\frac{14748773 \log (x)}{47416320}+\frac{14748773 \gamma
   }{23708160}-\frac{3969300389840021}{729042989875200}+\frac{14748773 \log (2)}{23708160}\right)x^{11/2}\\
   \nonumber\\
\rho_{31}^{\sigma}&=\frac{5 \sigma }{6}x^{3/2}-\frac{29 \sigma }{36}x^{5/2}-\frac{129079 \sigma }{42768}x^{7/2}
+\sigma  \left(-\frac{65 \log (x)}{378}-\frac{65 \gamma }{189}-\frac{1842715099}{668697120}-\frac{65 \log (2)}{189}\right)x^{9/2}\nonumber\\
&+\sigma  \left(\frac{377 \log (x)}{2268}+\frac{377 \gamma }{1134}-\frac{1848295232183}{359610451200}+\frac{377 \log
   (2)}{1134}\right)x^{11/2}\\
   \nonumber\\
\rho_{32}^{\sigma}&=\frac{2 \sigma }{3}x^{1/2}-\frac{1613 \sigma }{810}x^{3/2}-\frac{363103 \sigma }{962280}x^{5/2}+\sigma  \left(-\frac{104 \log (x)}{189}-\frac{208 \gamma }{189}+\frac{2565868645267}{496507611600}-\frac{416 \log
   (2)}{189}\right)x^{7/2}\nonumber\\
   &+\sigma  \left(\frac{41938 \log (x)}{25515}+\frac{83876 \gamma }{25515}-\frac{10180707438904003}{1179702085161600}+\frac{167752
   \log (2)}{25515}\right)x^{9/2}\nonumber\\
   &+\sigma  \left(\frac{4720339 \log (x)}{15155910}+\frac{4720339 \gamma
   }{7577955}-\frac{55888392652159222117}{7962989074840800000}+\frac{9440678 \log (2)}{7577955}\right)x^{11/2}\\
\nonumber\\
\rho_{33}^{\sigma}&=-\frac{\sigma }{2}x^{3/2}+\frac{7 \sigma }{20}x^{5/2}+\frac{15247 \sigma }{7920}x^{7/2}+\sigma  \left(\frac{13 \log (x)}{14}+\frac{13 \gamma }{7}-\frac{2131142749}{454053600}+\frac{13 \log (3)}{7}+\frac{13 \log
   (2)}{7}\right)x^{9/2}\nonumber\\
   &+\sigma  \left(-\frac{13 \log (x)}{20}-\frac{13 \gamma }{10}+\frac{97710564283}{28540512000}-\frac{13 \log (3)}{10}-\frac{13 \log
   (2)}{10}\right)x^{11/2}\\
\nonumber\\
\rho_{41}^{\sigma}&=\frac{5 \sigma }{8}x^{1/2}-\frac{1187 \sigma }{704}x^{3/2}-\frac{29908639 \sigma }{169128960}x^{5/2}+\sigma  \left(-\frac{1571 \log (x)}{22176}-\frac{1571 \gamma }{11088}+\frac{1310663510743}{2812952862720}-\frac{1571 \log
   (2)}{11088}\right)x^{7/2}\nonumber\\
   &+\sigma  \left(\frac{1864777 \log (x)}{9757440}+\frac{1864777 \gamma
   }{4878720}-\frac{30823552696653631}{19893202645155840}+\frac{1864777 \log (2)}{4878720}\right)x^{9/2}\nonumber\\
   &+\sigma  \left(\frac{46986471869 \log (x)}{2344127385600}+\frac{46986471869 \gamma
   }{1172063692800}-\frac{26722270856299940280173}{17462253281917796352000}+\frac{46986471869 \log (2)}{1172063692800}\right)x^{11/2}\\
\nonumber\\
\rho_{42}^{\sigma}&=\frac{3 \sigma }{5}x^{3/2}-\frac{403 \sigma }{550}x^{5/2}-\frac{83093993 \sigma }{44044000}x^{7/2}
+\sigma  \left(-\frac{1571 \log (x)}{5775}-\frac{3142 \gamma }{5775}-\frac{10890814807}{57229672500}-\frac{6284 \log
   (2)}{5775}\right)x^{9/2}\nonumber\\
   &+\sigma  \left(\frac{633113 \log (x)}{1905750}+\frac{633113 \gamma
   }{952875}-\frac{61976095519654871}{13876396934400000}+\frac{1266226 \log (2)}{952875}\right)x^{11/2}\\
\nonumber\\
\rho_{43}^{\sigma}&=\frac{5 \sigma }{8}x^{1/2}-\frac{1443 \sigma }{704}x^{3/2}-\frac{11933877 \sigma }{56376320}x^{5/2}+\sigma  \left(-\frac{1571 \log (x)}{2464}-\frac{1571 \gamma }{1232}+\frac{423522768659}{62510063616}-\frac{1571 \log
   (3)}{1232}-\frac{1571 \log (2)}{1232}\right)x^{7/2}\nonumber\\
   &+\sigma  \left(\frac{2266953 \log (x)}{1084160}+\frac{2266953 \gamma
   }{542080}-\frac{9259292611008173}{736785283153920}+\frac{2266953 \log (3)}{542080}+\frac{2266953 \log (2)}{542080}\right)x^{9/2}\nonumber\\
   &+\sigma  \left(\frac{18748120767 \log (x)}{86819532800}+\frac{18748120767 \gamma
   }{43409766400}-\frac{7723130549892884353}{975490379415552000}+\frac{18748120767 \log (3)}{43409766400}+\frac{18748120767 \log
   (2)}{43409766400}\right)x^{11/2}
  \end{align}
  \begin{align}
\rho_{44}^{\sigma}&=-\frac{3 \sigma }{5}x^{3/2}+\frac{146 \sigma }{275}x^{5/2}+\frac{87298781 \sigma }{44044000}x^{7/2}+\sigma  \left(\frac{6284 \log (x)}{5775}+\frac{12568 \gamma }{5775}-\frac{6129600606719}{915674760000}+\frac{12568 \log
   (2)}{1925}\right)x^{9/2}\nonumber\\
   &+\sigma  \left(-\frac{917464 \log (x)}{952875}-\frac{1834928 \gamma
   }{952875}+\frac{576017315427752921}{97134778540800000}-\frac{1834928 \log (2)}{317625}\right)x^{11/2}\\
\nonumber\\
\rho_{51}^{\sigma}&=\frac{14 \sigma }{15}x^{3/2}-\frac{4034 \sigma }{2925}x^{5/2}-\frac{5744501 \sigma }{2281500}x^{7/2}+\sigma  \left(-\frac{10822 \log (x)}{160875}-\frac{21644 \gamma }{160875}-\frac{16213030784959}{6406007107500}-\frac{21644 \log
   (2)}{160875}\right)x^{9/2}\nonumber\\
   &+\sigma  \left(\frac{3118282 \log (x)}{31370625}+\frac{6236564 \gamma
   }{31370625}-\frac{2272557953476553377}{531647341865640000}+\frac{6236564 \log (2)}{31370625}\right)x^{11/2}\\
\nonumber\\
\rho_{52}^{\sigma}&=\frac{3 \sigma }{5}x^{1/2}-\frac{39413 \sigma }{22750}x^{3/2}-\frac{5349527 \sigma }{47775000}x^{5/2}+\sigma  \left(-\frac{3092 \log (x)}{17875}-\frac{6184 \gamma }{17875}+\frac{4330589662282661}{2491224986250000}-\frac{12368 \log
   (2)}{17875}\right)x^{7/2}\nonumber\\
   &+\sigma  \left(\frac{5539318 \log (x)}{11090625}+\frac{11078636 \gamma
   }{11090625}-\frac{796581754627562940017}{234945163703250000000}+\frac{22157272 \log (2)}{11090625}\right)x^{9/2}\nonumber\\
   &+\sigma  \left(\frac{4135184371 \log (x)}{128096718750}+\frac{4135184371 \gamma
   }{64048359375}-\frac{6099355932458608484088427}{3094475116670437500000000}+\frac{8270368742 \log (2)}{64048359375}\right)x^{11/2}\\
\nonumber\\
\rho_{53}^{\sigma}&=\frac{2 \sigma }{5}x^{3/2}-\frac{1182 \sigma }{2275}x^{5/2}-\frac{300929 \sigma }{253500}x^{7/2}+\sigma  \left(-\frac{4638 \log (x)}{17875}-\frac{9276 \gamma }{17875}+\frac{181745950091}{237259522500}-\frac{9276 \log
   (3)}{17875}-\frac{9276 \log (2)}{17875}\right)x^{9/2}\nonumber\\
   &+\sigma  \left(\frac{2741058 \log (x)}{8133125}+\frac{5482116 \gamma
   }{8133125}-\frac{844630470658099711}{229724160065400000}+\frac{5482116 \log (3)}{8133125}+\frac{5482116 \log
   (2)}{8133125}\right)x^{11/2}\\
\nonumber\\
\rho_{54}^{\sigma}&=\frac{3 \sigma }{5}x^{1/2}-\frac{47633 \sigma }{22750}x^{3/2}-\frac{4565087 \sigma }{47775000}x^{5/2}+\sigma  \left(-\frac{12368 \log (x)}{17875}-\frac{24736 \gamma }{17875}+\frac{19650484510197881}{2491224986250000}-\frac{74208
   \log (2)}{17875}\right)x^{7/2}\nonumber\\
   &+\sigma  \left(\frac{294562472 \log (x)}{121996875}+\frac{589124944 \gamma
   }{121996875}-\frac{41115919165624337072107}{2584396800735750000000}+\frac{589124944 \log (2)}{40665625}\right)x^{9/2}\nonumber\\
   &+\sigma  \left(\frac{7057624502 \log (x)}{64048359375}+\frac{14115249004 \gamma
   }{64048359375}-\frac{488557785750184868183083213}{58795027216738312500000000}+\frac{14115249004 \log (2)}{21349453125}\right)x^{11/2}\\
\nonumber\\
\rho_{55}^{\sigma}&=-\frac{2 \sigma }{3}x^{3/2}+\frac{514 \sigma }{819}x^{5/2}+\frac{929219 \sigma }{456300}x^{7/2}+\sigma  \left(\frac{1546 \log (x)}{1287}+\frac{3092 \gamma }{1287}-\frac{73675207595933}{8968409950500}+\frac{3092 \log
   (5)}{1287}+\frac{3092 \log (2)}{1287}\right)x^{9/2}\nonumber\\
   &+\sigma  \left(-\frac{397322 \log (x)}{351351}-\frac{794644 \gamma
   }{351351}+\frac{27897212127015034469}{3721531393059480000}-\frac{794644 \log (5)}{351351}-\frac{794644 \log
   (2)}{351351}\right)x^{11/2}\\
\nonumber\\
\rho_{61}^{\sigma}&=\frac{7 \sigma }{12}x^{1/2}-\frac{2789 \sigma }{1728}x^{3/2}-\frac{50175395 \sigma }{651442176}x^{5/2}+\sigma  \left(-\frac{901 \log (x)}{30888}-\frac{901 \gamma }{15444}+\frac{6036361379999233}{49700712452751360}-\frac{901 \log
   (2)}{15444}\right)x^{7/2}\nonumber\\
   &+\sigma  \left(\frac{2512889 \log (x)}{31135104}+\frac{2512889 \gamma
   }{15567552}-\frac{331771917464745022019}{347620983098100940800}+\frac{2512889 \log (2)}{15567552}\right)x^{9/2}\nonumber\\
   &+\sigma  \left(\frac{2659295935 \log (x)}{690452066304}+\frac{2659295935 \gamma
   }{345226033152}-\frac{246291614146386340935625099}{228628051899310243263283200}+\frac{2659295935 \log
   (2)}{345226033152}\right)x^{11/2}\\
\nonumber\\
\rho_{62}^{\sigma}&=\frac{17 \sigma }{21}x^{3/2}-\frac{1819 \sigma }{1470}x^{5/2}-\frac{574085 \sigma }{271656}x^{7/2}+\sigma  \left(-\frac{30634 \log (x)}{189189}-\frac{61268 \gamma }{189189}-\frac{16770732413683}{15112379482200}-\frac{122536 \log
   (2)}{189189}\right)x^{9/2}\nonumber\\
   &+\sigma  \left(\frac{1638919 \log (x)}{6621615}+\frac{3277838 \gamma
   }{6621615}-\frac{2006995067160111247}{431609558011632000}+\frac{6555676 \log (2)}{6621615}\right)x^{11/2}
\end{align}

\begin{align}
\rho_{63}^{\sigma}&=\frac{7 \sigma }{12}x^{1/2}-\frac{3085 \sigma }{1728}x^{3/2}-\frac{35390867 \sigma }{651442176}x^{5/2}\nonumber\\
&+\sigma  \left(-\frac{901 \log (x)}{3432}-\frac{901 \gamma }{1716}+\frac{145753300581274073}{49700712452751360}-\frac{901 \log
   (3)}{1716}-\frac{901 \log (2)}{1716}\right)x^{7/2}\nonumber\\
   &+\sigma  \left(\frac{2779585 \log (x)}{3459456}+\frac{2779585 \gamma
   }{1729728}-\frac{94487187434422183760819}{17033428171806946099200}+\frac{2779585 \log (3)}{1729728}+\frac{2779585 \log
   (2)}{1729728}\right)x^{9/2}\nonumber\\
   &+\sigma  \biggl(\frac{1875715951 \log (x)}{76716896256}+\frac{1875715951 \gamma
   }{38358448128}\nonumber\\
   &-\frac{10689835353210782026889725049}{4343932986086894622002380800}+\frac{1875715951 \log
   (3)}{38358448128}+\frac{1875715951 \log (2)}{38358448128}\biggr)x^{11/2}\\
\nonumber\\
\rho_{64}^{\sigma}&=\frac{5 \sigma }{21}x^{3/2}-\frac{46 \sigma }{147}x^{5/2}-\frac{21617 \sigma }{31416}x^{7/2}+\sigma  \left(-\frac{36040 \log (x)}{189189}-\frac{72080 \gamma }{189189}+\frac{5008174606007}{5709121137720}-\frac{72080 \log
   (2)}{63063}\right)x^{9/2}\nonumber\\
   &+\sigma  \left(\frac{331568 \log (x)}{1324323}+\frac{663136 \gamma
   }{1324323}-\frac{3751539084367382743}{1467472497239548800}+\frac{663136 \log (2)}{441441}\right)x^{11/2}\\
\nonumber\\
\rho_{65}^{\sigma}&=\frac{7 \sigma }{12}x^{1/2}-\frac{3677 \sigma }{1728}x^{3/2}-\frac{8157683 \sigma }{651442176}x^{5/2}\nonumber\\
&+\sigma  \left(-\frac{22525 \log (x)}{30888}-\frac{22525 \gamma }{15444}+\frac{86699363941370165}{9940142490550272}-\frac{22525
   \log (5)}{15444}-\frac{22525 \log (2)}{15444}\right)x^{7/2}\nonumber\\
   &+\sigma  \left(\frac{82824425 \log (x)}{31135104}+\frac{82824425 \gamma
   }{15567552}-\frac{1160763233925296604961}{61939738806570713088}+\frac{82824425 \log (5)}{15567552}+\frac{82824425 \log
   (2)}{15567552}\right)x^{9/2}\nonumber\\
   &+\sigma  \biggl(\frac{10808929975 \log (x)}{690452066304}+\frac{10808929975 \gamma
   }{345226033152}\nonumber\\
   &-\frac{1471477155641855359574205745}{173757319443475784880095232}+\frac{10808929975 \log
   (5)}{345226033152}+\frac{10808929975 \log (2)}{345226033152}\biggr)x^{11/2}\\
\nonumber\\
\rho_{66}^{\sigma}&=-\frac{5 \sigma }{7}x^{3/2}+\frac{67 \sigma }{98}x^{5/2}+\frac{9605147 \sigma }{4618152}x^{7/2}\nonumber\\
&+\sigma  \left(\frac{9010 \log (x)}{7007}+\frac{18020 \gamma }{7007}-\frac{9897759433831}{1048614086520}+\frac{18020 \log
   (3)}{7007}+\frac{36040 \log (2)}{7007}\right)x^{9/2}\nonumber\\
   &+\sigma  \left(-\frac{60367 \log (x)}{49049}-\frac{120734 \gamma
   }{49049}+\frac{1394113205906052901}{163052499693283200}-\frac{120734 \log (3)}{49049}-\frac{241468 \log (2)}{49049}\right)x^{11/2}\\
\nonumber\\
\rho_{71}^{\sigma}&=\frac{27 \sigma }{28}x^{3/2}-\frac{10617 \sigma }{6664}x^{5/2}-\frac{1800329721 \sigma }{783499808}x^{7/2}+\sigma  \left(-\frac{26883 \log (x)}{758030}-\frac{26883 \gamma }{379015}-\frac{9712009819849013}{3999844869820800}-\frac{26883
   \log (2)}{379015}\right)x^{9/2}\nonumber\\
   &+\sigma  \left(\frac{10570993 \log (x)}{180411140}+\frac{10570993 \gamma
   }{90205570}-\frac{61582104147771940639}{16457454020990062080}+\frac{10570993 \log (2)}{90205570}\right)x^{11/2}\\
\nonumber\\
\rho_{72}^{\sigma}&=\frac{4 \sigma }{7}x^{1/2}-\frac{57473 \sigma }{34986}x^{3/2}-\frac{10380062389 \sigma }{259142561496}x^{5/2}\nonumber\\
&+\sigma  \left(-\frac{95584 \log (x)}{1137045}-\frac{191168 \gamma
   }{1137045}+\frac{34452124473358455517}{41672883756836732400}-\frac{382336 \log (2)}{1137045}\right)x^{7/2}\nonumber\\
   &+\sigma  \left(\frac{52822108 \log (x)}{218575035}+\frac{105644216 \gamma
   }{218575035}-\frac{241527439302784910158781377}{121358438544379713319843200}+\frac{211288432 \log (2)}{218575035}\right)x^{9/2}\nonumber\\
   &+\sigma  \biggl(\frac{31005246355943 \log (x)}{5261727747075345}+\frac{62010492711886 \gamma
   }{5261727747075345}\nonumber\\
   &-\frac{143944886406436413621998058101657}{118277147789737912398652381152000}+\frac{124020985423772 \log
   (2)}{5261727747075345}\biggr)x^{11/2}
\end{align}

\begin{align}
\rho_{73}^{\sigma}&=\frac{19 \sigma }{28}x^{3/2}-\frac{20843 \sigma }{19992}x^{5/2}-\frac{217718161 \sigma }{123710496}x^{7/2}\nonumber\\
&+\sigma  \left(-\frac{170259 \log (x)}{758030}-\frac{170259 \gamma }{379015}-\frac{12906790690613}{631554453129600}-\frac{170259
   \log (3)}{379015}-\frac{170259 \log (2)}{379015}\right)x^{9/2}\nonumber\\
   &+\sigma  \left(\frac{62258041 \log (x)}{180411140}+\frac{62258041 \gamma
   }{90205570}-\frac{1918909190748718903547}{394112714713183065600}+\frac{62258041 \log (3)}{90205570}+\frac{62258041 \log
   (2)}{90205570}\right)x^{11/2}\\
\nonumber\\
\rho_{74}^{\sigma}&=\frac{4 \sigma }{7}x^{1/2}-\frac{64193 \sigma }{34986}x^{3/2}-\frac{1192699477 \sigma }{259142561496}x^{5/2}\nonumber\\
&+\sigma  \left(-\frac{382336 \log (x)}{1137045}-\frac{764672 \gamma
   }{1137045}+\frac{165797414938111965193}{41672883756836732400}-\frac{764672 \log (2)}{379015}\right)x^{7/2}\nonumber\\
   &+\sigma  \left(\frac{3067911856 \log (x)}{2841475455}+\frac{6135823712 \gamma
   }{2841475455}-\frac{941107749182924698786607473}{121358438544379713319843200}+\frac{6135823712 \log (2)}{947158485}\right)x^{9/2}\nonumber\\
   &+\sigma  \biggl(\frac{14250373351196 \log (x)}{5261727747075345}+\frac{28500746702392 \gamma
   }{5261727747075345}\nonumber\\
   &-\frac{343849685570966735940482698960253}{118277147789737912398652381152000}+\frac{28500746702392 \log
   (2)}{1753909249025115}\biggr)x^{11/2}\\
\nonumber\\
\rho_{75}^{\sigma}&=\frac{3 \sigma }{28}x^{3/2}-\frac{2795 \sigma }{19992}x^{5/2}-\frac{717495539 \sigma }{2350499424}x^{7/2}\nonumber\\
&+\sigma  \left(-\frac{14935 \log (x)}{151606}-\frac{14935 \gamma }{75803}+\frac{792909572981707}{1439944153135488}-\frac{14935
   \log (5)}{75803}-\frac{14935 \log (2)}{75803}\right)x^{9/2}\nonumber\\
   &+\sigma  \left(\frac{3211025 \log (x)}{24980004}+\frac{3211025 \gamma
   }{12490002}-\frac{3461511047071262080289}{2695730968638172168704}+\frac{3211025 \log (5)}{12490002}+\frac{3211025 \log
   (2)}{12490002}\right)x^{11/2}\\
\nonumber\\
\rho_{76}^{\sigma}&=\frac{4 \sigma }{7}x^{1/2}-\frac{25131 \sigma }{11662}x^{3/2}+\frac{466355089 \sigma }{9597872648}x^{5/2}\nonumber\\
&+\sigma  \left(-\frac{286752 \log (x)}{379015}-\frac{573504 \gamma
   }{379015}+\frac{1608805429614769871}{171493348793566800}-\frac{573504 \log (3)}{379015}-\frac{1147008 \log (2)}{379015}\right)x^{7/2}\nonumber\\
   &+\sigma  \left(\frac{900795564 \log (x)}{315719495}+\frac{1801591128 \gamma
   }{315719495}-\frac{3523335266452814138505897}{166472480856487946940800}+\frac{1801591128 \log (3)}{315719495}+\frac{3603182256
   \log (2)}{315719495}\right)x^{9/2}\nonumber\\
   &+\sigma  \biggl(-\frac{4179007952529 \log (x)}{64959601815745}-\frac{8358015905058 \gamma
   }{64959601815745}\nonumber\\
   &-\frac{460728741621406041051059518219}{54081914855847239322657696000}-\frac{8358015905058 \log
   (3)}{64959601815745}-\frac{16716031810116 \log (2)}{64959601815745}\biggr)x^{11/2}\\
\nonumber\\
\rho_{77}^{\sigma}&=-\frac{3 \sigma }{4}x^{3/2}+\frac{293 \sigma }{408}x^{5/2}+\frac{709636883 \sigma }{335785632}x^{7/2}\nonumber\\
&+\sigma  \left(\frac{2987 \log (x)}{2210}+\frac{2987 \gamma }{1105}-\frac{53784383538559963}{5142657689769600}+\frac{2987 \log
   (7)}{1105}+\frac{2987 \log (2)}{1105}\right)x^{9/2}\nonumber\\
   &+\sigma  \left(-\frac{875191 \log (x)}{676260}-\frac{875191 \gamma
   }{338130}+\frac{89536447352345626970969}{9627610602279186316800}-\frac{875191 \log (7)}{338130}-\frac{875191 \log
   (2)}{338130}\right)x^{11/2}
\end{align}

\end{widetext}

\bibliographystyle{apsrev4-1}
\bibliography{refs}

\end{document}